\documentclass[pdftex]{pucthesis}	

\usepackage{graphicx}
\usepackage{float}
\floatstyle{boxed}
\restylefloat{figure}
\usepackage{subfigure}
\usepackage{tkz-euclide}
\usepackage{color}

\usepackage{amsmath}
\usepackage{amsfonts}
\usepackage{amssymb}
\usepackage{braket}
\usepackage{slashed}
\usepackage{nicefrac}
\usepackage{dsfont}

\usepackage{times}

\usepackage{booktabs}
\usepackage{multirow}

\usepackage{etoolbox}
\apptocmd{\sloppy}{\hbadness 10000\relax}{}{}

\usepackage{ragged2e} 
\usepackage{pdfpages}

\usepackage[colorlinks=true,linkcolor=black,urlcolor=magenta,citecolor=blue,pagebackref=true]{hyperref}

\newcommand{\bea}{\begin{eqnarray}}
\newcommand{\eea}{\end{eqnarray}}
\newcommand{\be}{\begin{equation}}
\newcommand{\ee}{\end{equation}}

\DeclareFontFamily{U}{calligra}{}
\DeclareFontShape{U}{calligra}{m}{n}{<->callig15}{}

\begin{document}

\date{\today} 
\version{1} 

\title[Delving into the Phenomenology of Very Special Relativity:\\From Subatomic Particles to Binary Stars]
   {\bf Delving into the Phenomenology of Very Special Relativity:\\From Subatomic Particles to Binary Stars}       
\author[Alessandro Santoni]{Alessandro Santoni}

\address{Facultad de F\'isica\\
                   Pontificia Universidad Cat\'olica de Chile\\ 
                   Vicu\~na Mackenna 4860\\
                  Santiago, Chile\\
                  {\it Tel.\/} : 56 (2) 354-2000}
\email{asantoni@uc.cl}

\facultyto    {the Faculty of Physics}
\faculty                             {Faculty of Physics}
\degree                  {Doctor of Philosophy (Ph.D.) in Physics}  
\advisor                            {Enrique Muñoz\\
Benjamin Koch}
\committeememberA {Jorge Alfaro}
\guestmemberA            {Marco Aurelio Díaz}
\ogrsmember                 { }
\subject                            {Theoretical Physics}
\date                                 {8 August 2024}
\copyrightname             {Alessandro Santoni}
\copyrightyear               {2024}
\dedication                      {\raggedleft A modo mio\\ 
$\,\,\,\;\;\;\;\,\,\,\;\;\;$quel che sono l'ho voluto io}


\NoChapterPageNumber
\pagenumbering{roman}
\maketitle






\phantomsection \label{acknowledgements} 
\chapter*{ACKNOWLEDGEMENTS}           
Writing the “acknowledgement” section in a thesis is often very challenging, partly because Ph.D. journeys usually last so long that too many people enter and leave our lives, and partly because we are probably not used enough to expressing gratitude and acknowledging the importance of the presence of others in our lives. Nevertheless, in the following lines, I will try my best to do it while keeping it short.\\
First of all, I would like to thank my Ph.D. advisors, professors Enrique Muñoz and Benjamin Koch, who backed me throughout the whole Ph.D. duration, teaching me many precious lessons and guiding my creative mind into fruitful directions. I am also very grateful to professor Jorge Alfaro, with whom I moved my first steps as an independent researcher, for his kindness and for all the interesting discussions we had during these years. Another thanks goes to the UC and TU Wien community and staff, which helped me keep going with all the bureaucratic stuff we all try to escape from. In particular, I should thank the UC coordinator of doctoral studies, Emilio Bravo, for his incredible commitment.\\
In seconda istanza, vorrei ringraziare tutte le persone che dall'Italia hanno sempre fatto il tifo per me, come la mia famiglia e i miei amici. Durante questi anni di lontananza, infatti, abbiamo affrontato eventi globali estremamente impattanti, tra i quali la pandemia di Covid-19, che hanno cercato di allontanarci ancor più di quanto già non fossimo geograficamente. Tuttavia, esattamente come accade a una casa durante un terremoto, dove quello che conta non è la bellezza della sua facciata ma la solidità delle sue fondamenta, è proprio in queste situazioni estreme che le relazioni vengono messe alla prova e si rivelano per quello che sono davvero. Un ringraziamento unico va ai “Marco al quadrato” per aver alleggerito molte delle mie serate, e ad Albert per rallegrare anche i giorni più bui.\\
Me gustaría agradecer también a todos los compañeros que encontré a lo largo de este camino, como Sebastián, Felipe y Rodrigo, juntos a la asombrosa cantidad de otras personas que me han apoyado en Chile. Entre ellas, un pensamiento especial va a Cecilia, la mamá de mi esposa, que me acogió en su casa desde el inicio y que, dolorosamente, tuvimos que despedir hace algunos años.\\
A la misma manera de la cual se deja el postre, o el mejor plato, para el final de una comida, quiero agradecer finalmente a mi esposa, el amor de mi vida, mi mejor amiga, mi compañera de viajes y cómplice de mil aventuras: Ignacia. En los últimos cuatro años, he vivido con ella algunos de los momentos más felices de mi vida, como nuestro matrimonio, y al mismo tiempo algunos de los más difíciles, como el cierre de este doctorado. Su apoyo a lo largo de este periodo fue inestimable, tanto que no podría encontrar palabras para lograr encerrarlo en unas pocas frases. Ella fue y seguirá siendo mi única constante en un mundo lleno de variables, mi estrella polar en el cielo nocturno, y por eso nunca terminaré de agradecerla.\\
I hope I did not forgot anyone while writing this text and if I did, keep in mind that it was not intentional at all.

\cleardoublepage



\pdfbookmark{\contentsname}{toc}
\tableofcontents
\phantomsection \label{listoffigures}
\listoffigures




\phantomsection \label{abstract}
\chapter*{ABSTRACT}
\hfill\begin{minipage}{14.5cm}
\begin{quote}
\justifying
In this thesis, we investigate the implications of Lorentz-violating (LV) theories, focusing on Very Special Relativity (VSR) and its phenomenological consequences. Initially presented as an alternative mechanism for neutrino masses, VSR has since become a significant part of the general LV framework, distinguished by its unique group structure and non-local operators. After a comprehensive introduction to the principles of LV and VSR, we present the corresponding modifications to the Dirac equation. A significant part of the thesis is dedicated to the development of a Hamiltonian formalism within the VSR context, addressing its inherent non-localities. This approach is further extended to the non-relativistic limit, connecting it to the conventional Schrödinger picture. We then set upper bounds on the VSR parameters by examining its corrections to a wide range of physical systems and scenarios, such as Landau levels of charged particles, the $\mathsf g-$factor of electrons, the energy spectrum of ultracold neutrons in Earth's gravitational field, and the gravitational emission from binary stars. The latter analysis led us to the construction of a VSR field theory for spin-2 fields in flat space, which was surprisingly found to accommodate a gauge-invariant graviton mass. Through this comprehensive study, we bridged theoretical predictions with experimental data, paving the way for future explorations in Lorentz-violating theories and highlighting their potential to address unresolved questions in modern physics.

\end{quote}
\end{minipage}

\vfill
\noindent {\bf Keywords}: \textit{Very Special Relativity}, \textit{Lorentz Violation}, \textit{Quantum Gravity}, \textit{Effective Field Theory}, \textit{Massive Gravity}, \textit{Non-Localities}, \textit{Phenomenology}

\cleardoublepage




\pagenumbering{arabic}

\chapter[INTRODUCTION]{Introduction} \label{intro}
%
If we were to use a single word to represent most of the theoretical advances in modern physics, one of the first coming to mind would surely be “Symmetry”. During the last century, in fact, symmetry served as a guiding principle for formulating and understanding physical laws, while simplifying complex systems by revealing underlying invariances. Moreover, symmetries lead to conserved quantities via Noether's theorem and have been crucial in the development of fundamental theories such as General Relativity (GR) and the Standard Model of Particle Physics (SM). They also provide predictive power, allowing for the categorization of particles as irreducible group representations, and the identification of interactions thanks to the concept of “gauge invariance”. Even in exploring extensions beyond established frameworks, symmetry arguments remain pivotal for uncovering new physical phenomena and guiding theoretical progress. Rephrasing Archimedes, who once said: “Give me a lever long enough and a place to stand, and I will move the world”, a modern theoretical physicist might say: “Show me the symmetries, and I will describe you how Nature works”. \\
When discussing symmetries, we cannot overlook talking about Lorentz Invariance (LI), one of the main pillars of modern physics. Since its introduction at the beginning of the twentieth century, in fact, LI has made its way into (almost) every fundamental theory of Nature that we have written down. That is the case because LI is among the most fundamental backbones of Quantum Field Theory (QFT), the mathematical framework through which Particle Physics is formulated nowadays. At the core of LI lies the realization of the existence of a class of equivalent observers, namely “inertial observers”, which experience the physical world in the same way. This concept, which goes by the name of “Relativity principle”, dates back to Galileo Galilei \cite{Galilei+1962}. Much ahead of his time, Galileo was the first to understand the importance of this idea, which would become a staple of Classical Mechanics, along with the concept of spatial isotropy \cite{mamone2011spatial}. Later on, combining both the Relativity Principle and the constancy of the speed of light (in vacuum), Albert Einstein gave birth to the theory of Special Relativity (SR), where the Lorentz group becomes one of the main actors \cite{Einstein:1905ve}. With that revolution, he laid down half of the theoretical foundations for the construction of the QFT framework.

\section{Particle and Coordinate Lorentz Transformations}

Now, a crucial distinction must be made to avoid confusion. When talking about Lorentz Transformations (LT), we should distinguish between Particle Lorentz Transformations (PLT) and Observer Lorentz Transformations (OLT). The former are related to transformations that link different physical systems from the point of view of the same observer, while the latter connect different observers studying the same physical system and are therefore just a subset of coordinate transformations.\footnote{In flat space, coordinate and frame transformations confuse into each other. That is because for flat spaces the local tangent structure coincides globally with the manifold itself. As a consequence, in this context, the positions of points are regarded, for all purposes, as vectors. Then, coordinate transformations naturally induce frame transformations, and vice versa. For curved manifolds, the same is no longer generally true.} 
Symmetry under OLT is a specific realization of the “Covariance Principle”, according to which the form of physics laws should not depend on the coordinate system used to describe them. However, to some extent, this is more a property of the underlying mathematical formulation of the theory rather than its physical content \cite{Angel1973}. Symmetry under PLT, instead, is related to the Relativity Principle and really encodes profound knowledge on the physical reality, such as conserved quantities. Indeed, an equivalent viewpoint is to regard PLT as spacetime symmetries, defining the allowed background structure on top of which phenomena occur. \\
Let us further clarify the nature of the above transformations by discussing the example of a charged particle living in $2D-$space. In the following, we adopt Einstein convention for repeated indices and represent the observer frame by the orthonormal basis vectors $\hat e_{(i)}$.
\begin{itemize}
    \item \textbf{No background}: We start by considering the motion of the particle “in vacuum” from the perspective of an observer $O$. Here, the particle is just free streaming. A particle rotation $R (\theta)$ will modify the direction of its magnetic moment $ \vec \mu $ 
    \begin{equation}
        \vec \mu= v^i \hat e_{(i)} \,\,\to\,\, \vec \mu_p = \mu_p^{i} \hat  e_{(i)} = {R(\theta)^i}_j \, \mu^j \hat e_{(i)} \neq \vec \mu \,.
    \end{equation}
    From there we can perform an observer rotation with the same angle $\theta $ to arrive in a new frame $O'$, defined by the basis vectors $\hat e'_{(j)} =  \hat e_{(i)} {R(\theta)^i}_j $, where the new components of the vector $\vec \mu_p$ are identical to the ones of the original vector
    \begin{equation}
       \vec \mu_p = \mu'^{\,i}_p \,\hat e'_{(i)}  \,\,\to\,\,  \mu'^{\,i}_p = \mu^i\,.
    \end{equation}
    Thus, the effect of the observer rotation on the magnetic moment components was to rotate them backwards
    \begin{equation}
        \vec \mu_p = \mu^{i}_p \,\hat e_{(i)} = \mu^{i}_p  \, \hat e'_{(j)} {R^{-1}(\theta)^j }_i \,\, \to \,\, \mu'^{\,j}_p = {R(-\theta)^j}_i \, \mu^{i}_p \,.
    \end{equation}
    In other words, for dynamic tensor components, particle transformations are equivalent to observer transformation “in the opposite direction”
    \be
    (PLT) = (OLT)^{-1} \,.
    \ee
    Since the spatial metric $\delta_{ij}$ is invariant under observer rotations $ \delta_{ij}= \hat e'_{(i)} \cdot \hat e'_{(j)} =  \hat e_{(i)} \cdot \hat e_{(j)}  $, scalar products, which are trivially invariant under observer transformations, are automatically invariant under particle rotations
    \begin{equation}
        \vec \mu_p \cdot \vec \mu_p = \mu^i_p \mu^j_p \delta_{ij} = \mu^k \mu^l {R(\theta)^i}_k {R(\theta)^j}_l\delta_{ij} = \mu^k \mu^l \delta_{kl} = \vec \mu \cdot \vec \mu \,.
    \end{equation}
    That is the reason why in the SR books we usually do not find any distinction among the two kinds of transformations. In fact, scalar products in SR are carried out with the Minkoswki metric $\eta_{\mu\nu}$, which is invariant under OLT and does not transform at all under PLT. Hence, scalar products involving only dynamical quantities are automatically PLT-invariant, making the spacetime symmetries of SR manifest. This also means that normally every PLT-symmetric theory can easily be expressed in a form which is manifestly OLT-covariant by writing it in function of OLT-covariant objects,\footnote{A good example of a PLT-symmetric theory which was only later formulated in a manifestly Lorentz-covariant fashion is electromagnetism.} while the inverse is not generally true: OLT-covariance does not ensure PLT-invariance, as we will see.
    \begin{figure}[ht]
	\begin{center}
	\includegraphics[width=0.5\textwidth]{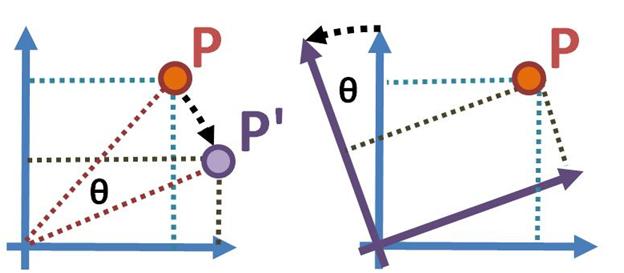}
	\caption[Particle vs Observer transformation]{Sketch of the difference between a Particle rotation (left) of a point $P$ and an Observer rotation (right) in a bi-dimensional space. \hyperlink{https://en.wikipedia.org/wiki/File:PassiveActive.JPG}{[Credit]}}
	\end{center}
    \end{figure}
    \vspace{1mm}
    \item \textbf{With background}: Now, suppose to be surrounded by some directional background, such as an external magnetic field $\vec B$ , which by definition is invariant under particle rotations, because we have no direct “control” over it. In this case, the above equivalence would be no longer valid 
    \be
    (PLT) \neq (OLT)^{-1}\,.
    \ee
    That is because by rotating just the magnetic moment of the particle, we change its orientation with respect to the magnetic field, meaning that we are really describing a different physical configuration with a different magnetic dipole energy $U^B$ 
    \begin{equation}
       U^B_p = -\vec \mu_p \cdot \vec B = - {R(\theta )^i}_k \mu^k B^j \delta_{ij} \neq - \vec \mu \cdot \vec B = U^B \,.
    \end{equation}
    However, the magnetic dipole energy would still be invariant under observer rotations since it has been expressed through a scalar product. Therefore, in the presence directional backgrounds, we cannot “undo” the physical effects of the PLT by an equivalent OLT. Their combined effect would, in fact, lead to a new kind of transformation, Background Lorentz Transformations (BLT), which in practice relate inequivalent backgrounds configurations
    \be
    (PLT)  (OLT) = (BLT)^{-1} \,.
    \ee
    For practical purposes, BLTs are  equivalent to situations in which we rigidly rotate our lab together with our experiment and instrumentations. This perspective is at the core of the search for sidereal LV variations due to Earth's rotation. Then, the existence of spacetime preferred directions essentially shrink the group of spacetime symmetries.
\end{itemize}
In the rest of this thesis, when generally talking about LV, independently of its origin, we will be referring to PLT symmetry breaking. In such a way, we focus on parametrizing the real physical differences with respect to PLT invariant theories. At the same time we avoid conceptual problems eventually arising with OLT violations \cite{PhysRevD.93.045011}, which may not directly translate into observable consequences. \\
Another important point to stress here is that PLT violation can be associated not only with more fundamental theories violating the SR principle, and thus reducing spacetime symmetries (more global point of view), but also with the presence of unknown and hidden degrees of freedom not contemplated in our current models of Nature yet (more local point of view). In this sense, PLT violation can be thought of as just another way to parametrize our ignorance, since LV might simply emerge as an apparent property of the particular physical scenario we are trying to describe. The latter philosophy is easily embraced, for example, when spontaneously breaking LI as we will better understand in the following section.

\section{Lorentz-Violating Models}

In the last decades, the literature on LV extensions of quantum field theories has raised significantly, both in the theoretical and experimental sectors. The quest for quantum gravity is likely one of the primary drivers behind this increased interest. In fact, trying to deal with spacetime fluctuations and/or discretized spacetimes often leads to certain degrees of LV \cite{collins2004lorentz} as a result, even in the same SM \cite{PhysRevLett.94.221302}. Other contexts leading to LV include String Theory \cite{PhysRevD.39.683,PhysRevD.69.105009,mavromatos2007lorentz}, Bumblebee models \cite{PhysRevD.71.065008,Seifert:2009gi,Alfaro:2006dd}, spacetime-varying couplings \cite{Kostelecky:2002ca} and Non-Commutative Field Theory \cite{carroll2001noncommutative}. \\
Starting from the pioneering work of Kosteleck\'y \cite{PhysRevD.51.3923,Colladay:1996iz}, 
significant efforts have been devoted to extend the current particle physics models to account for and possibly test the effects of LV \cite{PhysRevD.59.116008,mattingly2005modern,liberati2013tests}. One standard way to do that was to parametrize all possible Lorentz violating terms in the Lagrangians of each particle sector. This approach finally gave rise to the Standard Model Extension (SME), an Effective Field Theory (EFT) that contains (up to some dimension in energy) all local operators breaking Lorentz Invariance, and even the celebrated CPT symmetry \cite{Colladay:1998fq,PhysRevD.63.065008,lehnert2004dirac}. Today, the SME framework is constantly employed to set new bounds on LV parameters for different particles, such as neutrinos \cite{PhysRevD.70.031902,PhysRevD.85.096005}, photons \cite{kostelecky2002signals} and muons \cite{PhysRevD.90.076009}. SME applications are even considered in a few contexts apart from particle physics, such as gravitation \cite{Kostelecky:2004dv,Kostelecky:2020hbb,PhysRevD.74.045001} and condensed matter \cite{kostelecky2022lorentz}. Periodically, the results of these analyses are collected and updated in \cite{kostelecky2011data} for consultation convenience. \\
However, there are at least two ways to think about the novel LV operators in the SME Lagrangians: they could either be sourced by Spontaneous Symmetry Breaking (SSB) of Lorentz invariance or be explicit LV contributions. In the first scenario, the LV coefficients are usually dynamical quantities solving the equations of motions of the underlying theory. In contrast, an explicit background is a predetermined entity. Even if in most situations both approaches deliver the same effects, they can be distinguished by the presence in the former case of dynamical fluctuations, such as Nambu-Goldstone bosons and massive modes \cite{PhysRevD.71.065008}. Moreover, the SSB approach has the ability to easily avoid no-go constraints in curved spacetimes \cite{PhysRevD.91.065034,Kostelecky:2021tdf}. Another important aspect of the SSB viewpoint lies in its “effective” nature. Here, in fact, LV emerges as an “artifact” of a more fundamental theory that preserves LI and contains only dynamical building blocks. That would be analogous to how at low energies the Higgs field gets frozen around its vacuum expectation value (VEV), spontaneously “breaking” the gauge group $SU(2)_L \times U(1)$ of weak interactions. \\
The standard philosophy for constructing SME contributions involves the use of spurionic fields \cite{penco2020introduction, sheikh2008realization} and is quite straightforward: Imagine that we want to write all the LV operators for a spinor field $\psi$ involving a constant vector $v^\mu$. Then, we promote the vector $v^\mu$ to a legitimate dynamic quantum field $V^\mu$ that transforms covariantly under LTs. The next step is to construct all the interactions with $\psi$ allowed by Lorentz symmetry, such as
\be
\sum_\mu V_\mu \bar \psi \gamma^\mu \psi \,.
\ee
Through the SSB mechanism, the field $V^\mu$ can select a specific constant VEV $\braket{V^\mu} = v^\mu $ so that around the VEV we can expand it as $V^\mu = v^\mu + \delta V^\mu $. At this point, Lorentz symmetry is already spontaneously broken and, neglecting the fluctuations $\delta V^\mu$, we obtain the respective Lorentz-violating operators we were searching for
\be
\sum_\mu V_\mu \bar \psi \gamma^\mu \psi \,\, \to \,\, \sum_\mu \braket {V_\mu} \bar \psi \gamma^\mu \psi  = \sum_\mu v_\mu \bar \psi \gamma^\mu \psi  \,.
\ee
Note that in this way we have only undermined the invariance under PLT, while the symmetry under OLT has remained unaffected, hence the importance of the above clarifications. To summarize, the golden rule is to write contractions of terms that are scalar under OLT but lack PLT-symmetry, since they contain some object invariant under PLT.\\
However, there is a subtle yet fundamental detail that we omitted in the above discussion: All SME contributions constructed in this way are local contributions in the sense that no differential operators ever appear in the denominator of those terms! This is actually quite standard in the context of Lorentz-Invariant EFTs, where non-local terms are usually obtained by integrating out heavy states \cite{burgess2007introduction,davidson2020effective}. In that case, since the aim of the EFT is to be valid at low energies, meaning much below the cut-off imposed by the mass of the heavy particles, we can Taylor expand the non-local operators and get back again a local model. Non-analytic, nonlocal operators would thus need to emerge in a different way \cite{background}. Nevertheless, at the moment of giving up on LI, things are not that trivial anymore, and stranger landscapes can hide around the corner. That is exactly the case for the peculiar LV realizations we focus on in this thesis, Very Special Relativity. As we will appreciate in the next pages, in fact, one of the fundamental features that differentiates it from other LV models is exactly its non-local nature. 

\section{Neutrino Masses and Very Special Relativity}

The initial version of the SM for the electroweak sector, polished by Weinberg and others in the early 1970s \cite{weinberg2004making}, featured only massless (left-handed) neutrinos. Back in 1957, Goldhaber's experiment had already shown that neutrinos seemed to be exclusively left-handed \cite{PhysRev.109.1015}. Consequently, neutrinos could not acquire mass by coupling to the Higgs boson, as the other SM particles did, unless we assumed the existence of an unobserved right-handed (and much heavier) counterpart. It was therefore somewhat surprising that, around the transition between the 19th and 20th centuries, various experiments provided compelling evidence in favor of neutrino oscillations, clearly pointing towards their massive nature. Since then, physicists have come up with all sorts of proposals to “craft” tiny masses for neutrinos. Currently, the most popular idea is the so-called “see-saw” mechanism, in all its different realizations \cite{Gell-Mann:1979vob,Glashow:1979nm,Mohapatra:1979ia}, which predicts the existence of an additional heavy and sterile neutrino. Within such models, neutrinos may also come as Majorana particles (fermions that are identical to their own antiparticles), resulting in lepton number violation \cite{Zee:1980ai}. \\
Given the lack of experimental evidence in those directions, Cohen and Glashow suggested in 2006 an alternative path: Lorentz violation! \cite{vsr1} As we have already mentioned, Lorentz symmetry is at least very well realized in Nature, implying that any violation of it should be small, and thus could source small neutrino masses \cite{vsr2}. The Lorentz subgroups considered by Cohen and Glashow for the construction of their LV model, namely Very Special Relativity (VSR), all share the special property to be enlarged to the whole (proper and orthochronous) Lorentz group when adjoining to them one of the discrete symmetries $P$, $T$ or $CP$. It follows that deviations with respect to the standard LI theories are absent if any one of those discrete symmetries is conserved. Furthermore, the necessity of the smallness of the LV effects in VSR can be directly deduced from the smallness of the $CP-$violating effects observed in Nature \cite{PhysRevLett.13.138}.

\section{Algebras and Features of Very Special Relativity Subgroups} \label{introvsr}

Let us begin by introducing the Lorentz generators for rotations $\vec J$ and boosts $\vec K$
\be \label{lorentzgen}
\vec J = (J_x ,J_y , J_z) \,,\,\, \vec K = (K_x ,K_y , K_z) \,,
\ee
with their respective commutation relations
\be \label{lorentzcommutation}
[J_i,J_j]= i \, \epsilon_{ijk} \, J_k \,\,,\,\,\, [K_i,K_j]= -i \, \epsilon_{ijk} \, J_k \,\,,\,\,\, [J_i,K_j]= i \, \epsilon_{ijk} \, K_k \,,
\ee
with $\epsilon_{ijk}$ being the totally antisymmetric tensor and where repeated indices are intended to be summed. Then, we define two linear combinations of the above generators as follows
\bea
T_1 &\equiv& K_x + J_y  \,,\\
T_2 &\equiv& K_y - J_x  \,,
\eea
which are found to satisfy abelian commutation rules between each other
\be
[T_1, T_2] = 0 \,.
\ee
Hence, Lorentz subgroups are constructed by taking subsets of Lorentz generators that form a closed algebra (or subalgebra) under commutation, which means that the commutator among elements of the subset only results in generators within the same subset. One trivial example are spatial rotations, which are characterized by the three rotation generators $O(3) = \{ J_x, J_y, J_z \}$ and their commutation relations in \eqref{lorentzcommutation}. \\
The subgroups identified by Cohen and Glashow to construct VSR models can be expressed by listing the relative subalgebras as follows:
\begin{enumerate}
    \item $T(2) = \{ T_1, T_2 \}$, isomorphic to translations in two dimensions
    \item $E(2) = \{ T_1, T_2 , J_z \}$, stabilizer of a null vector in $4D-$spacetime
    \item $HOM(2) = \{ T_1, T_2 , K_z \}$, equivalent to two-dimensional homotheties 
    \item $SIM(2) = \{ T_1, T_2 , J_z , K_z\}$, stabilizer of a null direction in $4D-$spacetime and largest proper subgroup of the Lorentz group
\end{enumerate}
As mentioned earlier, all these subgroups have the peculiar property to be enlarged to the full Lorentz algebra when adjoined with at least one of the discrete transformations $P$, $T$ or $CP$. This can be easily shown by observing their action on the Lorentz generators in \eqref{lorentzgen}. Let us take the case of Parity $P$ as an example. Its action on the generators is
\be
 P J_i P^{-1}= J_i \,,\,\, P K_i P^{-1} = -K_i\,,
\ee
which implies that, by combining $T_1$ and $T_2$ with their $P-$transformed versions, we get back the standard boost and rotation generators in the $x-y$ plane
\bea
&& T_1 + P \, T_1 P^{-1} \propto J_y \,,\,\, T_1 - P \, T_1 P^{-1} \propto K_x \,, \\
&& T_2 + P \, T_2 P^{-1} \propto -J_x \,,\,\, T_2 - P \, T_2 P^{-1} \propto K_y \,.
\eea
Further commutation among the above combinations leads to boosts and rotations in the $z-$direction, according to the rules in \eqref{lorentzcommutation}. Then, just by starting with $T_1$ and $T_2$ and one of the discrete transformations, we can get back the full algebra of the Lorentz group, as previously stated.\\
However, even if all the VSR groups share the above property, some of them are even more “special” than others. In fact, while the first two subgroups, $T(2)$ and $E(2)$, allow for an additional invariant null vector in their geometric structure, the same is not true for the other two, $HOM(2)$ and $SIM(2)$, which leave unaltered only the Minkowski metric $\eta_{\mu\nu}$. The primary consequence of the absence of additional invariants is that the kinematics of free particles in VSR does not change with respect to SR. The constancy and isotropy of the velocity of light, time dilation, length contraction, all of those phenomena work the same when restricting the symmetry group from Lorentz to $SIM(2)$ or $HOM(2)$. As we shall better understand later, the lack of formal modifications to the dispersion relation of particles in VSR is also a direct result of this feature. In the literature, $SIM(2)-$invariant models are by far the most common and the reason is twofold: First, being $SIM(2)$ the largest Lorentz subgroups, a $SIM(2)-$invariant theory can be regarded, in some sense, as the smallest deviation from a Lorentz-invariant one. Second, contrary to its partner $HOM(2)$, the $SIM(2)$ subalgebra has the nice property of ensuring $CPT-$symmetry of the field theory to be constructed, in an equivalent manner as the Lorentz symmetry does \cite{vsr1}. In virtue of those reasons, in the rest of this thesis, when talking about VSR and VSR theories, we will always refer to $SIM(2)-$invariant formulations. It is worth mentioning that the massive $SIM(2)$ one-particle irreducible representations are all one-dimensional, labeled by spin along the preferred axis \cite{santoni-munoz-koch-2023,Fan:2006nd}. Consequently, as anticipated, theories built upon this symmetry can accommodate lepton-number conserving masses for left-handed neutrinos without requiring the introduction of sterile states \cite{vsr1}.
\\
Before we move on, let us point out that while we were playing with the Lorentz group generators to form smaller and larger subsets, the four generators of spacetime translations were left intact. Therefore, energy-momentum conservation is not affected and each VSR group combined with translation symmetry will return a subgroup of the connected Poincar\'e group. For example, the Poincar\'e subgroup concerning the $SIM(2)-$formulation of VSR is usually denoted $ISIM(2)$ \cite{grav1}.\\
At this point, you might be wondering how we can practically modify the field theories in question if, as we said, there are no additional $SIM(2)-$invariant building blocks to construct Lagrangian operators. The answer is that, despite the absence of new truly invariant tensors, there is still a preferred lightlike direction encoded in the VSR algebra. With our choice of generators, it can be identified by a null four-vector $n^\mu=(1,0,0,1)$. It is easy to see that, under VSR transformations, $n^\mu$ can only be rescaled as
\be
n^\mu \underset{SIM(2)}{ \Longrightarrow} \Omega\; n^\mu \,,
\ee
where the real scaling factor $\Omega$ depends on the nature of the boost in the $z-$direction. Hence, if we construct ratios of spacetime scalar products $n\cdot p \equiv \sum_\mu n^\mu p_\mu $ involving $n^\mu$ in both the numerator and denominator together with other dynamical variables
\be  \label{vsrratios}
\frac{n \cdot p}{ n\cdot q} \;\;\underset{SIM(2)}{ \Longrightarrow} \;\; \frac{ \Omega \, n \cdot p}{ \Omega \, n\cdot q} = \frac{  n \cdot p}{ n\cdot q}\,,
\ee
those quantities would be invariant under $SIM(2)$ but not under the full set of (Particle) Lorentz transformations, since here $n^\mu$ is the equivalent of a non-dynamical background. It is exactly from such constructions that non-localities will emerge in the VSR models. That is because, in the end, we will always utilize the four-vector operator $N^\mu$, defined as the following combination
\be \label{bignop}
N^\mu \equiv \frac{n^\mu }{n \cdot \partial } \,,
\ee
where the fraction $\frac{1}{n \cdot \partial}$ has to be interpreted as an inverse operator 
\be
\frac{1}{n \cdot \partial} \equiv (n \cdot \partial)^{-1} \,.
\ee
In the rest of this manuscript, we will often make use of the following integral representation for \eqref{bignop} to approach several calculations
\be \label{nintrep}
\frac{1}{n \cdot\partial} = \int _0^\infty d\alpha \, e^{-\alpha \, n \cdot \partial} \,.
\ee
Furthermore, thanks to the divergence theorem, when dealing with “well-behaved” spacetime functions $f(x)$ we have that
\be
0=\int d^4x \; n\cdot\partial \, \left ( \, \frac{1}{n\cdot\partial} f(x) \; \frac{1}{n\cdot\partial}\, g(x)\, \right) = \int d^4x \; \left ( \,  f \frac{1}{n\cdot\partial}\, g + \frac{1}{n\cdot\partial} f \, g \, \right) , 
\ee
implying that $(n\cdot \partial )^{-1}$ satisfies the convenient integration-by-parts relation
\be \label{bypartwithn}
\int d^4x \;   \bar \psi \left ( \, \frac{1}{n\cdot\partial} \psi \, \right) 
= - \int d^4x \; \left ( \, \frac{1}{n\cdot\partial} \bar \psi \, \right) 
  \psi \,. 
\ee
To conclude this section, we note that the magnitude of the four-vector $n^\mu$ is not really relevant for VSR effects. In fact, the fractional form \eqref{bignop} in which $n ^\mu$ appears allows us to rescale it as we like. The combination of this possibility with its lightlike nature $(n^0)^2 = -\vec n^2$, implies that it is always possible (in inertial frames) to normalize the VSR preferred direction as
\be \label{rescalednintro}
n^\mu =(1,\hat n)\,, \,\, \text{with }\,\, \hat n^2 = 1\,,
\ee
that is the expression we assume in most calculations.



\section{Purpose and Outline of the Thesis}

Although the birth of VSR was deeply related to the neutrino mass problem \cite{vsr2}, the unique features of VSR led researchers to investigate its viability and signatures in other contexts. At the moment, the VSR principles have been coherently implemented in many different areas related to theoretical physics. Both fermionic \cite{Fan:2006nd,dunn2006} and gauge sectors \cite{PhysRevD.91.105007,PhysRevD.91.129904}, including non-abelian ones \cite{Alfaro:2013uva}, have been extended to include VSR. Interestingly enough, all the modifications related to gauge theories seem to introduce some kind of correlation length into the system, which normally shows up as a new mass term in the dispersion relation. This also happens for long-range interactions, in particular for the photon \cite{PhysRevD.100.055029} and the graviton \cite{grav3}, which is one of the main topics investigated in this project and will have a dedicated chapter. The high-energy phenomenology of VSR has also been extensively explored. Calculations related to Compton and Bhabha scattering \cite{Bufalo:2019kea}, anomalies \cite{PhysRevD.103.075011,Bufalo:2020znk} and many more \cite{Alfaro:2020njh,Alfaro:2019snr,PhysRevD.100.065024} have already been worked out over the years. Nevertheless, not much has been done in the opposite regime, low energies. Many modestly sized or tabletop experiments have reached very high precision with their measurements and could therefore be exploited to constrain VSR-like violations in this complementary energy range. For this work, we decided to focus particularly on two of them: Electron $\mathsf g-2$ experiments and gravitational spectroscopy with ultracold neutrons. Thus, the purpose of this thesis is to further explore the properties and features of VSR while paying particular attention to its uncharted phenomenological consequences in the directions cited above. This will ideally lead to placing constraints on the VSR parameters in each different context.\\
With all that said, we are finally ready for a brief outline of this thesis.
In Chapter \ref{ch1}, we start by analyzing the VSR extension of Dirac theory, first in the free scenario and then including electromagnetic interactions. Chapter \ref{ch2} is dedicated to the study of the energy spectrum of charged VSR fermions in an external electromagnetic background, to mimic the configurations of $\mathsf g-$factor experiments with electrons. After that, use Chapter \ref{chnrlimit} to demonstrate the effectiveness of the VSR Hamiltonian formulation in the context of non-relativistic limits, deriving the non-relativistic Hamiltonian for a charged particle moving in an electromagnetic field. In Chapter \ref{ch3}, we include the findings we obtained when considering VSR in the context of accelerated frames, together with their relation to experiments measuring the eigenenergies of neutrons in the Earth's gravitational potential. Chapter \ref{ch4} is devoted to the discussion and implementation of VSR in the field of linearized gravity, which allowed the use of binary pulsar measurements to constrain the VSR graviton mass parameter. Finally, in the seventh and last chapter of this thesis, we summarize the results achieved while outlining various open directions for the investigation of VSR in the future. To avoid clutter in the main text of the thesis, many of the longer calculations will be included in the appendices at the end of the manuscript and will be accordingly cited along with the discussion of the different topics.\\
As a conclusive note, there are several well-written textbooks which already address all the foundational knowledge needed to go through this thesis. Consequently, assuming that I would not do a better job than them at covering those topics, I personally regard unnecessary providing yet another exhaustive summary of all the theoretical concepts connected to the work here presented, unless directly beneficial to the discussion. For anyone needing to fill in some knowledge gaps, I refer you to a few of my favorite books regarding the covered subjects: Special Relativity \cite{Landau:1975pou,10.1088/978-1-6817-4254-0} and Group Theory \cite{Zee:2016fuk,Hamermesh:1123140,booknadir,Schwichtenberg:2018dri}, Quantum Field Theory \cite{Peskin:1995ev,Schwartz:2014sze,Weinberg:1995mt,Weinberg:1996kr}, General Relativity \cite{weinberg1972gravitation,Misner:1973prb,Ferrari:2020nzo} and Gravitational Waves \cite{Maggiore:2007ulw,Maggiore:2018sht}.



\section{Notations and Conventions}

Before we get our hands dirty, allow us to clarify and highlight some of the most relevant notations and conventions adopted throughout the text.
\begin{itemize}
    \item \textit{Natural Units:} Unless differently specified, all the calculations will be carried out using natural units $c=\hbar =1$.
    \item \textit{Index summation}: Unless otherwise specified, depending on the context, repeated indices of any origin are always intended to be summed.
    \item \textit{Metric Signature:} We chose to work with the mostly-negative (or West coast) signature for the spacetime metric $g$ and the Minkowski tensor $\eta$
    \be 
    \eta_{\mu\nu}= diag\{1,-1,-1,-1\} \,.
    \ee
    Occasionally we will exploit the fact that $\eta^{ij} = - \delta^{ij}$, with $\delta^{ij}= diag \{ 1,1,1\}$ being the euclidean metric in three spatial dimensions.
    \item \textit{Scalar products}: Two different notations for scalar products will be used throughout the thesis. Scalar products between two Cartesian three-dimensional vectors $\vec v$ and $\vec w$ will be indicated as $\vec v \cdot \vec w = \delta^{ij} v^i w^j = v^i w^i$. Scalar products between two spacetime four-dimensional vectors $ v$ and $w$ will be indicated as $ v \cdot w = g_{\mu\nu} v^\mu w^\nu =v^\mu w_\mu$. Furthermore, in the context of Dirac theories, we will use the standard “slashed” notation $\slashed v$ to indicate scalar products with gamma matrices $\slashed v = \gamma^\mu v_\mu$.
    \item \textit{Gamma Matrices}: Being often interested in low-energy limits, we picked for the gamma matrices the standard Dirac representation.
    \begin{eqnarray}
    \gamma^0 = \left( \begin{array}{cc}\mathbf{1} & 0\\0 & -\mathbf{1} \end{array}\right) \,,\,\, \vec \gamma = \left( \begin{array}{cc}0 & \vec \sigma\\-\vec \sigma & 0 \end{array}\right) \,,\,\, \vec \Sigma = \left( \begin{array}{cc}\vec \sigma & 0 \\ 0 & -\vec \sigma\end{array}\right) \,,
    \label{gammadef}
    \end{eqnarray}
    where here $\mathbf{1}$ represents the two-by-two identity and $\sigma^i$ are the Pauli matrices 
    \begin{eqnarray}
    \sigma^1 = \left( \begin{array}{cc} 0 & 1 \\ 1 & 0 \end{array} \right) ,\,\, 
    \sigma^2 = \left( \begin{array}{cc}0 & -i\\i & 0 \end{array}\right) ,\,\, 
    \sigma^3 = \left( \begin{array}{cc} 1 & 0\\ 0 & -1 \end{array}\right) \,.
    \label{sigmadef}
    \end{eqnarray}
    Moreover, we define the antisymmetric commutator of two gamma matrices as 
    \begin{equation}
    \sigma^{\mu\nu} \equiv \frac12[\gamma^\mu ,\gamma^\nu]\,.
    \end{equation}
    \item \textit{Preferred direction $n^\mu$:} For the group theory treatment we aligned the $z-$axis with the spatial part of $n^\mu$. However, to keep the following calculations more general, we will try, whenever possible, to not make a specific choice for the frame orientation respect to $\hat n$.
    \item \textit{Fourier transform}: we will indicate the Fourier transform of objects by a “tilde” over them. The numerical convention we choose for the transform and anti-transform are the following
    \begin{eqnarray}
        F(x) &=& \int \frac{d^4 q}{(2\pi)^4}\,  e^{-iqx} \tilde F(q)   \,,  \\
        \tilde F (q) &=& \int d^4 x \, e^{iqx} F(x) \,. \nonumber
    \end{eqnarray}
    \item \textit{Directionality of Operators:} Standard operators (e.g. differential operators $\partial$) with no arrows on top are intended to be applied to the right. Operators applied to the left are instead distinguished by the presence of a left arrow on them, as for left derivatives $\overset{\leftarrow}{ \partial }$. 
\end{itemize}

\chapter[DIRAC FERMIONS WITHIN VERY SPECIAL RELATIVITY]{Dirac Fermions within Very Special Relativity} \label{ch1}
When talking about field theories, the easiest example we can think of is for sure the one for scalar particles, since they lack spin. In the VSR framework, however, when considering scalar fields, there are no new Lagrangian operators that can be written. That is, in short, because there are no dynamical objects to contract $n$ with other than partial derivatives (or, respectively, momentum), implying that non-trivial ratios such as the one in \eqref{vsrratios} do not exist in this case.\\
The next “rung in the ladder” are spin$-\frac12$ fields, or fermions, which, thanks to their internal spin structure, allow for VSR modifications. In particular, in the following we will deal with Dirac spinors $\psi$, leaving aside the more elementary Weyl analogues. The simplest VSR field theory is obtained by imposing $C-$invariance on the most general free Dirac Lagrangian in the VSR framework. Its expression reads \cite{dunn2006}
\be \label{diracvsrlagr}
    \mathcal{L} = \bar \psi \,  \left ( i \slashed \partial -m + i \lambda \frac{\slashed n }{n \cdot \partial} \right ) \,  \psi \,.
\ee
As we can see, only one VSR operator is introduced and, at this point, its non-local nature should be no surprise, as anticipated in Section \ref{introvsr}. The single parameter governing the magnitude of the LV is $\lambda$, which can be seen by dimensional analysis to have the dimension of a mass squared $[\lambda] = [m]^2$. In the literature, this parameter is often indicated as a square $\lambda \to \frac{M^2}{2}$ to ensure its positivity. Nevertheless, as we will see in the following pages, in the case $m\neq0$, there is no reason to assume that $\lambda$ cannot be negative. For this reason, we prefer to stick to the $\lambda-$notation.\\
As expected, the new term does not satisfy all the discrete symmetries as the standard Dirac counterparts. In fact, while preserving $C$ and the combined $CPT$ transformations, it breaks parity $P$ and time reversal $T$ as can be seen from the conventional transformation rules for Dirac spinors. Note that this additional particle-antiparticle asymmetry enabled by $CP-$violation could be relevant in many contexts, starting from baryogenesis, the physical process that is hypothesized to have taken place during the early universe to produce the matter-antimatter imbalance observed in our Universe \cite{Sakharov:1967dj}.

\section{Equations of Motion}

The first thing we would like to derive from the above Lagrangian \eqref{diracvsrlagr} are the equations of motion (EOM). Because of the property \eqref{bypartwithn}, the VSR Lagrangian is still Hermitian up to a total derivative, having that
\be
(i\lambda \bar \psi \, \slashed n \frac{1 }{n \cdot \partial} \psi)^\dagger = - i\lambda \frac{1 }{n \cdot \partial} \psi ^\dagger \slashed n ^\dagger \gamma^0 \psi = - i \lambda \frac{1 }{n \cdot \partial} \bar \psi \slashed n \psi \, \to \,
i \lambda \bar \psi \slashed N \psi + \text{surface term} \,.
\ee
Thus, the EOMs for $\psi$ and $\bar \psi \equiv \psi^\dagger \gamma^0$ are the Hermitian conjugate (or adjoint) of each other. In particular, the EOM for $\psi$ can be easily read off from \eqref{diracvsrlagr} to be 
\be  \label{diracvsreom}
( i \slashed \partial -m + i \lambda \frac{\slashed n }{n \cdot \partial} ) \, \psi = 0 \, .
\ee
Recalling the anticommutation property of gamma matrices $\{\gamma^\mu ,\gamma^\nu\}= 2\, \eta^{\mu\nu}$ and that $\slashed n \slashed n = n^2 = 0$, the new dispersion relation for the VSR fermions is readily derived by multiplying the EOM by $( -i \slashed \partial -m - i \lambda \frac{\slashed n }{n \cdot \partial} ) $ from the left
\be \label{vsrkleingordon}
((\slashed \partial + \lambda \frac{\slashed n}{n\cdot \partial})^2 +m^2) \,\psi = ( \partial^2 + 2\lambda +m^2) \,\psi  = 0 \,,
\ee
which is a standard Klein-Gordon (KG) equation with shifted mass squared
\be m_f^2 \equiv m^2 +2\lambda \,,\ee 
leading to the usual SR dispersion relation. Observe that for $ \lambda < -\frac{m^2}{2}$, the appearance of imaginary energies at low momentum implies the possibility of runaway modes, an issue already recognized in SME models \cite{PhysRevD.90.021701,alexandre2015foldy}. Then, to avoid tachyonic degrees of freedom (DOF), we should restrict ourselves to $ \lambda > - \frac{m^2}{2} $. After all, in an EFT context, we should expect small values of $\lambda << m^2$ to avoid large LV effects anyway. However, this restriction displays its relevance in the case of neutrinos, where having $m=0$ would imply a positive value for $\lambda$.

\section{Vector and Axial Currents} \label{sec:sec1.1}

Due to the complex nature of Dirac spinors $\psi$, quadratic Dirac Lagrangians automatically enjoy a global symmetry under phase transformations $\psi' = e^{i \phi } \, \psi$. That is trivially the case also for the VSR Lagrangian \eqref{diracvsrlagr}, implying the existence of a conserved vector current $j^\mu$, as in the LI case. The way to find it is straightforward. We start by taking the EOM for $\psi$ and $\bar \psi$ and multiplying them, respectively, by $\bar \psi$ from the left and by $\psi$ from the right, obtaining
\bea
    && \bar \psi(i\slashed \partial - m + i \lambda \slashed N )\psi =0 \,,\\
    && \bar \psi (-i\overset{\leftarrow}{ \slashed \partial} - m - i \lambda \overset{\leftarrow}{\slashed N}) \psi =0 \,. \nonumber
\eea
At this point, simply resting the two above equations (and inserting $1=N\cdot\partial$) we get
\begin{equation}
    \partial_\mu (\bar \psi \gamma^\mu \psi +\lambda N^\mu \bar \psi \slashed N \psi  ) =0\,,
\end{equation}
from which we directly deduce the form of the vector current 
\be
J^\mu = \bar \psi \gamma^\mu \psi +\lambda N^\mu \bar \psi \slashed N \psi \, . 
\ee
Interestingly enough, we can follow a similar path for axial transformations $\psi ' = e^{i \theta \gamma^5 }$ and see that, contrary to the Dirac mass term, the VSR operator does not contribute to the non-conservation of the axial current $j^\mu_5$. In fact, since axial transformations anticommute with both $\slashed \partial$ and $\slashed N$, we obtain
\be
    \partial_\mu j ^\mu_5 = -2i m \bar \psi \gamma^5 \psi \,,
\ee
where we defined the VSR axial current as
\be
j^\mu_5 = \bar \psi \gamma^5 \gamma^\mu \psi +\lambda N^\mu \bar \psi \gamma^5 \slashed N \psi \,.
\ee
Hence, as in the LI case, the axial current is conserved in the limit of $m \to 0$, although in the VSR context that does not coincide with massless fermions.

\section{Coupling to the Electromagnetic Field}

Due to its peculiar properties, VSR is believed to have no non-trivial consequences in the context of particle freely streaming through space, apart from the mass shift we already learned about \cite{vsr1}. Therefore, it becomes of fundamental importance, if we want to discriminate VSR from the LI scenario, to include interactions in our picture. \\
Since we are interested in describing the behavior of charged fermions, like electrons, the natural way to go is to extend the free VSR Lagrangian \eqref{diracvsrlagr} and EOM \eqref{diracvsreom} to account for interactions with the electromagnetic field. In the LI scenario, this is achieved by gauging the $U(1)$ group of phase transformations and introducing the covariant derivative
\be \label{covdevem}
D_\mu = \partial_\mu + i e A_\mu \,,
\ee
where $A_\mu$ geometrically represents the gauge group connection and behaves such that the transformation rule for $D_\mu \psi$ under local phase transformations is identical to the $\psi-$one. Thus, the minimal prescription teaches us that the gauge-invariant way to couple charged fermions to the electromagnetic field is simply to replace every partial derivative with the covariant one in \eqref{covdevem}. This simple instruction also works in the VSR case, as shown in Appendix \ref{gaugeinvarianceVSREM}. The resulting Lagrangian should look like
\be \label{emvsrlagr}
    \mathcal{L} = \bar \psi \,  ( i \slashed D -m + i \lambda \frac{\slashed n }{n \cdot D} ) \,  \psi \,.
\ee
The first thing you may note is the presence of the electromagnetic potential $A_\mu$ in the denominator of the non-local operator, which is quite an uncommon feature for a QFT. In the context of Feynman diagrams and perturbation theory, that implies the emergence of an infinite series of new vertex interactions with respect to the usual ones in Quantum Electrodynamics (QED) \cite{Alfaro:2020njh}. Nevertheless, the latter can be gauged away from the theory by working in the light-cone gauge
\be \label{lightconegaugeA}
n\cdot A = 0 \, ,
\ee
which will prove to be a useful choice in the calculations for this thesis. Here, however, we will not enter into the details of the full many-particle picture of QFT, since it goes beyond the scope of this manuscript. In fact, for us, it will often be sufficient to interpret the below EOM \eqref{emvsreom} as the effective EOM for a single test particle immersed in an external electromagnetic background
\be \label{emvsreomnogauge}
      ( i \slashed D -m + i \lambda \frac{\slashed n }{n \cdot D} ) \,  \psi = 0 \,.
\ee
Consequently, the electromagnetic potential $A_\mu$ should not be considered as a dynamical object, but as a fixed background depending on the situation under analysis. Finally, note that, when working with light-cone gauges \eqref{lightconegaugeA}, the interacting EOM reduces to
\be \label{emvsreom}
      ( i \slashed \partial -e \slashed A -m + i \lambda \frac{\slashed n }{n \cdot \partial} ) \,  \psi = 0 \,,
\ee
which is much easier to deal with in general. Before we start with some practical applications in the next chapters, let us further explore a few standard ways we have to more easily deal with VSR charged fermions in some specific contexts.

\section{Hamiltonian Approach\label{sec:sec1.2}}

Given any quantum system, the Hamiltonian is a fundamental tool for describing its evolution and behavior. This is particularly true in the non-relativistic limit, where the system's dynamics is well characterized by the Schr\"odinger equation. In Dirac systems, the standard strategy to find the Hamiltonian operator $\mathcal{H}$ is to write the EOM in a Scrh\"odinger-like form. For example, in the free LI case, multiplying the EOM by $\gamma^0$ we obtain
\be
(i \slashed \partial - m )\psi =0 \,\to\, i \partial_0 \psi= \mathcal{H_D} \psi \,,\,\, \text{with } \mathcal{H_D} = m \beta + \vec \alpha \cdot \vec p \,,
\ee
where we defined the Dirac $\alpha-$matrices and the momentum operator $\vec p$ as
\be \beta \equiv \gamma^0 \,\,,\,\,\, \alpha^i \equiv \gamma^0 \gamma^i \,\,,\,\,\, p^i \equiv - i \partial_i \,. \ee
After coupling the LI fermion to the electromagnetic interaction through the introduction of the covariant derivative \eqref{covdevem}, the picture is not much different. In fact, repeating the same process, we get to
\be
\mathcal{H_D}^{EM} = m \beta + \vec \alpha \cdot (\vec p - e \vec A) +e A^0 \,.
\ee
The VSR case is more involved. The presence of the non-local operator $(n \cdot \partial)^{-1}$ requires some ad hoc plan to be tackled. That is because normally, by using the integral representation \eqref{nintrep} and expanding the exponential, we would get an infinite series of time derivatives, which is clearly problematic. Let us start by analyzing the scenario without interactions.

\subsection{Free Hamiltonian} \label{freevsrhamiltoniancase}

First, we want to get rid of the non-locality in the EOM  \eqref{diracvsreom}. Multiplying it from the left by $- i n\cdot \partial$ we obtain the equation
\be \label{towardvsrham}
( \,n\cdot \partial \, \slashed \partial +i m \, n \cdot \partial +\lambda \slashed n \,) \,\psi = 0 \,.
\ee
Expanding the above scalar products, we naturally see the appearance of squared time derivatives. However, due to the KG equation \eqref{vsrkleingordon}, our $\psi$ should also satisfy the relation
\be \label{freesquaredtime}
\partial_0^2 \psi = -( \partial_i \partial^i +m^2_f) \psi \,,
\ee
which can be used in \eqref{towardvsrham} to eliminate the squared time derivatives in favor of the Laplacian and the squared fermion mass. Factorizing the remaining time derivatives on the left-hand side of the equation and pushing everything else on the right, we get
\be
\left (1 - \frac{i}{m} (\gamma^0 n^ i +\gamma^ i) \partial_ i \right ) \; i \partial_ 0 \psi =
    \left ( \frac{1}{m} \gamma^0 ( \partial_i \partial^i +m^2_f) - \frac{1}{m } \gamma^ j n^i \partial_ i \partial_ j - i \, n^ i \partial_i -\frac{\lambda}{m} \slashed n \right) \psi \,.
\ee
Inverting the operator on the left and defining for convenience
\be \Delta_\partial \equiv \left (1 - \frac{i}{m} (\gamma^0 n^ i +\gamma^ i) \partial_ i \right )^{-1} \,,\ee 
we are almost done with our derivation
\be
i \partial_ 0 \psi = \Delta_\partial
    \left ( m \gamma^0+\frac{1}{m} \gamma^0 \partial_i \partial^i - \frac{1}{m } \gamma^ j n^i \partial_ i \partial_ j - i \, n^ i \partial_i +\frac{2 \lambda}{m} \gamma^0 -\frac{\lambda}{m} \slashed n \right) \psi \,.
\ee
Since we expect the limit $\lambda \to 0$ to reproduce the LI case, we could guess that 
\be \Delta_\partial \left (m \gamma^0+\frac{1}{m} \gamma^0 \partial_i \partial^i- \frac{1}{m } \gamma^ j n^i \partial_ i \partial_ j - i \, n^ i \partial_i \right) = \mathcal{H}_D \,, \ee
which is exactly what we can prove by remembering $\gamma^i \gamma^j \partial_i\partial_j = \partial^i \partial_i$ and by adding and subtracting $ i \gamma^0 \gamma^i \partial_i $ in the above parenthesis
\bea \label{connectiontodiracfree}
&& \Delta_\partial \left (m \gamma^0 - i \, n^ i \partial_i +i \gamma^0 \gamma^i \partial_i - i \gamma^0 \gamma^i \partial_i  +\frac{1}{m} \gamma^0 \gamma^i \partial_i \gamma^j \partial_j - \frac{1}{m } n^i \partial_ i \gamma^ j  \partial_ j  \right) = \nonumber \\
&& = \Delta_\partial \left (\Delta_\partial^{-1} m \gamma^0 - i \Delta_\partial^{-1} \gamma^0 \gamma^i \partial_i  \right) = m \gamma^0 - i \gamma^0 \gamma^i \partial_i  = \mathcal H_D \,.
\eea
Therefore, the final expression of our VSR Hamiltonian operator is
\be
\mathcal{H}_{VSR} = m \gamma^0 -i \gamma^0 \gamma^i \partial_i + \frac{\lambda}{m} \Delta_\partial (\gamma^0 - \gamma^i n_i) \,.
\ee
Naturally, the presence of gamma matrices in the denominator of the VSR contribution is not the best for intuition and calculations. Let us then “rationalize” the denominator in the following way 
\be \label{freeration}
\Delta_\partial \; \frac{1 + \frac{i}{m} (\gamma^0 n^ i +\gamma^ i) \partial_ i }{1 + \frac{i}{m} (\gamma^0 n^ i +\gamma^ i) \partial_ i } = \frac{1 + \frac{i}{m} (\gamma^0 n^ i +\gamma^ i) \partial_ i }{1+\frac{1}{m^2} (n^i n^j + \eta^{ij}) \partial_ i \partial_j} = \frac{1 + \frac{i}{m} \gamma^0 n^ i \partial_ i + \frac{i}{m}\gamma^ i \partial_ i }{1+\frac{1}{m^2} (n^i n^j - \delta^{ij}) \partial_ i \partial_j} \,.
\ee
From here, defining
\be
\Box_\partial \equiv (1+\frac{1}{m^2} (n^i n^j - \delta^{ij}) \partial_ i \partial_j )^{-1} \,,
\ee
and replacing back \eqref{freeration} into the Hamiltonian, we get
\bea
\mathcal{H}_{VSR} &=& m \gamma^0 -i \gamma^0 \gamma^i \partial_i  \\
&& + \frac{\lambda \, \Box_\partial }{m} (\gamma^0 - \gamma^i n_i + \frac{i}{m} n^ i \partial_ i - \frac{i}{m} n^ i \partial_ i \gamma^j n_j - \frac{i}{m} \gamma^0 \gamma^ i \partial_ i - \frac{i}{m} \gamma^ i \gamma^j \partial_ i n_j )\nonumber \,,
\eea
that after some easy math can be expressed as
\be \label{definitiveHVSR}
\mathcal{H}_{VSR} = (m + \frac{\lambda \,\Box_\partial}{m}) \gamma^0 -i (1+ \frac{\lambda\,\Box_\partial}{m^2})\gamma^0 \gamma^i \partial_i - \frac{\lambda \, \Box_\partial}{m}( \gamma^i n_i + \frac{i}{m} n^ i \partial_ i \gamma^j n_j + \frac{i}{m} \sigma^{ij} \partial_ i n_j ) \,,
\ee
where only spatial non-localities survived. Note that the limit $m\to0$ of the above expression is actually “well-defined”, since 
\be
\lim_{m\to0} \;\frac{\Box_\partial}{m^2} = \left  ((\vec n \cdot \vec \partial)^2 -\vec\partial^{\,2} \right )^{-1} \,.
\ee
As we could have expected, the Hamiltonian density $H_{VSR}=  \psi^\dagger \, \mathcal{H}_{VSR} \, \psi$ is not equivalent to the one derived from the VSR Lagrangian through the standard Legendre transformation 
\be \label{vsrhamiltonianfreeeq}
H_{VSR} \neq \frac{\partial \mathcal L}{\partial(\partial_0 \psi)} \partial_0 \psi+\frac{\partial \mathcal L}{\partial(\partial_0 \psi^\dagger)} \partial_0 \psi^\dagger - \mathcal L \,, \ee
because of the presence of the additional $N-$operator terms in the Lagrangian. Another important fact to note is that, even if our starting Lagrangian was perfectly Hermitian, our new Hamiltonian is not. This can be directly seen from the Hamiltonian density in the low-energy limit $m>>|\vec p|$, where at leading order we can practically neglect the spatial non-locality $\Box \simeq \frac{1}{m}$. Already in this context, there appear a few “problematic” operators. For example, the term $ -\frac{\lambda}{m} n_i \, \psi^\dagger \gamma^i \psi $ is easily proved to be non-Hermitian
\be
(-\frac{\lambda}{m} n_i \, \psi^\dagger \gamma^i \psi)^\dagger = -\frac{\lambda}{m} n_i \, \psi^\dagger \gamma^0 \gamma^i \gamma^0 \psi =  \frac{\lambda}{m} n_i \, \psi^\dagger \gamma^i \psi\,.
\ee
The lack of Hermiticity could, in principle, be seen as a negative aspect of the VSR formulation. The Hermiticity of the Hamiltonian, in fact, ensures the realness of the energy spectrum and unitarity.\footnote{Actually the lack of unitarity, and thus the non-conservation of probabilities, is a feature sometimes required to describe certain open physical systems, such as decays.} Nevertheless, there already exists an extensive literature on the possibility of meaningfully working with non-hermitian Hamiltonians in both Quantum Mechanics (QM) and QFT \cite{Bender:2007nj}. An example for this are $PT-$symmetric theories, where Hermiticity is replaced by the requirement of $PT-$symmetry \cite{Bender:2023cem,Bender:2020gbh}. A clear hint of the consistency of the Hamiltonian approach in VSR is given by the form of its free spectrum, which by looking at \eqref{freesquaredtime} should directly be
\be
E_{VSR} = \pm \sqrt{ m^2_f + \vec p^2} \,.
\ee
The energy eigenvalues of $H_{VSR}$ are therefore real for any possible value of momentum and, as expected, positive and negative energies are related to particles and antiparticles solutions. Each of these eigenergies obviously shows twice due to spin degeneracy. All of that has been checked in Mathematica, by taking as a matrix input the operator \eqref{definitiveHVSR}.


\subsection{Interacting Hamiltonian}

The coupling to electromagnetism further complicates the Hamiltonian derivation. Starting from \eqref{emvsreomnogauge} and multiplying by $-i n \cdot D$ we arrive to
\begin{equation} \label{vsrDirac2}
    ( n\cdot D \slashed D +i m \, n\cdot D + \lambda \slashed n) \psi = 0 \,.
\end{equation}
Let us calculate separately
\begin{eqnarray} \label{calculationnDD}
    n\cdot D \slashed D &=& \gamma^0 D_0 ^2 + \gamma^0 n^ i D_ i D_ 0 + \gamma^ i D_ 0 D_i + \gamma^ j n^i D_ i D_ j \nonumber\\
    &=& \gamma^0 D_0 ^2 + (\gamma^0 n^ i +\gamma^ i) D_ i D_ 0 + i e \,
    \gamma^ i F_{0i} + \gamma^ j n^i D_ i D_ j \,,\nonumber \\
    i m \, n\cdot D &=& i m D_0 + i m \, n^ i D_i\,,
\end{eqnarray}
where we used the fact that
\begin{equation}
    [D_\mu, D_\nu] \psi = i e ( [\partial_\mu, A_\nu] +[A_\mu,\partial_\nu] )\psi = i e (\partial_\mu A_\nu - \partial_\nu A_\mu) \psi \equiv i e F_{\mu\nu} \psi\,.
\end{equation}
Putting everything together in \eqref{vsrDirac2}, we get
\begin{equation}
    \left ( \gamma^0 D_0 ^2 + (\gamma^0 n^ i +\gamma^ i) D_ i D_ 0 + i e \,
    \gamma^ i F_{0i} + \gamma^ j n^i D_ i D_ j + i m D_0 + i m \, n^ i D_i + \lambda \slashed n \right ) \psi = 0 \,.
\end{equation}
Factoring out the $D_0 -$terms and defining
\be \Delta_D \equiv \left (1 - \frac{i}{m} (\gamma^0 n^ i +\gamma^ i) D_ i \right )^{-1} ,\ee 
we can isolate the time derivative on the left-hand side to obtain
\begin{equation} \label{almostschem}
    i D_0 \psi = -\Delta_D \,  \left ( \frac{1}{m} \gamma^0 D_0 ^2 + i \frac{e}{m} \,
    \gamma^ i F_{0i} + \frac{1}{m } \gamma^ j n^i D_ i D_ j + i \, n^ i D_i + \frac{\lambda}{m} \slashed n \right ) \psi \,.
\end{equation}
At this point, we shall replace the $D_0^2$ using the squared EOM as in the free case. However, in the presence of interactions, we have more than just a KG equation. Multiplying the equation \eqref{emvsreomnogauge} by $(i \slashed D - m + i \lambda \frac{\slashed n }{n \cdot \partial})$ from the left, we get
\begin{eqnarray} \label{kgforemvsr1}
0 &=& \bigg( (\slashed D + \lambda \frac{\slashed n}{n \cdot D})^2 +m^2 \bigg) \psi= \bigg( \slashed D ^2 + \lambda \{ \slashed D , \frac{\slashed n}{n \cdot D} \} +m^2 \bigg) \psi \nonumber\\
    &=& \bigg( (\eta^{\mu\nu} +\sigma^{\mu\nu})D_\mu D_\nu +2\lambda - \lambda \slashed n [ \slashed D , \frac{1}{n \cdot D} ] +m^2 \bigg) \psi \\ 
    &=& \bigg( D^2 +m^2_f+\frac{i}{2} e \, \sigma^{\mu\nu} F_{\mu\nu} -\lambda \sigma^{\mu\nu} n_\mu [ D_\nu , \frac{1}{n\cdot D} ] \bigg) \psi \nonumber\,.
\end{eqnarray}
To continue the calculation, it is fundamental to find the commutator in \eqref{kgforemvsr1}
\begin{equation}\label{commutatorDnDv1}
   [D_\nu , \frac{1}{n\cdot D}] \psi = \int d\alpha [D_\nu, e^{-\alpha n\cdot D}] \psi = \sum_{n=0} \frac{(-1)^n}{n!} \int d\alpha \,\alpha^n [D_\nu,(n\cdot D )^n] \psi \,.
\end{equation}
Expanding for the first few $n$, we can identify the following pattern 
\begin{equation}
    [D_\nu , \frac{1}{n\cdot D}] \psi = -i e \, n^\rho F_{\nu \rho} \frac{1}{(n\cdot D)^2} \psi  + i e \, n^\rho (n\cdot \partial F_{\nu \rho}  ) \frac{1}{(n\cdot D)^3} \psi - ... \,,
\end{equation}
which can be condensed as (see also Appendix \ref{DnDcommutator})
\begin{equation} \label{commutator1nDwithD}
    [D_\nu , \frac{1}{n\cdot D}] \psi = -i e \,n^\rho \sum_{n =0} ((-n\cdot \partial)^n F_{\nu\rho}) \frac{1}{(n\cdot D)^{n+2}} \psi \,.
\end{equation}
Replacing it back in \eqref{kgforemvsr1}, we obtain
\begin{equation} \label{kgforemvsr2}
    \bigg( D^2 +m^2_f+\frac{i}{2} e \, \sigma^{\mu\nu} F_{\mu\nu} +i e \lambda \sigma^{\mu\nu} n_\mu n^\rho \sum_{n =0} ((-n\cdot \partial)^n F_{\nu\rho}) \frac{1}{(n\cdot D)^{n+2}} \bigg) \psi =0 \,. 
\end{equation}
Returning to \eqref{almostschem} and eliminating the $D_0^2 $ thanks to \eqref{kgforemvsr2}, we arrive at
\begin{eqnarray}
    i D_0 \psi &=& \Delta_D \,  \left ( \frac{\gamma^0}{m} \bigg( D^i D_i +m^2_f+\frac{i}{2} e \, \sigma^{\mu\nu} F_{\mu\nu} \bigg) - i \frac{e}{m} \,
    \gamma^ i F_{0i} - \frac{1}{m } \gamma^ j n^i D_ i D_ j - i \, n^ i D_i  \nonumber \right.\\
    && \left. - \frac{\lambda}{m} \slashed n  +i e \frac{\lambda}{m} \gamma^0 \sigma^{\mu\nu} n_\mu n^\rho \sum_{n =0} ((-n\cdot \partial)^n F_{\nu\rho}) \frac{1}{(n\cdot D)^{n+2}} \right ) \psi \,.
\end{eqnarray}
Now, thanks to the connection with the $\lambda\to 0$ limit, as shown in the interaction-free case \eqref{connectiontodiracfree}, we can simplify this expression to
\begin{eqnarray} \label{schlikeeqhamapproach}
    i \partial_0 \psi &=&  \left ( m\gamma^0 -i \gamma^0 \gamma^i D_i + e A_0 +\frac{2 \lambda}{m} \Delta_D \gamma^0- \frac{\lambda}{m}\Delta_D \slashed n \right.\\
    && \;\;\;\left. + i e \frac{\lambda}{m} \Delta_D \gamma^0 \sigma^{\mu\nu} n_\mu n^\rho \sum_{n =0} ((-n\cdot \partial)^n F_{\nu\rho}) \frac{1}{(n\cdot D)^{n+2}} \right ) \psi \,. \nonumber
\end{eqnarray}
We can again try to “rationalize” the non-local $\Delta_D$ so that gamma matrices do not appear in the denominators. Nevertheless, due to the non-commutative nature of $D_i$ this trick no longer works. In fact, using $A^{-1} B^{-1} = (BA )^{-1}$, we have
\begin{equation}
    \Delta_D = \Box_D (1 + \frac{i}{m} (\gamma^0 n ^i +\gamma^i) D_i)\,,
\end{equation}
where we defined 
\begin{equation}
    \Box _D \equiv \left (1 + \frac{1}{m^2} (n^i n^j - \delta^{ij } )D_i D_j + \frac{i e}{2 m^2} ( \sigma^{ij } -2 \gamma^0 \gamma^i n^j) F_{ij} \right )^{-1} .
\end{equation}
However, we can still exploit this expression in the Schr\"odinger-like equation \eqref{schlikeeqhamapproach} to get
\begin{eqnarray} \label{almostfinalschem2}
    i \partial_0 \psi &=&  \left ( (m + \frac{\lambda \Box_D}{m})\gamma^0 -i (1+ \frac{\lambda \Box_D}{m^2})\gamma^0 \gamma^i D_i + e A_0 \right.\\
    && \;\;\;- \frac{\lambda \Box_D}{m} \gamma^i n_i -i \frac{\lambda\Box_D}{m^2} \sigma^{ij} n_j D_i - i \frac{\lambda \Box_D}{m^2} \gamma^0\gamma^j n^i n_j D_i \nonumber\\
    && \;\;\; + i e \frac{\lambda \Box_D}{m}  \gamma^0 \sigma^{\mu\nu} n_\mu n^\rho \sum_{n =0} ((-n\cdot \partial)^n F_{\nu\rho}) \frac{1}{(n\cdot D)^{n+2}} \nonumber\\
    && \;\;\;\left. - e \frac{\lambda \Box_D}{m^2}  (n^i -\gamma^0\gamma^i ) \sigma^{\mu\nu} n_\mu n^\rho D_i \sum_{n =0} ((-n\cdot \partial)^n F_{\nu\rho}) \frac{1}{(n\cdot D)^{n+2}} \right ) \psi \,. \nonumber
\end{eqnarray}
Because of the presence of new $\frac{1}{n\cdot D}-$terms we cannot directly identify the above expression with the Hamiltonian. However, when working in the non-relativistic limit, where $m$ becomes the largest energy scale, equation \eqref{almostfinalschem2} is extremely useful since, as shown in Appendix \ref{app1divnDlimit}, at leading order we have
\begin{equation}
    \frac{1}{(n\cdot D)^2 }\psi \sim - \frac{1}{m^2}\psi \,.
\end{equation}
Thus, supposing we want our Hamiltonian to be expanded up to order $\frac{1}{m^3}$, we could retain only the first term in the $n-$sums and expand accordingly $\Box_D$, leading to
\begin{eqnarray} \label{almostfinalschem3}
    i \partial_0 \psi &\simeq&  \left ( (m + \frac{\lambda}{m})\gamma^0 -i (1+ \frac{\lambda }{m^2})\gamma^0 \gamma^i D_i + e A_0 \right.\\
    && \;\;\;- \frac{\lambda }{m} \gamma^i n_i -i \frac{\lambda}{m^2} \sigma^{ij} n_j D_i - i \frac{\lambda}{m^2} \gamma^0\gamma^j n^i n_j D_i - \frac{i e \lambda}{m^3} \gamma^0 \sigma^{\mu\nu} n_\mu n^\rho F_{\nu\rho} \nonumber \\
    && \;\;\; -\frac{\lambda}{m^3} \gamma^0 (n^i n^j - \delta^{ij } )D_i D_j -\frac{i e \lambda}{2m^3} (\gamma^0 \sigma^{ij } +2 \gamma^i n^j) F_{ij}  \nonumber\\
    && \;\;\; \left. + \frac{\lambda}{m^3} \gamma^k n_k (n^i n^j - \delta^{ij } )D_i D_j + \frac{i e \lambda}{2m^3} (\sigma^{ij } -2 \gamma^0 \gamma^i n^j) \gamma^k n_k F_{ij} \right ) \psi \,,\nonumber
\end{eqnarray}
which, at this point, can finally be regarded as an Hamiltonian operator
\begin{eqnarray} \label{almostfinalschem4}
   \mathcal H _{VSR}^{EM} &\simeq&   (m + \frac{\lambda}{m})\gamma^0 -i (1+ \frac{\lambda }{m^2})\gamma^0 \gamma^i D_i + e A_0 - \frac{\lambda }{m} \gamma^i n_i  -i \frac{\lambda}{m^2} \sigma^{ij} n_j D_i \nonumber \\
    && \;\;\; - i \frac{\lambda}{m^2} \gamma^0\gamma^j n^i n_j D_i  - \frac{i e \lambda}{m^3} \gamma^i ( F_{i0}+ n^j F_{ij} +n_i n^j F_{j0}) \\
    &&  \;\;\; - \frac{i e \lambda}{2 m^3} \gamma^0 \sigma^{ij} ( F_{ij} + 2 n_i F_{j0}) -\frac{e \lambda}{2m^3} \gamma^0 \gamma^5 \epsilon^{ijk} n_k F_{ij} \nonumber\\
    && \;\;\; -\frac{\lambda}{m^3} \gamma^0 (n^i n^j - \delta^{ij } )D_i D_j +\frac{\lambda}{m^3} \gamma^k n_k (n^i n^j - \delta^{ij } ) D_i D_j  \,,\nonumber
\end{eqnarray}
where we used the following property of gamma matrices to simplify the expressions
\begin{equation}
    \gamma^\mu\gamma^\nu\gamma^\rho = \gamma^\mu \eta^{\rho \nu} + \gamma^\rho \eta^{\mu\nu} - \gamma^\nu \eta^{\mu \rho} + i \epsilon^{\sigma\mu\nu\rho} \gamma_\sigma \gamma^5\,.
\end{equation}

\section{Field Redefinitions} \label{sec:fieldredefchi}

In general, not all possible LV terms actually lead to physical consequences. This is very well known for fermions in the context of the SME \cite{PhysRevD.74.125001,Colladay:2002eh}, where field redefinitions are used to reduce the number of parameters to the smallest possible meaningful set. That naturally leads to the question of whether field redefinitions might be beneficial for VSR. In particular, for the free case, VSR is usually thought to have no other consequence apart from the shifted mass, meaning we should expect to be able to redefine our spinors in such a way as to re-absorb the VSR term.

\subsection{Free Scenario}

Let us first start by examining how the free Lagrangian in \eqref{diracvsrlagr} would transform under a field redefinition. If we define the kinetic operator $\mathcal{Q}$ so that
\begin{equation}
    \mathcal{L}_{VSR}= \bar \psi \mathcal Q_\psi \psi = \bar \psi ( i\slashed \partial -m +i \lambda \slashed N) \psi \,.
\end{equation}
Then, under the field redefinition $\psi = \mathcal R \chi $ we have
\begin{equation}
    \mathcal{L}_{VSR} = \chi^\dagger \mathcal R ^\dagger \gamma^0 \mathcal Q_\psi \mathcal R \chi = \bar \chi \bar {\mathcal R} \mathcal Q_\psi \mathcal R \chi \,,
\end{equation}
from which we can identify the new kinetic operator as
\begin{equation}\label{Qchidef}
    \mathcal Q_\chi \equiv \bar {\mathcal R} \mathcal Q_\psi \mathcal R \,,
\end{equation}
with $\mathcal{\bar R} = \gamma^0 \mathcal R^\dagger \gamma^0$. Due to the form of the additional VSR term, an educated guess for a field redefinition able to re-absorb the VSR contribution is 
\begin{equation}
    \psi = \mathcal R \chi \,\,\text{with }\,\, \mathcal R = 1 + i\delta \slashed N\,,
\end{equation}
where $\delta$ is some to-be-defined constant. We should highlight that this transformation is not a projector, since it has a well-defined inverse 
\begin{equation}
    \mathcal R ^{-1} = 1 - i \delta \slashed N \,\,\to \,\,\, \mathcal R  \mathcal R ^{-1 }=1\,. 
\end{equation}
Moreover, thanks to the fact that $\slashed N^2= N^2=0$, the transformation can be represented in an exact exponential form as
\begin{equation}
    \mathcal R = e^{i \delta \slashed N} \,.
\end{equation}
Let us therefore calculate the new kinetic operator. Using $(\gamma^\mu)^\dagger = \gamma^0\gamma^\mu \gamma^0$ and the definition \eqref{Qchidef}, we obtain\footnote{Observe that when performing the adjoint-operation on $\mathcal R$ we should also consider the additional minus sign in the $\slashed N-$contribution coming from the integration by part of $\frac{1}{n\cdot \partial}$ which would otherwise act on $\bar \chi$.}
\begin{eqnarray}
    \mathcal Q_\chi = \gamma^0 (1+i\delta \slashed N ^\dagger) \gamma^0 \mathcal Q_\psi  (1+i\delta \slashed N) = (1+i\delta\slashed N) ( i\slashed \partial -m +i \lambda \slashed N)  (1+i\delta \slashed N)\,,
\end{eqnarray}
which after some algebra becomes
\begin{eqnarray}
    \mathcal Q_\chi = i\slashed \partial -(m+2\delta) +i(\lambda-2m\delta -2\delta^2) \slashed N\,.
\end{eqnarray}
Clearly, the right way to go here is to choose $\delta$ such that the parenthesis multiplying $\slashed N$ vanish. Solving the quadratic equation we found the following two solutions
\begin{equation}
    \delta_\pm = - \frac{m}{2} \pm \frac{1}{2} \sqrt{m^2+2\lambda}\,,
\end{equation}
and replacing these values into the kinetic operator we obtain
\begin{eqnarray}
     \mathcal Q_\chi = i\slashed \partial \mp \sqrt{m^2 +2\lambda}\,,
\end{eqnarray}
from which we see that we need to pick the positive solution $\delta_+$ to not have sign problems with the Dirac mass term
\begin{equation} \label{freeRtransf}
    \mathcal R = 1+i\delta_+ \slashed N = 1+ \frac{i}{2}( \sqrt{m^2 + 2\lambda}-m) \slashed N\,.
\end{equation}
Hence, after this field redefinition we found that the Lagrangian gets transformed to a standard Dirac Lagrangian but now with the shifted mass value $m_f$ instead of the old $m$
\begin{equation}
    \mathcal L_{VSR} = \bar \chi (i\slashed \partial - m_f) \chi \,,
\end{equation}
in agreement with what also commented in \cite{dunn2006} for its momentum space counterpart. This derivation reflects the absence of non-trivial VSR effects in the context of free-particle propagation and kinematics. Clearly, that changes when coupling fermions to sources or external fields, as we will see for the case of the electromagnetic one.\\
In general, field redefinitions can mess up the usual transformation properties of Dirac spinors like $\psi$. However, it is not difficult to see that this is not the case for the one defined here. In fact, assuming that the starting spinor $\psi$ transformed as a Dirac one, the new spinor $\chi$ will transform accordingly, as shown in Appendix \ref{apptransfproperties}. 

\subsection{Switching-on the Electromagnetic Interaction} 

When considering interactions with the electromagnetic field, the game changes: First, replacing the partial derivatives with the covariant ones would imply the presence of the four-potential $A_\mu$ also in our transformation $\mathcal R$ through the denominator of $\slashed N$. That would lead to a much more complex transformation, which we shall avoid by working in the light-cone gauge $n\cdot A =0$. Being 
\begin{equation}
    Q^{EM}_\psi = i\slashed \partial - m  +i\lambda\slashed N -e \slashed A\,,
\end{equation}
with this gauge choice we clearly have that
\begin{equation}\label{QEMchi1}
    Q^{EM}_\chi = \bar {\mathcal R} \mathcal Q^{EM}_\psi \mathcal R= Q_\chi - e \bar {\mathcal R} \slashed A \mathcal R \,.
\end{equation}
The calculation of the new term is not trivial, because $\slashed N$ in the transformation \eqref{freeRtransf} would not only act on the spinor $\chi$. However, it can be tackled with some caution
\begin{equation} \label{RbarAR1}
     \bar {\mathcal R} \slashed A \mathcal R = (1+i\delta_+ \slashed N) \slashed A (1+i\delta_+ \slashed N) = \slashed A +i \delta_+ \{\slashed N, \slashed A\}-\delta^2_+ \slashed N \slashed A \slashed N\,.
\end{equation}
To carry on the calculation, we observe that the above anticommutator is equal to
\begin{eqnarray}
     \{\slashed N , \slashed A\} &=& \{ \frac{1}{n\cdot \partial} \slashed n, \slashed A\} = \frac{1}{n\cdot\partial}\{\slashed n, \slashed A\} + [\slashed A, \frac{1}{n\cdot\partial}] \slashed n  \\
     &=& -\sum_{n =1} ((-n\cdot \partial)^{n} \slashed A) \frac{1}{(n\cdot \partial)^{n+1}}  \slashed n \nonumber \,,
\end{eqnarray}
where we used the expression of \eqref{commutator1nDwithD} in the light-cone gauge
\begin{equation}
    [A_\nu , \frac{1}{n\cdot \partial}] \psi = - \sum_{n =1} ((-n\cdot \partial)^{n} A_{\nu}) \frac{1}{(n\cdot \partial)^{n+1}} \psi \,.
\end{equation}
From the above considerations, we trivially see that the last term in \eqref{RbarAR1} is zero 
\begin{equation}
    \slashed N \slashed A \slashed N = \{ \slashed N, \slashed A\} \slashed N - \slashed A \slashed N^2 = 0\,,
\end{equation}
since $\slashed n^2 = 0$. Thus, the new contribution in \eqref{QEMchi1} becomes 
\begin{equation} \label{RbarAR2}
     \bar {\mathcal R} \slashed A \mathcal R = \slashed A -i \delta_+ \sigma^{\mu\nu} n_\mu \sum_{n =1} ((-n\cdot \partial)^{n} A_\nu) \frac{1}{(n\cdot \partial)^{n+1}} \,.
\end{equation}
Finally, the Lagrangian expressed in terms of the spinor $\chi$ would look like a standard interacting Dirac Lagrangian adjoined with new interaction terms
\begin{equation}
    \mathcal{L} = \bar \chi \left ( i \slashed D - m_f + i e \delta_+ \sigma^{\mu\nu} n_\mu \sum_{n =1} ((-n\cdot \partial)^{n} A_\nu) \frac{1}{(n\cdot \partial)^{n+1}} \right ) \chi \,,
\end{equation}
where we should remind ourselves that we are working in the light-cone gauge. What we showed, then, is that the theory of a VSR spinor $\psi$ minimally coupled to $A_\mu$ is equivalent, at this level, to the theory of a Dirac spinor $\chi$ interacting with the electromagnetic field through a peculiar non-minimal prescription. This analogy may prove to be particularly useful in certain context, like the diagrammatic calculation of specific processes. Nevertheless, being mostly interested in low-energy limits, is not the approach we will undertake in this thesis. In fact, as already mentioned, there may be further issues involved with field redefinitions \cite{Colladay:2002eh} which we leave for future work.

\chapter[LANDAU LEVELS AND ELECTRON GYROMAGNETIC FACTOR IN VSR]{Landau Levels and Electron Gyromagnetic Factor in VSR} \label{ch2}

The first physical calculation that we decided to tackle within the Dirac VSR context was the corrections to the relativistic Landau energy levels, that is, the spectrum of a charged fermion surrounded by a constant and homogeneous magnetic field $\vec B$. Besides the theoretical significance, the pragmatic reason for this calculation is to connect VSR to electron $\mathsf g-2$ experiments and use them to constrain the LV parameters. We will explore this phenomenological link in Subsection \ref{sec:egfactor}. \\
Let us start the analysis from the easiest possible case, which is when $\hat n$ is parallel to the quantization direction determined by the magnetic field, $\hat n // \vec B$. Later, we will relax this parallelism condition and allow for a more generic $\hat n $, while keeping the magnetic field fixed along $\hat e_3$
\be
\vec B = (0,0,B)\,.
\ee
In both cases, we will work in the following gauge
\begin{eqnarray}
A^\mu =( 0,0,B x^1,0) \, ,
\label{eq_gauge}
\end{eqnarray}
such that translational symmetry along $ {\hat e}_2$ and ${\hat e}_3$ is preserved
\begin{eqnarray}
\left[A^{\mu}(x),p_3 \right]_{-} = \left[A^{\mu}(x),p_2 \right]_{-} = 0 \,,
\label{eq_comm}
\end{eqnarray}
with $p_j = -p^j = i\partial_{j}$ representing here the components of the momentum operator. Since in this context the electromagnetic potential depends only on $x^1$, we can factorize a solution of the VSR interacting EOM \eqref{emvsreom} in the following form
\begin{equation}
\psi(x) = e^{-i E t} e^{i\left( k^3 x^3 + k^2 x^2 \right)}\left(\begin{array}{c}\varphi(x^1)\\\chi(x^1)\end{array}\right) .
\label{eq_eig1}
\end{equation}
In light of this, we will replace the operator components $p_2$ and $p_3$ by their relative eigenvalues, that is, $p_3 \rightarrow k_3$ and $p_2 \rightarrow k_2$. Moreover, to finally describe electrons, we will take $e =-|e|$, where $|e|$ is the absolute value of the electron's charge.

\section{$\hat n \, // \vec B $ Scenario \label{sec:sub2.1.1}}

Let us start with the simpler case $\hat n \, // \vec B$, where we have
\be n^\mu = (1,0,0,1) \,. \ee
Having chosen the gauge \eqref{eq_gauge} implies $n\cdot A =0$ automatically. Hence, the VSR operator for an eigenstate of the form \eqref{eq_eig1} simplifies to
\begin{eqnarray}
i\lambda\slashed{N} \rightarrow -\lambda \frac{\gamma^0 - \gamma^3}{E - k^3}\, .
\label{eq_VSR2}
\end{eqnarray}
Inserting Eq.~\eqref{eq_eig1} and Eq.~\eqref{eq_VSR2} into Eq.~\eqref{emvsreom}, we obtain the following system of equations for the $x^1-$dependent part of the bi-spinor $\psi$
\begin{eqnarray}
\left[\begin{array}{cc} \left(E - m - \frac{\lambda}{ E - k^3}  \right)\mathbf{1} & e B x^1 \sigma^2 +
\frac{\lambda}{ E - k^3}\sigma^3 - \sigma^i p^i\\-e B x^1 \sigma^2 -
\frac{\lambda}{E - k^3}\sigma^3 + \sigma^i p^i & -\left(E + m - \frac{\lambda}{ E - k^3}  \right)\mathbf{1}
\end{array} \right] \left(\begin{array}{c}\varphi(x^1)\\\chi(x^1) \end{array}\right) = 0 \,. 
\label{eq_Dirac1}
\end{eqnarray}
Solving for the lower spinor $\chi$ in terms of the upper one $\varphi$, leads to the expression
\begin{eqnarray}
\chi(x^1) = -\frac{eB x^1\sigma^2 + \frac{\lambda}{E - k^3}\sigma^3 -\sigma^i p^i}{E + m - \frac{\lambda}{E - k^3}}\varphi(x^1) \, .
\label{eq_chi}
\end{eqnarray}
Replacing Eq.~\eqref{eq_chi} back into the first system of equations \eqref{eq_Dirac1}, we get an equation for the upper spinor $\varphi$ only
\begin{eqnarray} \label{eq_phi}
\left[\left( E - \frac{\lambda}{ E - k^3} \right)^2 - m^2 - \left(eB x^1 \sigma^2 + \frac{\lambda}{E - k^3}\sigma^3 - \sigma^i p^i  \right)^2  \right] \varphi(x^1) = 0 \, . 
\end{eqnarray}
Defining the coefficient
\begin{eqnarray}
a(k^3,E) = \frac{1}{|e| B} \left[  \left(E - \frac{\lambda}{E-k^3}  \right)^2 - m^2- \left(k^3 - \frac{\lambda}{E-k^3}  \right)^2  \, \right] ,
\label{eq_a}
\end{eqnarray}
and exploiting the standard properties of the $SU(2)$ algebra generated by Pauli matrices, we can calculate the square differential operators in Eq.~\eqref{eq_phi}, obtaining
\begin{eqnarray}
\left[|e| B a(k^3,E)-\left(e B x^1 - k^2  \right)^2  +  e B\sigma^3 - p_1^2 \right] \varphi(x^1) = 0 \, .
\label{eq_phi2}
\end{eqnarray}
The above equation is diagonal in the two components of the spinor $\varphi(x^1)$
\begin{eqnarray}
\varphi(x^1) = \left(\begin{array}{c} f^{1}(x^1)\\f^{2}(x^1) \end{array} \right) \, ,
\end{eqnarray}
and will therefore have two independent solutions $f_\alpha (x^1)$
\begin{equation}
    \vec f_{+1} (x^1) \equiv \left(\begin{array}{c} f_{+1}(x^1)\\0 \end{array} \right) \, , \,\,\,
    \vec f_{-1} (x^1) \equiv  \left(\begin{array}{c} 0\\f_{-1}(x^1) \end{array} \right)\,,
\end{equation}
determined by the solutions to the two independent differential equations
\begin{eqnarray}
&&\left[ -\partial_{1}^2 + (eB)^2\left( x^1 - \frac{k^2}{e B}  \right)^2- \alpha e B - |e| B a(k_3,E) \right]f_{\alpha}(x^1)=0\, ,
\label{eq_fa1}
\end{eqnarray}
with $\alpha=\pm 1$ representing the two eigenvalues of $\sigma^3$. Now, it is convenient to define the auxiliary dimensionless coordinate
\begin{eqnarray} \label{xicoord}
\xi \equiv \sqrt{|e| B}\left(x^1 - \frac{k^2}{e B} \right) \,,
\end{eqnarray}
such that dividing Eq.~\eqref{eq_fa1} by $|e| B$, we get\footnote{This calculation could be easily generalized to an electric charge with indefinite sign by just replacing everywhere $|e|$ with $s \,e$ where $s= sign(e)$ is the sign of the charge.}
\begin{eqnarray}
\left[-\frac{d^2}{d\xi^2} + \xi^2 + \alpha \right]f_{\alpha}(\xi) = a(k^3,E) f_{\alpha}(\xi)\,.
\label{eq_falpha}
\end{eqnarray}
The only $L^2$-normalizable solutions of Eq.~\eqref{eq_falpha} are the functions
\begin{eqnarray}\label{def_falpha}
f_{n,\alpha}(\xi) = C e^{-\xi^2/2} H_{n}(\xi)\,,
\end{eqnarray}
where $H_n(\xi)$ are the Hermite polynomials of order $n\in\mathbb{N}_0$, provided the following quantization condition is satisfied
\begin{eqnarray}
a(k^3,E) - \alpha = 2 n + 1 \,,\,\, n\in\mathbb{N}_0 \,,\,\,\alpha=\pm 1\,,
\label{eq_aq}
\end{eqnarray}
while, the normalization coefficient $C$ in Eq.~\eqref{def_falpha} is obtained from the orthonormality condition for the Hermite polynomials
\begin{equation}
    \int _{-\infty}^{+\infty} d\xi \; e^{-\xi^2} H_n(\xi) H_m(\xi) = 2^n n! \sqrt{\pi} \delta_{nm}\,.
    \label{eq_normHerm}
\end{equation}
Following the quantization condition \eqref{eq_aq}, the energy spectrum $E \equiv E_{\pm}\left(k^3,n,\alpha \right)$ is defined by the roots of the following algebraic equation
\begin{eqnarray}
\left(E - \frac{\lambda}{E - k^3}\right)^2 - \left(k^3-\frac{\lambda}{E-k^3}\right)^2 = |e| B\left(2 n + 1 + \alpha\right) + m^2 \,,
\end{eqnarray}
which can be solved explicitly, to obtain the exact energy eigenvalues
\begin{eqnarray}
E^{(0)}_{\pm}(k^3, n,\alpha) &=& \pm \sqrt{|e| B (2n + 1 + \alpha) + (k^3)^2 + m^2_f } \; .
\label{eq_VSRspectrum}
\end{eqnarray}
As expected, in the limit $\lambda \rightarrow 0$ where the full Lorentz symmetry is restored, the spectrum in Eq.~\eqref{eq_VSRspectrum} reduces to the well known “unperturbed” relativistic Landau levels
\begin{eqnarray} \label{E0}
\left.E^{(0)}_{\pm} (k^3, n,\alpha)\right|_{\lambda=0} = \pm\,  E^{(u)}(k^3, n,\alpha) \equiv \pm \sqrt{|e| B (2n + 1 + \alpha) + (k^3)^2 + m^2} \,.
\end{eqnarray}
Furthermore, it is also clear from Eq.~\eqref{eq_VSRspectrum} that for this configuration, where $\hat n$ is parallel to the field $\vec{B}$, the sole effect of the VSR term in the single-particle spectrum is to shift the particle mass $m \rightarrow m_f$.\\ 
Following the previous analysis, we observe that each energy eigenvalue, excluding the ground state one for $n=0$ and $\alpha=-1$, can be obtained with the two combinations $(n\, ,\alpha= +1)$ and $(n+1 \,, \alpha=-1)$ and is therefore doubly degenerate. Introducing the system eigenstates $\ket{\vec f^{(0)} , n ,\alpha } $ such that $\vec f^{(0)}_{n ,\alpha} (\xi) = \langle\xi|\vec{f} ^{(0)} ,n, \alpha\rangle$, our eigenvectors' basis will look like
\begin{eqnarray}
    \bigg \{
    \ket{\vec{f}^{(0)}, 0,- 1} , \left [\ket{\vec{f}^{(0)},0,+ 1}, \ket{\vec{f}^{(0)},1,- 1} \right ]\, , ...\, ,  \left [\ket{\vec{f}^{(0)},n,+ 1}, \ket{\vec{f}^{(0)},n+1,- 1} \right ] , ...
    \bigg \} \nonumber \,.
\end{eqnarray}
Here, the square brackets are highlighting the 2-dimensional degenerate eigenspaces. To simplify the notation in the rest of the calculations and to directly label each degenerate eigenspace, we reorder the eigenstates by relabeling $\ket{\vec f^{(0)} , n , \alpha} \to \ket{\vec f^{(0)} , \bar n, \alpha} $, where
\begin{equation}
    \left\{\begin{array}{l} 
    \bar n = n \;\;\;\;\;\;\;\;\; \text{for } \; \alpha =-1 \,,\\ 
    \bar n= n+1 \;\;\; \text{for } \; \alpha =+1 \, . 
    \end{array} \right.  
\end{equation}
The eigenvectors' basis then becomes
\begin{eqnarray}
    \bigg \{
    \ket{\vec{f}^{(0)}, \bar 0,-1  } , \left [\ket{\vec{f}^{(0)},\bar 1,+1 }, \ket{\vec{f}^{(0)},\bar 1,- 1} \right ]\, , ...\, , \left [\ket{\vec{f}^{(0)}, \bar n,+1 }, \ket{\vec{f}^{(0)}, \bar n,-1 } \right ] , ...
    \bigg \} \nonumber \, , 
\end{eqnarray}
with each degenerate eigenspace ($\bar n \geq1$) spanned by the two orthogonal eigenvectors
\begin{eqnarray} \label{eq_unpertspinor}
\langle\xi|\vec{f}^{(0)},\bar n,+1\rangle &=& \frac{1}{\pi^{1/4}2^{\frac{\bar n-1}{2}}\sqrt{(\bar n-1)!}}\left( \begin{array}{c}e^{-\frac{\xi^2}{2}}H_{\bar n-1}(\xi)\\0 \end{array}\right)\,, \\
\langle\xi|\vec{f}^{(0)}, \bar n,-1\rangle &=& \frac{1}{\pi^{1/4}2^{\frac{\bar n}{2}}\sqrt{\bar n!}}\left( \begin{array}{c}0\\e^{-\frac{\xi^2}{2}}H_{\bar n}(\xi) \end{array}\right) \,, \nonumber
\end{eqnarray}
while the ground state $\bar n=n=0$ is defined by
\begin{equation}\label{ground0}
\langle\xi|\vec{f}^{(0)},\bar 0,-1\rangle = \frac{1}{\pi^{1/4}}\left( \begin{array}{c} 0 \\ e^{-\frac{\xi^2}{2}}H_{0}(\xi)  \end{array}\right)\,,
\end{equation}
and its energy is not degenerate.

\section{Generic $\hat n$-Orientation \label{sec:sub2.1.2}}

Let us now consider the generic case of the VSR unit vector $\hat n$ oriented with an angle $\theta \in [0,\pi]$ with respect to the magnetic field $  {{\hat n}}$, that is $\vec{B}\cdot  {{\hat n}} = B\cos\theta$. Therefore, without loss of generality, we select our coordinate system such that
\begin{eqnarray} \label{genericnorientation}
n^{\mu} = (1,  {{\hat n}}) = (1,\sin\theta,0,\cos\theta) \, .
\end{eqnarray}
We stick to the same gauge as in the previous case, $A^{\mu} = (0,0,B x^1,0)$, so that the translational invariance along the $x^2$ and $x^3$ directions allows us to choose the same separation of variables as in Eq.~\eqref{eq_eig1}, while preserving the light-cone condition $n \cdot A=0$. \\
In this case, however, the VSR term in the equation of motion reduces to the form
\begin{eqnarray}
i \lambda \slashed{N} \rightarrow -\lambda \frac{\gamma^0 - \gamma^1\sin\theta - \gamma^3\cos\theta }{E - k^3\cos\theta -  p^1 \sin\theta} \, .
\label{eq_VSR2-2}
\end{eqnarray}
Substituting Eq.~\eqref{eq_VSR2-2} into Eq.~\eqref{emvsreom}, we obtain for this general case the expression
\begin{eqnarray}
&&\left[\gamma^0 E + \gamma^1 p_1 - \gamma^2\left( k^2 - e B x^1 \right) - \gamma^3 k^3 - m\right.\\
&& \;\;\;\;\;\;\;\;\;\;\; \left.-\lambda \frac{\gamma^0 - \gamma^1\sin\theta - \gamma^3\cos\theta }{E - k^3\cos\theta -  p^1 \sin\theta}
\right]\left(\begin{array}{c}\varphi(x^1)\\\chi(x^1) \end{array}\right) = 0 \,.\nonumber
\label{eq_EOM2}
\end{eqnarray}
From this linear system, the lower spinor $\chi(x^1)$ can be solved in terms of the upper component $\varphi(x^1)$, as done in the case $\vec{B} \parallel {\hat n}$, and then replaced again to obtain the following single equation for the upper spinor $\varphi(x^1)$
\begin{eqnarray}
&& \left( \left[ E - \frac{\lambda}{E - k^3\cos\theta -  p^1 \sin\theta} \right]^2 - m^2 - \left[ \left(p^1 - \frac{\lambda \sin\theta}{E - k^3\cos\theta -  p^1 \sin\theta}\right)\, \sigma^1 \right.\right. \nonumber \\
&& \left.\left. \;\;\; + \left(k^2 - e B x^1 \right)\sigma^2 + \left(k^3 - \frac{\lambda \cos\theta}{E - k^3\cos\theta -  p^1 \sin\theta} \right)\sigma^3 \right]^2 \right ) \varphi(x^1) = 0 \, .
\label{eq_phi3}
\end{eqnarray}
By expanding the squares, we observe that, due to the anti-commutation relations satisfied by the Pauli matrices, among the terms with mixed $\sigma$'s only the ones that involve operators acting on $x^1 \varphi(x^1)$ can have a chance to generate a surviving part after summing up the sigma products with interchanged indices. For example, for terms involving $ p^1$, we can exploit the fact that, for any function $g(x^{1})$
\begin{equation} \label{anticom2}
\left[\sigma^{1} p^1,g(x^1)\sigma^2 \right]_{+} \varphi (x^1) = -\sigma^{3}\partial_{1}g(x^1) \varphi (x^1)\, .
\end{equation}
Bearing this in mind and defining the auxiliary operator
\begin{equation} \label{defD}
      \mathcal P \equiv (\tilde E -  p^1 \sin\theta)^{-1}  \,\text{ with }\,\, \tilde E \equiv E-k^3\cos\theta\,,
\end{equation}
we expand the square brackets in \eqref{eq_phi3}, obtaining
\begin{eqnarray} \label{app1.1}
&&\left[\left( E - \lambda  \, \mathcal P \right)^2 - m^2 -( {p}^1 - \lambda \sin\theta   \mathcal P)^2-(k^2 - e B x^1)^2- (k^3 - \lambda \cos\theta  \mathcal P)^2 \right. \nonumber\\
&& \;\;\; - i e B   \sigma^1   \sigma^2 - e B \lambda \sin\theta \mathcal P x^1   \sigma^1   \sigma ^2 - e B \lambda \sin\theta x^1   \mathcal P    \sigma^2   \sigma ^1   \nonumber \\
&& \;\;\; \left.- e B \lambda \cos\theta   \mathcal P x^1   \sigma^3   \sigma^2 -e B \lambda \cos\theta  x^1   \mathcal P   \sigma^2   \sigma^3\right]\varphi(x^1)  = 0 \, . 
\end{eqnarray}
Therefore, we now have to calculate $  \mathcal P (x^1)$. To do that, we will use the integral representation \eqref{intrepr}, implying
\begin{eqnarray}
        \mathcal P (x^1 \varphi) = \int_0^\infty dt \; e^{-t (\tilde E -  p^1 \sin\theta)} \, x^1 \varphi =\int_0^\infty dt \; e^{-t \tilde E} \sum_{i=0}^\infty (t \sin\theta   p^1)^i x^1 \varphi  \,,
\end{eqnarray}    
that can be expanded as
\begin{eqnarray}
    && \mathcal P (x^1 \varphi) =  \int_0^\infty dt \; e^{-t \tilde E} \left ( x^1 \sum_{i=0}^\infty (t \sin\theta   p^1)^i \, \varphi  -i t \sin\theta \sum_{i=1}^\infty (t \sin\theta   p^1)^{i-1}  \right ) \nonumber .\\
\end{eqnarray}
Sending the index of the second sum to $i \rightarrow i-1$, and “reversing” the use of the integral representation, we have the relation
\bea
       \mathcal P(x^1 \varphi ) &=& x^1   \mathcal P \varphi +i\sin\theta \frac{d}{d\tilde E}(  \mathcal P) \varphi =x^1   \mathcal P \varphi -i\sin\theta   \mathcal P ^2 \varphi  \,.
\eea
Thus, Eq.\eqref{app1.1} becomes
\begin{eqnarray}\label{Aeq1}
    &&\left[\left( E - \lambda   \mathcal P \right)^2 - m^2 -( {p}^1 - \lambda \sin\theta   \mathcal P)^2-(k^2 - e B x^1)^2- (k^3 - \lambda \cos\theta \mathcal P)^2 \right. \nonumber\\
    && \;\;\;\left. + e B   \sigma^3 - e B \lambda \sin^2\theta   \mathcal P^2    \sigma^3  +e B \lambda \sin\theta \cos\theta   \mathcal P^2    \sigma^1 \right]\varphi(x^1)  = 0 \, ,
\end{eqnarray}
which has the correct limits, in fact, for $B\to 0$, we re-obtain, as expected, the usual VSR dispersion relation 
\begin{equation}
    [E^2 -p^2 -m^2 -2\lambda ]\, \varphi = 0\,,
\end{equation}
while, for $\theta \to 0$ we find again Eq.~\eqref{eq_phi2}, the equation of motion for the case $\vec B \parallel \vec n$.\\
Finally, expanding the squares in \eqref{Aeq1} and using definitions \eqref{defD}, we get to
\begin{eqnarray} \label{eq_phi4}
&& \left[
E^2 - (k^3)^2 - m^2  -M^2 - ( p^1)^2 + e B  \sigma^3 - (k^2 -e B x^1)^2  \right. \\
&& \left. \;\;\;\;\;\; - \frac{ \lambda e B\sin^2 \theta}{(\tilde E -  p^1 \sin\theta)^2}  \sigma^3 + \frac{ \lambda e B\sin\theta \cos\theta}{(\tilde E -  p^1 \sin\theta)^2}  \sigma^1
\right]\varphi(x^1) = 0 \, . \nonumber
\end{eqnarray}
At this point, we introduce again
\begin{eqnarray} \label{aEdef}
 |e| B \,a(k^3, E) \equiv E^2 - (k^3)^2 - m^2 - 2\lambda \, ,  
\end{eqnarray}
along with the change of variable $x^1 \to \xi$ already seen in \eqref{xicoord}. Dividing Eq.~\eqref{eq_phi4} by $|e| B$, we obtain
\bea \label{eqref111}
&& \left[ -\partial_{\xi}^2 + \xi^2 + \sigma ^3  -a -  \frac{\lambda \sin\theta}{(\tilde E - \sqrt{|e| B} \sin\theta\, p^\xi )^2} (\sin\theta  \, \sigma^3 - \cos\theta \, \sigma^1) \right ] \varphi(x^1) = 0 \, . \nonumber \\
\eea
The above Eq.\eqref{eqref111} can be expressed as an eigenvalue equation
\begin{eqnarray}\label{eq_f1}
    \left[ -\partial_{\xi}^2 + \xi^2 + \sigma ^3 - \frac{\lambda \sin \theta }{  P^2_\xi} (\sin\theta\,  \sigma^3 - \cos\theta \,  \sigma^1  ) \right ] \left(\begin{array}{c}f^{1}(\xi)\\f^{2}(\xi) \end{array}\right)= a \left(\begin{array}{c}f^{1}(\xi)\\f^{2}(\xi) \end{array}\right) ,
\end{eqnarray}
where for convenience we defined the operator 
\begin{equation} \label{Pxidef}
    P_\xi \equiv \tilde E - \sqrt{|e| B}  \sin\theta \,p^\xi = \tilde E + i \sqrt{|e| B}  \sin\theta \,\partial_\xi \,.
\end{equation}
In contrast to the previous case, the $f^1$ and $f^2$ components are now mixed, which makes finding an exact solution much more complicated. However, thinking of VSR as a slight deviation from special relativity, we can consider $\lambda$ much smaller than the other system's energy scales, like $m^2$ or $eB$, and thus apply perturbation theory.

\section{Perturbative Scheme in VSR}

In terms of the eigenstates $\ket {\vec f}$ of the operator on the left-hand side of Eq.~\eqref{eq_f1}, we can interpret the equation \eqref{eq_f1} as a standard eigenvalue equation
\begin{equation} \label{firstperteq}
    (H_0 + \lambda V ) \ket{\vec f}= a \ket{\vec f}\, ,\,\, \text{with } \, \braket{\xi|\vec f} =\left(\begin{array}{c}f^{1}(\xi)\\f^{2}(\xi) \end{array}\right) ,
\end{equation}
where, for the sake of the perturbative treatment, we identified an unperturbed operator $H_0$ and a perturbation operator $V$ as follows
\begin{eqnarray}
&& H_0 = -\partial_{\xi}^2 + \xi^2 + \sigma ^3 \,, \nonumber \\
&& V = -\frac{\sin\theta}{   P^2_\xi} (\sin\theta  \,\sigma^3 - \cos\theta \, \sigma^1) \, .
\end{eqnarray} 
We can then approach the problem in a perturbative scheme, with $\lambda$ as our small parameter. Let us start by expanding $\ket {\vec f}$ and $a$ in a power series up to first order in $\lambda $, so that
\begin{eqnarray}
 a &=& a^{(0)} + \lambda \, a^{(1)} + O(\lambda^2) \, ,\nonumber \\
\ket {\vec f} &=&  \ket{\vec f^{(0)}} + \lambda \ket{\vec f^{(1)}} + O(\lambda^2)  \, .
\end{eqnarray}
Replacing the expansions in the original eigenvalue problem \eqref{firstperteq}, at zero order, we obtain again the equation solved in the previous section
\begin{equation}
    H_0 \ket{\vec f^{(0)},\bar n,\alpha} = a^{(0)}_{\bar n,\alpha} \ket{\vec f^{(0)}, \bar n,\alpha} \, ,
\end{equation}
where the eigenvalues are $a^{(0)}_{n,\alpha} = 2n +1+\alpha$ or $a^{(0)}_{\bar n,\alpha} =2 \bar n$ in the $\bar n -$basis, and the unperturbed eigenvectors are listed in the Eqs.~\eqref{eq_unpertspinor} and \eqref{ground0}. At first order in the perturbation, we instead find
\begin{equation}
    (H_0 -a^{(0)}_{\bar n,\alpha } \mathbf{1}) \ket{\vec f^{(1)} , \bar n ,\alpha} = (a^{(1)}_{\bar n, \alpha} \mathbf{1} - V) \ket{\vec f^{(0)} ,\bar n,\alpha} \,,
    \label{eq_pert1}
\end{equation}
which must be treated differently depending on the value of $\bar n$ considered.

\subsection{Perturbative Correction to the State $\bar n=0$}

For the case $\bar n=0$, we necessarily have $\alpha=-1$ and no degenerate eigenvectors, which allows us to use ordinary perturbation theory. Multiplying \eqref{eq_pert1} by $\bra{f^{(0)}, \bar 0,-1 }$ and applying the zero-order property $\bra{\vec f^{(0)}, \bar n,\alpha}(H_0-a^{(0)}_{ \bar n,\alpha}     \mathbf{1}) = 0$, we see that
\begin{eqnarray}
    a^{(1)}_{\bar 0,-1} &=& \bra{\vec f^{(0)} ,\bar 0 ,-1}{V}\ket{ \vec f^{(0)} , \bar 0 ,-1} \,.
\end{eqnarray}
Inserting two identities through the completeness relation $1= \int d\xi \ket{\xi}\bra{\xi}$ and using the expression \eqref{ground0} of the ground state, we obtain
\begin{eqnarray}\label{a1n0.1}
a^{(1)} _{\bar 0,-1} &=&   \frac{1}{\sqrt{\pi} } \int_{-\infty}^{\infty} d\xi \; e^{-\xi^2/2} H_0(\xi) \frac{\sin^2\theta}{P^2_\xi} (e^{-\xi^2/2} H_0(\xi))  \\
&=& - \frac{\sin^2\theta}{ \sqrt{\pi} \tilde E^2 } \frac{d}{d A }\int_{-\infty}^{\infty} d\xi \; e^{-\xi^2/2} H_0(\xi)  \frac{1}{A+i \eta\sin\theta \partial_\xi} (e^{-\xi^2/2} H_0(\xi)) \bigg |_{A=1} \, , \nonumber
\end{eqnarray}
where we have introduced the new dimensionless quantity
\begin{equation}
    \eta \equiv \sqrt{|e| B}/{\tilde E} \,.
\end{equation}
For calculation purposes, we introduce a Schwinger-type integral representation for the inverse operator in Eq.~\eqref{a1n0.1} (which is valid for $A>0$)
\begin{eqnarray}
\frac{1}{A+i \eta\sin\theta \partial_\xi} = \int_{0}^{\infty}dt\,e^{-t\left(A+i \eta\sin\theta \partial_\xi\right)} \,.
\label{intrepr}
\end{eqnarray}
With the integral form in Eq.~\eqref{intrepr}, we get
\begin{eqnarray}
a^{(1)} _{\bar 0,-1} &=& - \frac{ \sin^2\theta}{\sqrt{\pi} \tilde E^2 } \frac{d}{d A}\int_{0}^\infty dt \int_{-\infty}^{\infty} d\xi \; e^{-\xi^2/2} H_0(\xi) \;  e^{- A t (1+i \eta\sin\theta \partial_\xi)} (e^{-\xi^2/2} H_0(\xi)) \bigg |_{A=1} \nonumber\\
&=& - \frac{\sin^2\theta}{\sqrt{\pi} \tilde E^2 } \frac{d}{dA}\int_{0}^\infty e^{-At} dt \int_{-\infty}^{\infty} d\xi \; e^{-\xi^2/2} H_0(\xi) \;  e^{-i t \eta\sin\theta \partial_\xi} (e^{-\xi^2/2} H_0(\xi)) \bigg |_{A=1} \,, \nonumber
\\
\end{eqnarray}
where clearly we take the limit $A\rightarrow 1$ at the end. Defining the integrals
\begin{equation}\label{I1nk}
   I_1(\bar n,k)=  \int_{-\infty}^{\infty} d\xi \; e^{-\xi^2/2} H_{\bar n}(\xi) \,
   \partial_\xi^k (e^{-\xi^2/2} H_{\bar n}(\xi)) \,,
\end{equation}
and expanding the exponential operator, we get
\begin{eqnarray}\label{a1n0.2}
a^{(1)} _{\bar 0,-1} &=& - \frac{\sin^2\theta}{ \sqrt{\pi} \tilde E^2 }  \sum_{k=0}^\infty \frac{(-i \eta\sin\theta )^k}{k!} \frac{d}{dA} \int_{0}^\infty dt \; e^{-At} \, t^k \, I_1(0,k)  \nonumber \\
&=&  \frac{ \sin^2\theta}{ \sqrt{\pi} \tilde E^2 }  \sum_{k=0}^\infty (k+1) (i \eta\sin\theta )^k  I_1(0,k) \, .
\end{eqnarray}
The parity properties of the $I_1(\bar n ,k)-$integrand, which are shown in Appendix \ref{app2} imply the vanishing of those integrals for odd values of $k$. Moreover, as shown in detail in Appendix \ref{app2}, the analytical expression of the integrals $I_1(\bar n, k)$ for even $k$ is 
\begin{eqnarray} \label{I1n2kexpr}
I_1(\bar n,2k)=\sqrt{\pi} (-1)^k \bar n! \, 2^{\bar n-2k}\frac{\Gamma(2k+1)}{\Gamma(k+1)} F(-k,-\bar n;1;2) \,,
\end{eqnarray}
where $F(a,b;c;z)$ is the Hypergeometric function. For $\bar n =0$, \eqref{I1n2kexpr} reduces to
\begin{eqnarray}
I_1(0,2k) = \sqrt{\pi} \, 2^{-2k}(-1)^k \frac{\Gamma(2k+1)}{\Gamma(k+1)}\,.
\end{eqnarray}
Therefore, we can rewrite Eq.~\eqref{a1n0.2} as
\begin{eqnarray} \label{a1n0sumk}
    a^{(1)} _{\bar 0,-1} &=& \frac{ \sin^2\theta}{ \sqrt{\pi} \tilde E^2 }  \sum_{k=0}^\infty (2k+1)(-1)^k \left(\eta\sin\theta \right)^{2k}  I_1(0,2k)\nonumber\\
    &=& \frac{ \sin^2\theta}{ \tilde E^2 } \sum_{k=0}^\infty \left(\frac{\eta\sin\theta}{2}\right)^{2k}\frac{\Gamma(2k + 2)}{\Gamma(k+1)} \,,
\end{eqnarray}
that is a completely real expression, as one would expect. The $k-$sum in Eq.~\eqref{a1n0sumk} is not convergent in the standard sense. However, it can be regularized, for example, by the Borel prescription, to obtain a closed form in terms of the incomplete Gamma function $\Gamma(-1/2,z)$ as follows (see Appendix \ref{Borel} for details)
\begin{eqnarray}\label{aln0borel}
a^{(1)} _{\bar 0,-1}= \frac{\sin^2\theta}{\tilde{E}}\frac{e^{-\frac{1}{\eta^2\sin^2\theta}}}{\left(- \eta^2\sin^2\theta\right)^{3/2}}  \,\Gamma \left( -\frac{1}{2},-\frac{1}{\eta^2\sin^2\theta}\right) .
\end{eqnarray}
At the lowest orders for $\eta\ll1$, both the power series Eq.~\eqref{a1n0sumk} and the regularized Borel Eq.~\eqref{aln0borel} reduce to
\begin{eqnarray}
a^{(1)} _{\bar 0,-1}\simeq\frac{ \sin^2\theta}{ \tilde E^2 }\left[1 + \frac{3}{2}\eta^2\sin^2\theta + O(\eta^4)\right] .
\end{eqnarray}
We note that, at this order $\eta^2$, the result is unique regardless of the regularization prescription. To go further in our analysis, we assume to be in a situation where the magnetic field is small with respect to other energy scales $\eta = \frac{\sqrt{eB}}{\tilde E}  \ll1$. In this weak field approximation, we have that at first order in $\lambda$
\begin{equation}\label{eqa10p}
    a_{0,-1} = a^{(0)} _{0,-1} + \lambda \, a^{(1)} _{0,-1} \approx \lambda \frac{\sin^2\theta}{\tilde E^2}\left[1 + \frac{3}{2}\eta^2\sin^2\theta  \right] ,
\end{equation}
which is expressed in the $n- $notation since it makes no difference for the ground state.

\subsection{Perturbative Correction to the States $\bar n>0$}

For $\bar n > 0$, in the unperturbed case, we have two-dimensional degenerate eigenspaces spanned by the two degenerate spinors in Eq.~\eqref{eq_unpertspinor}. Hence, when including the perturbation, we must workout the calculations within each degenerate eigenspace. The first step is to express the first-order eigenvector corrections $\ket{\vec f^{(1)} ,\bar n ,\alpha} $ as a combination of the unperturbed contributions
\begin{eqnarray}
    \ket{\vec f^{(1)} ,\bar n ,\alpha} = \sum_{\alpha' = \pm} C^{\bar n}_{\alpha,\alpha'}\ket{\vec f^{(0)} ,\bar n ,\alpha'} \,.
    \label{eq_funcexp}
\end{eqnarray}
Then, projecting Eq.~\eqref{eq_pert1} over each of the unperturbed spinors $\bra{\vec f^{(0)}, \bar n, \alpha'}$, we obtain the linear eigenvalue system
\begin{equation}
    \left[\begin{array}{cc} V^{\bar n}_{+1,+1} & V^{\bar n}_{+1,-1}\\V^{\bar n}_{-1,+1} & V^{\bar n}_{-1,-1}\end{array}  \right]\left(\begin{array}{c}C^{\bar n}_{\alpha,+1}\\C^{\bar n}_{\alpha,-1}\end{array} \right) = a^{(1)} _{\bar n ,\alpha} \left(\begin{array}{c}C^{\bar n}_{\alpha,+1}\\C^{\bar n}_{\alpha,-1}\end{array} \right) ,
    \label{eq_Vsyst}
\end{equation}
where we defined the matrix elements of the perturbation as
\begin{eqnarray} \label{matrixeldef}
    V^{\bar n}_{\alpha,\alpha'} = \bra{\vec f^{(0)} ,\bar n ,\alpha}  {V}\ket{\vec f^{(0)} ,\bar n ,\alpha'} \,.
\end{eqnarray}
Up to second order in $\eta$, the matrix is explicitly given by
\begin{equation}
    V^{\bar n} = \frac{\sin^2\theta }{\tilde E^2} \left[\begin{array}{cc}  -(1+\frac{3({2\bar n-1})}{2}) \eta^2\sin^2\theta & -i \eta \sqrt{{2\, \bar n}} \cos\theta  \\ i \eta \sqrt{2\, \bar n} \cos\theta  & 1+\frac{3(2\bar n+1)}{2} \eta^2\sin^2\theta \end{array}  \right] .
\end{equation}
The calculation of the elements $ V^{\bar n}_{ \alpha,\alpha'}$ follows from a procedure similar to that of the previous case $\bar n=0$, and is shown explicitly in Appendix \ref{app3}.\\
At this point, the first-order correction $a^{(1)}_{\bar n,\alpha}$, defined up to order $\eta^2$, is obtained from the two eigenvalues of the linear system Eq.~\eqref{eq_Vsyst}, i.e. from the characteristic equation $\det(   V -      \mathbf{1} \, a_{\bar n}^{(1)}) = 0$, which reads
\begin{eqnarray}
\left( \frac{a^{(1)}_{\bar n,\alpha} \tilde E^2}{\sin^2\theta } - \frac32 \eta^2\sin^2\theta \right)^2 - \left ( 1+ 3 \, \bar n \,\eta^2\sin^2\theta \right)^2 -2\bar n \,\eta^2 \cos^2\theta = 0\,.
\end{eqnarray}
Solving for the two eigenvalues $a_{n,\alpha}^{(1)}$ up to order $\eta^2$, we obtain
\begin{eqnarray}
    a^{(1)}_{\bar n,\pm1} &=& \frac{\sin^2\theta}{\tilde E^2} \left ( \frac32 \eta^2 \sin^2\theta \pm \sqrt{\left (1+3 \,\bar n\, \eta^2\sin^2\theta \right )^2 + {2 \bar n} \, \eta^2  \cos^2\theta } \right) \nonumber \\
    &\simeq& \pm \frac{\sin^2\theta}{\tilde E^2} \left ( 1+ \frac32 (2 \bar n \pm 1) \eta^2\sin^2\theta + \bar n \,\eta^2 \cos^2\theta  \right ) .
\end{eqnarray}
In this perturbative scheme, we are able to identify which correction corresponds to each unperturbed eigenvector. Consider, for example, the negative eigenvalue $a^{(1)}_{\bar n,-1}$ of $V^{\bar n}$, that implies the equation
\begin{eqnarray}
   \left[\begin{array}{cc}  -(1+\frac{3(1+2\bar n)}{2} \eta^2\sin^2\theta) & -i \eta \sqrt{2 \,\bar n} \cos\theta  \\ i \eta \sqrt{2 \,\bar n} \cos\theta  & 1+\frac{3(1+2\bar n)}{2} \eta^2\sin^2\theta \end{array}  \right]  \left(\begin{array}{c}C^{\bar n}_{+1}\\C^{\bar n}_{-1}\end{array} \right)\nonumber \\
    \,\,\, = -\left ( 1+ \frac32 (2 \bar n - 1) \, \eta^2\sin^2\theta + \bar n \,\eta^2 \cos^2\theta  \right ) \left(\begin{array}{c}C^{\bar n}_{+1}\\C^{\bar n}_{-1}\end{array} \right),
\end{eqnarray}
from which we find the relation
\begin{eqnarray}
C_{-1}^{\bar n} = - i \eta \cos\theta \sqrt{\frac{\bar n}{2}} \,C_{+1}^{\bar n}\,,
\end{eqnarray}
clearly implying $|C_{-1}^{\bar n}|<<|C_{+1}^{\bar n}|$. Therefore, we can identify the negative eigenvalue correction as corresponding to the unperturbed state $\ket{\vec{f}^{(0)}, \bar n , +1}$, while the positive one corresponds to the unperturbed state $\ket{\vec{f}^{(0)}, \bar n , -1}$. \\
Going back from the $\bar n-$notation to the $n-$notation, we can write the eigenvalue first-order corrections in a compact way as
\begin{eqnarray}
    a^{(1)}_{n,\alpha} = -\alpha \frac{\sin^2\theta}{\tilde E^2} \left ( 1+ 3 (n-\frac{\alpha}{2}+ \delta_{\alpha,+1})\,\eta^2\sin^2\theta +( {n+\delta_{\alpha, +1}}) \,\eta^2 \cos^2\theta \right ) ,
    \label{eq_a_corr}
\end{eqnarray}
which actually also accounts for the ground state energy correction in the correct way. Note that since we are working at first order in $\lambda$, the expression for $\tilde E $ can be safely taken at zero order. Thus, in the following we will use 
\begin{equation}
    \tilde E \simeq E^{(0)} - k ^3 \cos \theta \,.
\end{equation}

\subsection{Corrections to the Energy Spectrum}
Remembering the definition \eqref{aEdef}, we can now solve for the eigenenergies to find
\begin{eqnarray} 
E^2 _{n,\alpha}(k^3) &=&   m^2_f +(k^3)^2 + |e| B \left (a^{(0)}_{n ,\alpha} + \lambda \, a^{(1)}_{n,\alpha} \right ) \\
&=& m^2_f +(k^3)^2 + |e| B \left (2 n +1 -\alpha + \lambda \,a^{(1)}_{n,\alpha} \right ) \nonumber = {{E^{(0)}_{n,\alpha}}^2(k^3)} + |e| B \lambda \, a^{(1)}_{n,\alpha} \,.
\end{eqnarray}
Substituting expression \eqref{eq_a_corr} into the above equation, we end up with the corrected VSR spectrum for the general configuration $\vec{B}\cdot  {{\hat n}} = B\cos\theta$
\begin{eqnarray}\label{eqEna}
E^{\pm }_{n,\alpha}(k^3) &=& \pm \left( 
{{E^{(0)}_{n,\alpha}}^2(k^3)}  - \lambda \frac{\alpha |e| B  \sin^2\theta}{({{E^{(0)}_{n,\alpha}}} -k^3 \cos\theta)^2} \times \right. \\
&&\left. \;\;\;\;\times\left [ 1+ \frac{3\,  |e| B  \sin^2\theta}{({{E^{(0)}_{n,\alpha}}} -k^3 \cos\theta)^2}  \left ( n-\frac{\alpha}{2} + \delta_{\alpha,+1} +\frac {n+\delta_{\alpha, +1}}{3} \cot^2\theta  \right ) \right ]
\right)^{\frac{1}{2}} .  \nonumber 
\end{eqnarray}
To prepare ourselves for the upcoming calculations, let us define two new perturbative parameters in order to fully perturbatively expand the eigenenergies \eqref{eqEna}, which are
\begin{equation}
    \mu \equiv \frac{2\lambda}{m^2_f} \,\,,\,\,\, \epsilon \equiv \frac{|e|B}{m^2_f}\,.
\end{equation}
Note that we are using the shifted mass $m_f$ since it is the value to which experiments would be sensible if VSR were valid. Picking the positive square root from \eqref{eqEna}, because we are only dealing with particles, and working in the particle's rest frame, where we can neglect its momentum, we obtain
\begin{eqnarray}
    E_{n,\alpha} &=& E^{(0)}_{n,\alpha} \bigg [  1 -  \frac{\alpha \mu \epsilon \sin^2\theta}{2(1+\epsilon (2n+1+\alpha))^2}  - \frac{\alpha \mu \epsilon^2 \sin^4 \theta}{2(1+\epsilon (2n+1+\alpha))^3} \times \nonumber  \\
    && \;\;\;\;\;\;\;\;
     \times \left (3 \,(n-\frac{\alpha}{2} + \delta_{\alpha,+1}) + (n+\delta_{\alpha, +1}) \cot^2\theta \right ) \bigg ]^\frac{1}{2}  , 
\end{eqnarray}
from which we expand in the parameter $\mu$ to first order
\begin{eqnarray}
    E_{n,\alpha} &=&  m_f ( 1+\epsilon (2n+1-\alpha))^\frac{1}{2} \bigg [ 1 -\frac{1}{4} \frac{ \alpha \mu \epsilon \sin^2 \theta}{(1+\epsilon(2n+1+\alpha))^2}  \\ 
     &&\;\;\; -\frac{n-\frac{\alpha}{2} + \delta_{\alpha,+1}}{4} \frac{3 \alpha \mu \epsilon^2 \sin^4 \theta}{(1+\epsilon (2n+1+\alpha))^3} -\frac{ n+\delta_{\alpha, +1} }{4} \frac{\alpha \mu \epsilon^2 \sin^2 \theta \cos^2\theta}{(1+\epsilon (2n+1+\alpha))^3} \bigg ] , \nonumber
\end{eqnarray}
and now in the parameter $\epsilon$ to second order
\begin{eqnarray}
    \frac{ E_{n,\alpha}}{m_f} &= & \,1+\frac{\epsilon}{2} (2n+1+\alpha) -\frac{\epsilon^2}{8} (2n+1+\alpha)^2  -\frac{\alpha}{4} \mu \epsilon \sin^2\theta \times  \, \\
     && \times \bigg (1-\frac{3}{2}\epsilon(2n+1+\alpha) +3 \,\epsilon (n -\frac{\alpha}{2} + \delta_{\alpha,-1})\sin^2\theta  +\epsilon (n+\delta_{\alpha, -1}) \cos^2 \theta  \bigg ) . \nonumber
\end{eqnarray}
In the following, to compare our result with the ones from the conventional LI framework, we will take into account LI contributions to the anomalous magnetic moment $\mathsf g-2$. Those are parametrized by adding another perturbative term $-i \mathsf a \frac{e}{4m} \sigma_{\mu \nu } F^{\mu\nu}$ into the VSR Dirac equation, from which we obtain, at first order in $\mathsf a$, an extra term in the spectrum \cite{koch2022}
\begin{eqnarray} \label{appSpectrum}
    \frac{ E_{n,\alpha}}{m_f} &= & \,1+\frac{\epsilon}{2} (2n+1+\alpha) -\frac{\epsilon^2}{8} (2n+1+\alpha)^2 + \frac{1}{2}\alpha \mathsf a \epsilon -\frac{\alpha}{4} \mu \epsilon \sin^2\theta \times  \, \\
     && \times \bigg (1-\frac{3}{2}\epsilon(2n+1+\alpha) +3 \,\epsilon (n-\frac{\alpha}{2} + \delta_{\alpha,-1})\sin^2\theta  +\epsilon (n+\delta_{\alpha, -1}) \cos^2 \theta  \bigg ) . \nonumber
\end{eqnarray}
Here, $\mathsf a \equiv (\mathsf g-2)/2$ where $\mathsf g-2 \neq 0$ can be thought to derive from standard QFT loops. In principle, VSR will also modify these corrections as shown in \cite{Alfaro:2023qib}. However, the additional radiative terms would be further suppressed by the smallness of the VSR parameter. Thus, in the subsequent analysis, we do not care about this additional source of perturbations.

\section{Penning Traps with Electrons} 

\begin{figure}[ht]
	\begin{center}
	\includegraphics[width=0.5\textwidth]{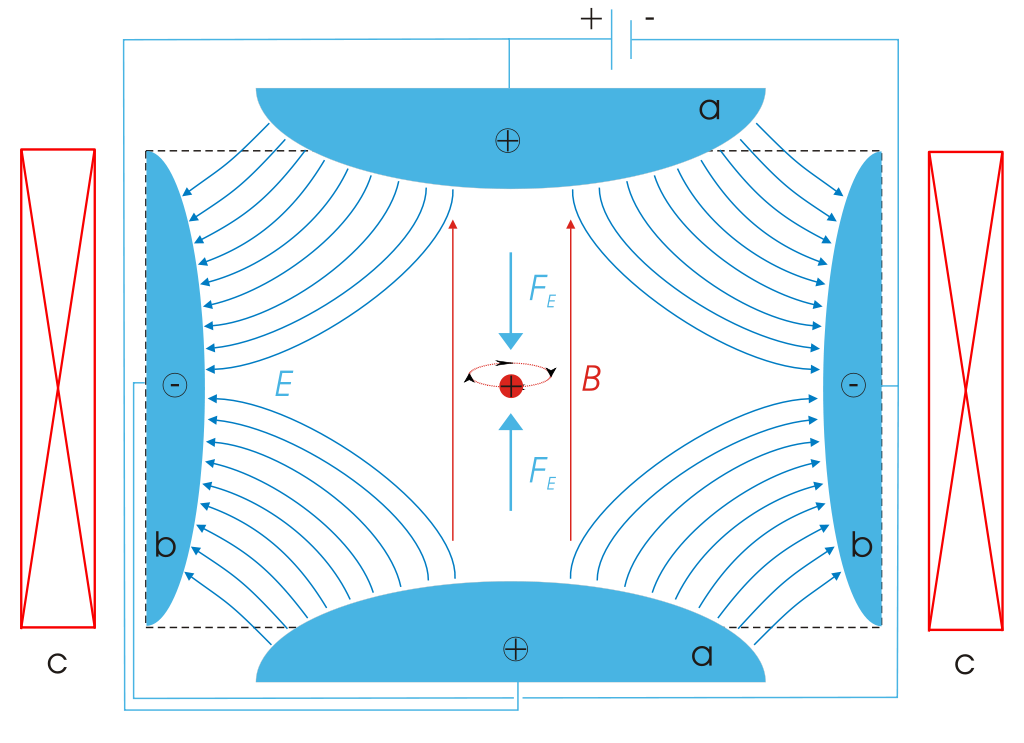}
	\caption[Penning trap schematic representation]{Schematic representation of the electromagnetic field inside a Penning trap. The constant and homogeneous axial magnetic field traps the electron radially, while the quadrupole electric field traps it axially. \hyperlink{https://upload.wikimedia.org/wikipedia/commons/b/b6/Penning_Trap.svg}{[Credit]}}
	\label{fig:pentrap}
	\end{center}
\end{figure}

The anomalous magnetic moment of the electron is probably the most precise theoretical prediction ever verified by observations. Since the first $\mathsf g-$factor experiments \cite{foley,kusch}, the measurements have been continuously improved, today reaching an outstanding precision of 13 significant digits~\cite{penning1,penning2,penning3}. Those experiments are based on cylindrical Penning traps, elaborate experimental setups in which electrons are trapped by a constant and homogeneous magnetic field and a quadrupolar electric field, as shown in Fig.\ref{fig:pentrap}.
\\
The extreme sensitivity reached in these contexts allows, in general, to put strong constraints on any fundamental physics extension affecting the electron $\mathsf g-$factor value, such as the introduction of new interactions \cite{NA64:2021xzo} or Lorentz violations \cite{Crivellin:2022idw}. Hence, it seems natural to put VSR at test within this context. For this purpose, the notation and results from~\cite{koch2022, penning1} will be used. Normally, the motion of an electron in a Penning trap has four eigenfrequencies, known as the spin-, cyclotron-, axial-, and magnetron-frequency. These four frequencies can be combined in a suitable ratio to extract an experimental value for the gyromagnetic factor of the electron, as explained in Eq.~$(10)$ from \cite{penning1}. However, for the calculations in this paper, we will consider a simplified setup without electric fields, magnetron motion, or cavity shifts effects.
\begin{figure}[ht]
	\begin{center}
	\includegraphics[width=0.4\textwidth]{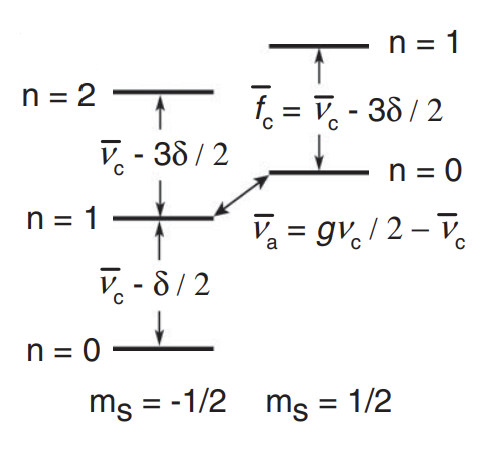}
	\caption[Cyclotron and spin energy levels]{Cyclotron and spin energy levels of an electron confined within a Penning trap. \cite{PhysRevLett.97.030801} Here, $n$ represents our same basic quantum number, while $\alpha $ is connected to the sign of the spin projection $m_S$.}
	\label{fig:penlevels}
	\end{center}
\end{figure}\\
It follows that, for small values of the magnetic field and defining the “free-space” cyclotron frequency $\nu_c = \frac{|e|B}{2 \pi m}$ \cite{penning1}, the energy eigenvalues of the unperturbed Dirac system are given by \cite{brown}
\begin{eqnarray} \label{EnergiesOdom}
E^{(u)}_{n, \alpha} \simeq \frac{1}{2}
\left( 2 n+ 1+\alpha \right)h \nu_c
+\alpha \frac{\mathsf a}{2}h \nu_c -\frac{1}{8}
\frac{h^2 \nu_c^2}{m c^2} \left( 2 n+ 1+\alpha  \right)^2 .
\end{eqnarray}
Higher orders in $|\vec B|$ are experimentally not relevant \cite{koch2022}, since the magnetic field strength, used in \cite{penning1,penning2,penning3}, and which can be calculated from the measured cyclotron frequency of about $\nu_c\approx 149~$GHz~\cite{penning1}, is too “weak”
\be\label{BB}
|\vec B|=\frac{2\pi \nu_c m}{|e|}\approx 5.3~T \to \frac{|e| B}{m^2} \sim 10^{-9}\,.
\ee
With our assumptions, the experimentally useful expression of $\mathsf a$ depends only on the anomaly frequency $\nu_a$ and the relativistic cyclotron frequency $f_c$ \cite{penning1}, which in our notation correspond respectively to the unperturbed transition energies $E^{(u)}_{0,-1} -E^{(u)}_{1,+1}$ and $E^{(u)}_{1,-1}-E^{(u)}_{0,-1} + \frac{3}{2} \frac{(eB)^2}{m^3} $ (see also Fig.\ref{fig:penlevels}). In fact, taking the ratios of these energy differences, we obtain exactly the theoretical $\mathsf a-$parameter
\begin{equation}\label{aexp}
    \frac{E_{0,+1}^{(u)} -E_{1,-1}^{(u)}}{E_{1,+1}^{(u)}-E_{0,+1}^{(u)} + \frac{3}{2 m} (h \nu_c)^2 } = \mathsf a \equiv\frac{\mathsf g-2}{2} \,.
\end{equation}
Thus, our goal is to see which effects are produced when evaluating the ratio in \eqref{aexp} with our new VSR energy spectrum \eqref{eqEna}. Apart from the mass shift $m\to m_f$, this replacement introduces non-trivial corrections encoded in the new expressions for the energy levels
\begin{equation}\label{avsr}
    \mathsf a_{VSR}\equiv \frac{\mathsf g_{VSR}-2}{2} \equiv\frac{E_{0,-1} -E_{1,+1}}{E_{1,-1}-E_{0,-1} + \frac{3}{2} m_f \epsilon^2}  \,.
\end{equation}
Assuming this VSR value to be the one measured by experiments, we will be able to calculate deviations with respect to the $\mathsf a-$value coming from the LI context and use them to constraint the VSR parameter. From \eqref{avsr}, keeping terms up to first order in $\mu$ and second order in $\epsilon$, we obtain the following
\begin{eqnarray}
    \mathsf a_{VSR} - \mathsf a = -\frac{\mu}{2} \sin^2\theta +\frac{\mu}{2}\epsilon \sin^2\theta \cos^2\theta \, (2-\mathsf a) \, , \nonumber
\end{eqnarray}
from which we see that the discrepancy between the gyromagnetic VSR value $\mathsf g_{VSR}$ and the one from LI radiative corrections would be
\begin{equation} \label{vsrdiscrgfactor}
    \mathsf g_{VSR} - \mathsf g = - \mu \left [ 1-\epsilon (2-\mathsf a)  \cos^2\theta \right ] \sin^2\theta \, . \nonumber
\end{equation}
Being $\epsilon$ already pretty small \eqref{BB}, in the following we just consider the leading order
\begin{equation} \label{gexp1}
    \mathsf g_{VSR} -\mathsf g \sim - \mu \sin^2 \theta \,.
\end{equation}

\section{Constraints on VSR from Electron $\mathsf g-$Factor} \label{sec:egfactor}

At this point, using \eqref{gexp1}, we are ready to establish experimental limits on the magnitude of the perturbative parameter $\mu$ and, consequently, the electronic VSR parameter $\lambda$. To do that, we (naively) assume that the entire discrepancy between the experimental $\mathsf g_{exp}$ and the theoretical value $\mathsf g$ is due to the VSR perturbation \eqref{gexp1}, i.e.
\begin{equation} \label{gexp2}
    \mathsf g_{exp} -\mathsf g \sim - \mu \sin^2 \theta \,.
\end{equation}
Using the most precise experimental value \cite{CODATA, penning1, penning2} and theoretical prediction \cite{gfactor} so far, that are respectively
\begin{eqnarray}
    && \mathsf g_{exp}/2 = 1.001159 652 180 73 (28) \,,\\
    && \mathsf g/2=1.001 159 652 182 032 (720) \,.
\end{eqnarray}
we first see that Eq.~\eqref{gexp2} is consistent with positive values of $\lambda$ since $\mathsf g_{exp}-\mathsf g<0$. Moreover, observing that $\sin^2 \theta \leq 1$, we obtain the following restriction for the $\mu-$parameter
\begin{equation}\label{boundmu}
    \mu \lesssim 2.7 \times 10^{-12}\, ,
\end{equation}
that is comparable or slightly stronger than other upper bounds found in literature \cite{dunn2006, maluf2014}. The only more stringent estimation would be $\mu < 9.7\times 10^{-19}$ obtained in~\cite{alfaro2}, which refers to the electric dipole-like interaction terms $\vec n \times \vec E \cdot \vec \sigma$, where $\vec E$ represents an electric field. Nevertheless, as already stated in \cite{dunn2006}, experiments in which the electric and magnetic fields are parallel, like the one \cite{edipole} used in \cite{alfaro2} to derive the upper bound for $\mu$, are insensitive to interaction terms of that type. Consequently, other experiments with non-parallel magnetic and electric field should be used in that situation to give a coherent upper limit from electric dipole-like VSR interactions.\\
Now, employing the value of electron mass $m_f \sim 0.51\, MeV$, we can translate the restriction in Eq.~\eqref{boundmu} to the following rough upper bound for the electron VSR parameter
\begin{equation} \label{Melimit}
   2 \lambda = { \mu \, m_f^2 } \lesssim 1 \, eV^2 \, .
\end{equation}
In principle, the electron VSR parameter and the electronic neutrino VSR mass $M_{\nu}$ are connected. That is basically because in the VSR extension of the Standard Model there is just one VSR parameter for each leptonic family \cite{dunn2006,alfaro2}. Therefore, our bound \eqref{Melimit} would directly imply an upper limit for the VSR electronic neutrino mass $M_\nu \lesssim 1 \, eV$, which is interestingly similar to other present bounds \cite{neutrinoup}. This leaves open the possibility for VSR to significantly contribute to neutrino masses. \\
Let us now spend some words on the nature of the VSR $\theta-$angle, which we have not discussed so far. In general, we can think of at least two different scenarios:
\begin{itemize}
    \item The VSR spatial preferred direction $\mathbf{\hat n}$ parameterizes the existence of spacetime anisotropies at large scales, due, for example, to primordial or cosmic magnetic fields \cite{Giovannini:2002sv} or quantum gravity arguments \cite{Das:2018umm,Mann:2020jcu}. In this case, it would be fundamental to take into account the orientation changes of $\vec n$ with respect to $\vec B$ due to Earth's rotation, allowing for sidereal-time analyses \cite{Ding:2019aox}. An interesting possibility opens for long-duration experiments of the order $\sim days$, for which anisotropic effects would average over the Earth's rotation period $ \to\int_T \sin^2\theta$. 
    \item If the VSR four-vector $n^\mu$ is instead emerging through some effective mechanism in a more local context \cite{background}, then the $\theta-$dependence could eventually be studied by rotations of the experimental setup. Once again, everything would depend on the supposed origin for the VSR corrections.
\end{itemize}
Independently of the scenarios we are interested in, changes in orientation can only introduce a numerical factor smaller than unity, meaning that our rough upper limit estimation \eqref{boundmu} for the VSR parameter should remain valid anyway. An exhaustive analysis of the angular behavior of \eqref{gexp1}, as the one done in \cite{kostel}, is left for future work.\\
In conclusion, we note that theoretically all calculations in this chapter could also apply to the case of muons, which recently received more attention due to the increasing discrepancy between the theoretical and experimental values of their $\mathsf g-$factor \cite{muon1}. However, since muonic $\mathsf g-2$ is measured using accelerators \cite{muon2} and described through the semi-classical approach of the Bargmann-Michel-Telegdi equation \cite{PhysRevLett.2.435}, a different link between the VSR scheme and the experimental one must be found to develop a consistent treatment. 




\chapter[NON-RELATIVISTIC LIMIT FOR VSR FERMIONS]{Non-Relativistic Limit for VSR Fermions} \label{chnrlimit}





In this chapter, we explore the properties of charged fermions in the non-relativistic limit within the VSR framework. To do that, we decide to use the established Hamiltonian formalism, which will allow us, thanks to the Foldy-Wouthuysen transformation, to more easily interpret our findings. 

\section{Foldy Wouthuysen Transformation \label{sec:sec1.3}}

In the Hamiltonian $\mathcal {H}_D$ of the standard Dirac theory, there is already a mixing between the positive and negative energy sectors, which we know are respectively related to particle and antiparticles. Therefore, to describe from there the non-relativistic behavior of Dirac systems, we always have to deal with their decoupling. One standard way to do that is through the Foldy-Wouthuysen (FW) transformation \cite{foldy1950dirac}, a fundamental tool when talking about the non-relativistic limit for fermions. The nice property of FW transformations is that they allow for a connection with the standard Schrodinger picture of non-relativistic quantum mechanics (QM) \cite{Bjorken:1965sts,Berestetskii:1982qgu}, which is very useful in many practical situations. In fact, historically, the FW transformation was important in achieving a correct interpretation of the electromagnetic contributions appearing in the low-energy limit of Dirac theory \cite{Costella:1995gt}. While in the interaction-free context it is possible to find an exact form of the FW transformation, in the interacting case, the FW procedure has to be carried out perturbatively in an iterative way. 

\subsection{Lorentz-Invariant Dirac} \label{FWLIcase}

Let us start by reviewing the FW machinery in the usual LI case to later tackle the VSR problem. In this context, as we shall see by starting from the free case, FW transformations can be regarded at the same time as representation changes, from the physical perspective, and as diagonalization problems, from the mathematical one.

\subsubsection*{Free Scenario}

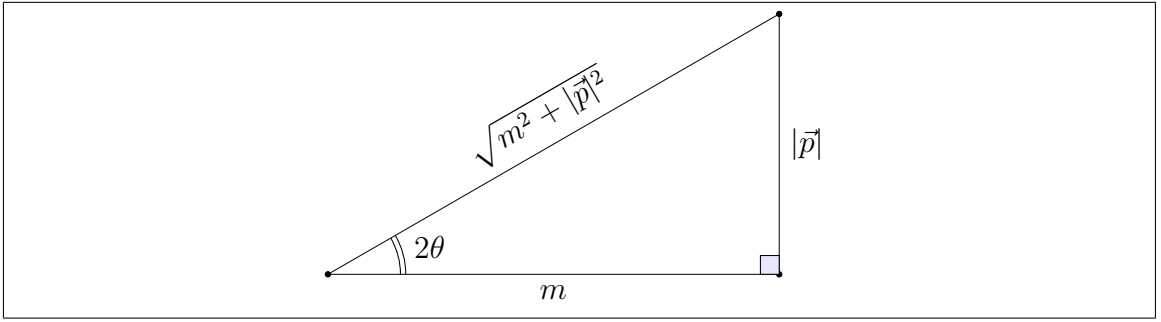
\begin{figure}[ht] 
\centering
\begin{tikzpicture}
\tkzDefPoint(0,0){A}
\tkzDefPoint(6,0){B}
\tkzDefTriangle[two angles = 30 and 90](A,B)
\tkzGetPoint{C}
\tkzDrawSegment(A,B)
\tkzDrawPoints(A,B)
\tkzDrawSegments(A,C B,C)
\tkzDrawPoints(C)
\tkzLabelAngle[pos=1.4](B,A,C){$2 \theta$}
\tkzMarkAngle[arc=ll, fill=blue!10](B,A,C)
\tkzMarkRightAngle[fill=blue!10](C,B,A)
\tkzLabelSegment[sloped,above](A,C){$\sqrt{m^2 + |\vec p|^2}$}
\tkzLabelSegment[sloped,below](A,B){$ m $}
\tkzLabelSegment[right](B,C){$|\vec p|$}
\end{tikzpicture}
\caption[Foldy-Wouthuysen triangle construction]{Schematic representation of the Foldy-Wouthuysen triangle construction for the interaction-free scenario.}
\label{Fig:FWtriangle}
\end{figure}

When interactions are turned off, the Hamiltonian operator $\mathcal H_D$ is time independent, which practically makes it a four-by-four matrix over the field of complex numbers. Its Hermiticity automatically ensures the possibility of diagonalization through a unitary transformation $U^{-1}=U^\dagger$ of $\psi$. Unitary transformations $\psi' = U \psi$ have the additional benefit of leaving the Hamiltonian density $H_D$ invariant
\be
 H_D = \psi^\dagger \mathcal H_D \psi = \psi^\dagger U^\dagger U \mathcal H_D U^\dagger U \psi = (\psi') ^\dagger \mathcal H'_D \psi' =H_D' \,.
\ee
The aim of the FW transformation is to eliminate the non-diagonal contribution from $\mathcal{H_D}$, which in the free LI case is just $ \vec\alpha \cdot \vec p$. The standard way to proceed is to propose the following form for the transformation
\be \label{diracfreefw}
U= e^{\beta \frac{\vec \alpha \cdot \vec p}{|p|} \theta} = \cos\theta + \beta \frac{\vec \alpha \cdot \vec p}{|\vec p|} \sin\theta \, , 
\ee
where we exploited the properties of the $\alpha-$matrices to expand the exponential in favor of trigonometric functions. Clearly, a four-by-four identity matrix is intended to be multiplied by the above $\cos\theta$. Note that since the exponent of \eqref{diracfreefw} is anti-Hermitian, the transformation is automatically unitary, as desired. Applying this transformation to $\mathcal{H}_D$ and considering $(\vec \alpha \cdot \vec p)^2 = \vec p^{\,2} $, we obtain
\bea
\mathcal H '_{D} &=& U \mathcal H_D U^\dagger \\
&=& (\cos\theta + \beta \frac{\vec \alpha \cdot \vec p}{|\vec p|} \sin\theta) (m \beta + \vec \alpha \cdot \vec p )(\cos\theta - \beta \frac{\vec \alpha \cdot \vec p}{|\vec p|} \sin\theta) \nonumber \\
&=& (m\beta + \vec\alpha \cdot \vec p) (\cos\theta - \beta \frac{\vec \alpha \cdot \vec p}{|\vec p|} \sin \theta)^2 \nonumber\\
&=& \beta \, (m \cos 2\theta + |\vec p| \sin 2\theta) + \vec \alpha \cdot \vec p \, (\cos 2\theta - \frac{m}{|\vec p|} \sin 2\theta) \,.\nonumber
\eea
From the last line, we can easily read the condition that $\theta$ must satisfy in order to get rid of the unwanted operator
\be \label{thetafreefw}
\tan 2\theta  = \frac{|\vec p|}{m} \,,
\ee
as also shown geometrically in Fig.\ref{Fig:FWtriangle}. Finally, replacing the condition \eqref{thetafreefw} in the transformed Hamiltonian, we end up with the diagonalized system
\be
\mathcal{H}_D' = \beta \sqrt{m^2+ \vec p^{\,2}} \,,
\ee 
where both positive and negative energies are present and doubly represented. The FW transformation makes clear how negative energies and four-component wave functions are the price to pay to allow for the existence of the Dirac representation, where the Hamiltonian is linear in momentum.

\subsubsection*{Interacting Scenario}

Including interactions, such as electromagnetic ones, makes things more complicated. In fact, the electromagnetic potential and hence the Hamiltonian
$\mathcal H_{D}^{EM}$ itself may depend on time. In this case, also the transformation $S $ is time-dependent, and it becomes generally impossible to construct it so that all off-diagonal terms are removed. Therefore, we content ourselves with a non-relativistic weak-field expansion of the transformed Hamiltonian in terms of $\nicefrac{1}{m} $. For such a calculation, we follow the Bjorken and Drell reference \cite{Bjorken:1965sts} and expand everything up to order $\nicefrac{1}{m^3}$ while maintaining only terms at most linear in $A^\mu$. An alternative approach would be to relate the physical quantities in play to some dimensionless reference parameter depending on the specific context. However, here we prefer to keep the discussion general by exploiting the above-mentioned formal expansion in $\nicefrac{1}{m}$. To proceed, we first observe that we can divide the Hamiltonian into different contributions depending on their (anti)commutation properties with $\beta\equiv \gamma^0$
\begin{equation}
    \mathcal H_D^{EM} = m +\mathcal E_D +\Theta _D \,,
\end{equation}
with the block-diagonal “even” operator $\mathcal E$ and the “odd” operator $\Theta$ defined as
\begin{eqnarray} \label{evenodddirac}
    \mathcal E &=& e A^0 \sim O(m^0)  \,, \, \\
    \Theta &=& -i \gamma^0 \gamma^i D_i = - i \alpha^i \partial_i +e \,\alpha^i A_i  \sim O(m^0) \,,
\end{eqnarray}
such that they satisfy the following properties
\begin{equation}
    [\mathcal E, \beta] =0 \,\,\,,\,\,\, \{\Theta , \beta\} =0\,.
\end{equation}
The typical unitary FW transformation $U=e^{iS}$, with $S$ hermitian, is designed to cancel the leading order “odd” contributions in the Hamiltonian and it reads
\begin{equation}
    S \equiv -\frac{i}{2m} \beta \Theta \sim {\mathcal{O}}(m^{-1}) \, .
\end{equation}
Expanding the transformed Hamiltonian up to the relevant order, we obtain
\begin{eqnarray}\label{HFWcommut}
    \mathcal{H'} &=& e^{iS} \mathcal H e^{-iS} - i e^{iS} \frac{\partial }{\partial t} (e^{-iS}) \\
    &=& \mathcal H + i [S,\mathcal H] + \frac{i^2}{2!} [S,[S,\mathcal H]] +\frac{i^3}{3!} [S,[S,[S,\mathcal H]]] +\frac{i^4}{4!} [S,[S,[S,[S,\mathcal H]]] \nonumber \\
    && - \dot S - \frac{i}{2} [S,\dot S] + \frac16 [S,[S,\dot S]] + {\mathcal{O}}(m^{-3}) \nonumber\,,
\end{eqnarray}
where $\dot S \equiv \frac{\partial S }{\partial t}$. In the following, we choose to work with stationary field configurations, so that all terms involving time derivatives vanish. Taking into consideration the order of the even and odd operators in \eqref{evenodddirac}, we can explicitly calculate the expressions of the relevant commutators to be
\begin{eqnarray}
    [S, \mathcal H_D]  &=& i \Theta -i\frac{\beta}{2m} [\Theta, \mathcal{E}] -i\frac{\beta \Theta^2}{m} \,, \\
    \,[S,[S, \mathcal H_D]] &=& \frac{\beta  \Theta^2}{m}  - \frac{1}{4 m^2} [[\Theta, \mathcal{E}] , \Theta] +\frac{\Theta^3}{m^2}  \,, \nonumber\\
    \,[S,[S,[S,\mathcal H_D]]] &\simeq& i\frac{\Theta^3}{ m^2 } -i \frac{\beta \Theta^4}{m^3 } \,, \nonumber \\
    \,[S,[S,[S,[S,\mathcal H_D]]]] &\simeq& \frac{\beta \Theta^4}{ m^3 } \, .\nonumber
\end{eqnarray}
Putting all these contributions together we get to
\begin{equation}
    \mathcal H' \simeq \beta m  + \mathcal{E} +\frac{1}{2m } \beta  [\Theta, \mathcal{E}]+ \frac{\beta \Theta^2}{2m }  +\frac{1}{8 m^2 } [[\Theta,\mathcal{E}],\Theta] -\frac{\Theta^3}{3m^2 } - \frac{ \beta \Theta^4}{8m^3 }   \,.
\end{equation}
We can see that the odd operators, corresponding to those with an odd number of spatial gamma matrices, are now of order $O(\frac{1}{m})$. By applying this procedure two more times, we end up with the expression
\begin{eqnarray} \label{HFWafter}
    \mathcal H'''\simeq \beta m  +\mathcal{E} + \frac{\beta \Theta^2}{2m }-\frac{1}{8 m^2 } [\Theta,[\Theta,\mathcal{E}]] - \frac{ \beta \Theta^4}{8m^3 } \,.
\end{eqnarray}
Thus, in the above Hamiltonian we managed to decouple the particle-antiparticle Dirac sectors up to order $O(\nicefrac{1}{m^4})$. Calculating these products of operators and inserting them back into \eqref{HFWafter}, we finally obtain the non-relativistic Hamiltonian $H^{EM}_{D}$ for a charged Dirac particle 
\begin{eqnarray}
     H_D^{EM} &=& \beta 
     \left (m + \frac{(\vec p - e \vec A)^2}{2m} -\frac{\vec p ^{\;4}}{8 m^3} \right ) + e A^0 - \frac{e}{2m} \beta \vec \Sigma \cdot \vec B \\
     && -\frac{i e }{ 8m^2} \vec \Sigma \cdot (\vec \nabla \times \vec E) - \frac{e}{4m^2} \vec \Sigma \cdot (\vec E \times \vec p) -\frac{e}{8m^2} \vec \nabla \cdot \vec E \nonumber\,,
\end{eqnarray}
where we used the standard definition for the momentum operator $p^ i =  -i \nabla^i = -i\partial_i$, and the electric and magnetic field are defined from their relation to the electromagnetic potential $A^\mu$ (or equivalently the electromagnetic strength $F_{\mu\nu}$) as
\begin{equation}
    F_{i0} = - E^ i \,\,\,,\,\,\, F_{ij} = - \epsilon^{ijk} B^k\,.
\end{equation}
As we can see, the physical interpretations of the perturbative contribution is straightforward. The terms in the first bracket give the expansion of relativistic square root with shifted momentum to the desired order. The second and third terms are the electrostatic and magnetic dipole energies. The next pair of terms, which taken together are hermitian, represent the spin-orbit energy. The last term is, instead, known as the Darwin term, and is usually attributed to the zitterbewegung, a German term for “trembling motion”, which refers to an apparent rapid oscillatory motion of Dirac particles, first theorized by Schrödinger in 1930 when analyzing the solutions to the Dirac equation. However, this effect would arise due to the interference between positive and negative energy states and, just like all the artifacts related with the interpretation of negative energy states, it disappears in a QFT setting.

\section{FW Transformation in VSR \label{sec:sub1.3.2}}

We now turn to the VSR case. The main problematic difference with the standard scenario is that now the Hamiltonian is no longer Hermitian. Nevertheless, simil-FW procedures have already been used in the literature \cite{alexandre2015foldy} to successfully deal with non-Hermitian Hamiltonians, obtaining meaningful equivalent descriptions in the FW (block-)diagonal representation. In fact, as already mentioned in Section \ref{freevsrhamiltoniancase}, where we discussed the free VSR Hamiltonian \eqref{vsrhamiltonianfreeeq}, its spectrum is real and having four independent eigenvectors it can always be diagonalized. Here, we focus our attention on the interacting case,\footnote{We are also working for an exact transformation in the free VSR case. However, we do not include it because more work is needed to express it in a nice compact form.} taking advantage of the iterative procedure presented in the previous section. Clearly, the presence of non-Hermitian terms in $\Theta$ will imply a non-unitary transformation, leading to a different Hamiltonian density. However, that should not affect the energy eigenvalues of the system in a time-independent scenario.\\
Let us start by defining the relevant odd and even operators up to order $\nicefrac{1}{m^3}$ from the interacting Hamiltonian \eqref{almostfinalschem4}
\begin{eqnarray} \label{EOvsrcase}
    \mathcal E_{VSR} &=& e A_0 + \frac{\lambda}{m} \gamma^0 -i \frac{\lambda}{m^2} \sigma^{ij} n_j D_i  - \frac{i e \lambda}{2 m^3} \gamma^0 \sigma^{ij} ( F_{ij} +2 n_i F_{j0}) \nonumber\\
    && -\frac{\lambda}{m^3} \gamma^0 (n^i n^j - \delta^{ij } )D_i D_j \,,  \\
    \Theta _{VSR} &=& -i (1+ \frac{\lambda }{m^2})\gamma^0 \gamma^i D_i -\frac{\lambda }{m} \gamma^i n_i - i \frac{\lambda}{m^2} \gamma^0\gamma^j n^i n_j D_i \nonumber \,.
\end{eqnarray}
In the above expression for $\Theta$ we have dropped the terms $O(\frac{1}{m^3})$ since they do not contribute to the transformed Hamiltonian up to the relevant order. It is easy to see that in the SR limit $\lambda \to 0$ we get back the LI operators \eqref{evenodddirac}. Being the leading contributions in \eqref{EOvsrcase} still of order $\sim O(m^0)$, we can follow the same procedure applied in Section \ref{FWLIcase} and still use \eqref{HFWafter} as our final non-relativistic decoupled Hamiltonian. Then, we are only left with the calculations of the three non-trivial operator products appearing in \eqref{HFWafter}. Let us start from the computation of $\Theta^2$. Here, we proceed up to order $\nicefrac{1}{m^2}$ since we already know that we will have to multiply it by an additional $\frac{1}{m}$ at the end. Considering only linear terms in the electromagnetic potential, we have
\begin{eqnarray} \label{Theta2EMv1}
    \Theta^2 &\simeq& \left (-i \gamma^0 \gamma^i D_i- \frac{\lambda }{m} \gamma^i n_i - i \frac{\lambda}{m^2} \gamma^0\gamma^j n^i n_j D_i -i  \frac{\lambda }{m^2} \gamma^0 \gamma^i D_i \right )^2  \\
     &=& \left ( -i \gamma^0 \gamma^i D_i - \frac{\lambda }{m} \gamma^i n_i \right) ^2 - \frac{\lambda}{m^2} \{ \gamma^0 \gamma^i D_i \, ,\,   \gamma^0\gamma^j n^i n_j D_i +  \gamma^0 \gamma^i D_i  \} \nonumber \\
     &=&  (1+\frac{2\lambda}{m^2})\gamma ^i \gamma^j D_i D_j - i \frac{2 \lambda}{m} \gamma^0 \sigma^{ij} n_i  D_j -\frac{\lambda^2}{m^2} +\frac{\lambda}{m^2} \{ \gamma^i D_i \, ,\, \gamma^j D_k \} \,n^k n_j \,. \nonumber
\end{eqnarray}
Using the relation $\gamma^i\gamma^j = \eta^{ij } +\sigma^{ij}$ and the fact that
\begin{eqnarray}
    \{ \gamma^i D_i \, ,\, \gamma^j D_k \}\, n^k n_j =(n^k n_j \gamma^i \gamma^j + n^i n_j \gamma^j \gamma^k ) D_i D_k \,,
\end{eqnarray}
we can rewrite \eqref{Theta2EMv1} as
\begin{eqnarray} \label{Theta2EMv2}
    \Theta^2 &=&  (1+\frac{2\lambda}{m^2}) D_i D^i + \frac{i e}{2}(1+\frac{2\lambda}{m^2}) \sigma^{ij} F_{ij} - i \frac{2 \lambda}{m} \gamma^0 \sigma^{ij} n_i  D_j -\frac{\lambda^2}{m^2}\nonumber \\
    && +\frac{2\lambda}{m^2} n^i n^k D_i D_k +i e \frac{\lambda}{m^2}  \sigma^{ij} n^k n_j F_{ik} \,. 
\end{eqnarray}
The calculation of $\Theta^4$ is trivial, since being multiplied by $\frac{1}{m^3}$ in the final Hamiltonian, it is sufficient to retain only $m^0-$terms
\begin{equation}
    \Theta ^4 \simeq (-i \gamma^0 \gamma^i D_i)^4 \simeq (\partial_i \partial^i)^2 \,,
\end{equation}
where we also discarded the contributions of the type $\sim (momentum)^3 (EM \,\,potential)$, consistently with \cite{Bjorken:1965sts} and in analogy with the LI case. To obtain the last relevant contribution, let us first go through the following preliminary calculation
\begin{eqnarray}
    [\Theta, \mathcal E] &=& [ -i \gamma^0 \gamma^i D_i - \frac{\lambda }{m} \gamma^i n_i ,  e A_0 + \frac{\lambda}{m} \gamma^0] +O(\frac{1}{m^2}) \nonumber \\
    &\simeq& -i \frac{\lambda}{m} [\gamma^0 \gamma^i D_i ,  \gamma^0] -i e [  \gamma^0 \gamma^i D_i  ,  A_0 ] \nonumber \\
    &=&  i\frac{2 \lambda}{m } \gamma^i D_i -i e \gamma^0 \gamma^i (\partial_i A_0) \,.
\end{eqnarray}
In this way, we can easily find the last contribution
\begin{eqnarray}
    [\Theta,[\Theta, \mathcal E]] &=& [ -i \gamma^0 \gamma^i D_i -\frac{\lambda }{m} \gamma^i n_i \,,\, i\frac{2 \lambda}{m } \gamma^i D_i -i e \gamma^0 \gamma^i (\partial_i A_0) ] \\
    &=& e\, [  \gamma^i \partial_i ,  \gamma^j (\partial_j A_0) ]+\frac{4 \lambda}{m }  \gamma^0 \gamma^i \gamma^j D_i D_j -i e \frac{2 \lambda }{m} \gamma^0  n^i (\partial_i A_0) \nonumber\\
    &=& e \, ( \partial^i F_{i0} +\sigma^{ij} \partial_i F_{j0} + 2 \sigma^{ij } F_{j0} \partial_i) \nonumber\\
    && + \frac{4 \lambda}{m }  \gamma^0 D^i D_i +i e \frac{2 \lambda}{m }  \gamma^0 \sigma^{ij} F_{ij} -i e \frac{2 \lambda }{m} \gamma^0  n^i F_{i0} \nonumber\,.
\end{eqnarray}
Finally, we can put everything together and obtain, after some cancellations, the non-relativistic VSR Hamiltonian $H_{VSR}^{EM}$ (for time-independent fields)
\begin{eqnarray} \label{HNREMwithF}
    H_{VSR}^{EM} &=& m( 1 + \frac{\lambda}{m^2} - \frac{\lambda^2}{2 m^4} ) \gamma^0 +\frac{1}{2m} (1-\frac{\lambda}{m^2}) \gamma^0 D_i D^i + e A_0  \\
    && + \frac{i e}{4m} (1 - \frac{\lambda}{m^2}) \gamma^0 \sigma^{ij} F_{ij} - \frac{e}{8 m^2} ( \partial^i F_{i0} +\sigma^{ij} \partial_i F_{j0} + 2 \sigma^{ij } F_{j0} \partial_i) - \gamma^0 \frac{(\partial_i \partial^i)^2}{8 m^3 }  \nonumber\\
    && -i e \frac{\lambda}{ m^3} \gamma^0 \sigma^{ij} n_i F_{j0}   +i e\frac{\lambda}{2m^3}  \gamma^0 \sigma^{ij} n^k n_j F_{ik} +i e \frac{\lambda }{4m^3} \gamma^0  n^i F_{i0} \nonumber\,.
\end{eqnarray}
The new non-trivial VSR effects in \eqref{HNREMwithF} are all proportional to $e$, in line with the fact that, when the interactions are switched-off $e \to 0$, the only surviving VSR signatures come from the expansion of the new fermion mass $m_f$
\begin{equation}
    m_f = \sqrt{m^2 + 2\lambda} \simeq m(1+ \frac{\lambda}{m^2} - \frac{\lambda^2}{2 m^4}) \,.
\end{equation}
The terms in the last line of \eqref{HNREMwithF} have no LI counterpart and, as expected, also depend on the preferred space direction $\hat n$, thus corresponding to entirely new VSR effects. Using the relations
\begin{eqnarray}
    \sigma^{ij} = -i \epsilon^{ijk} \Sigma^k \,\,\, ,\,\,\, \epsilon^{ijm} \epsilon^{klm}= \delta ^{ik}\delta^{jl} - \delta^{il} \delta^{jk} \,\,\,,\,\,\, \epsilon^{imn} \epsilon ^{jmn} = 2 \delta^{ij } \,,
\end{eqnarray}
we obtain a formulation for \eqref{HNREMwithF} where physical effects can be more easily identified
\begin{eqnarray} \label{HNREMwithEB1}
    H_{VSR}^{EM} &=& \beta m ( 1 + \frac{\lambda}{m^2} ) + \beta (1-\frac{\lambda}{m^2}) \frac{(\vec p - e \vec A)^2}{2m} - \beta \frac{\vec p^{\,4}}{8 m^3 }  + e A_0  \\
    &&- \frac{e \beta}{2m} (1-\frac{2\lambda}{m^2}) \vec \Sigma \cdot \vec B -\frac{i e }{ 8m^2} \vec \Sigma \cdot (\vec \nabla \times \vec E) - \frac{e}{4m^2} \vec \Sigma \cdot (\vec E \times \vec p) -\frac{e}{8m^2} \vec \nabla \cdot \vec E  \nonumber\\
    &&- \frac{e \lambda}{2 m^3} \beta  \, \vec n \cdot \vec \Sigma \,\vec n \cdot \vec B - i \frac{e\lambda}{4 m^3} \beta \,\vec n \cdot \vec E - \frac{e \lambda}{m^3} \beta \, \vec \Sigma \cdot (\vec n \times \vec E) +O(\lambda^2)\,, \nonumber
\end{eqnarray}
which, at this $\lambda-$order, can be rewritten in function of the shifted mass $m_f \simeq m +\frac{\lambda}{m}$ as
\begin{eqnarray} \label{HNREMwithEB2}
    H_{VSR}^{EM} &=& \beta  ( m_f+  \frac{(\vec p - e \vec A)^2}{2m_f} -\frac{\vec p^{\,4}}{8 m_f^3 }  )  + e A_0 - \frac{e }{2m_f} (1-\frac{\lambda}{m_f^2}) \beta \vec \Sigma \cdot \vec B \nonumber  \\
    && -\frac{i e }{ 8m_f^2} \vec \Sigma \cdot (\vec \nabla \times \vec E) - \frac{e}{4m_f^2} \vec \Sigma \cdot (\vec E \times \vec p) -\frac{e}{8m_f^2} \vec \nabla \cdot \vec E  \\
    &&- \frac{e \lambda}{2 m_f^3} \beta \, \vec n \cdot \vec \Sigma \,\vec n \cdot \vec B - i \frac{e\lambda}{4 m_f^3} \beta \, \vec n \cdot \vec E - \frac{e \lambda}{m_f^3} \beta \,\vec \Sigma \cdot (\vec n \times \vec E)\,. \nonumber
\end{eqnarray}
The above expressions \eqref{HNREMwithF} and \eqref{HNREMwithEB2} are of very practical use, and can be easily employed to calculate VSR corrections in many contexts, as we will show. It is also trivial to check that the Hamiltonian obtained in the FW representation is actually Hermitian.

\subsection{$\mathsf g-$Factor “Strikes Back”}

As an immediate example, we want to calculate the leading-order VSR corrections to the $\mathsf g-$factor emerging from the Hamiltonian approach. In this way, we can compare them with the ones obtained by directly expanding the VSR Dirac equation in Chapter \ref{ch2}. The relevant terms are clearly those involving the magnetic field, i.e.
\begin{equation} \label{gyromagnetichamiltonian}
    - \frac{e }{2m_f} (1-\frac{\lambda}{m_f^2}) \beta \vec \Sigma \cdot \vec B- \frac{e \lambda}{2 m_f^3} \beta \vec n \cdot \vec \Sigma \,\vec n \cdot \vec B \,.
\end{equation}
Let us take the magnetic field $\vec B$ in the $\hat z-$direction, such that $\vec \Sigma \cdot \vec B = B \Sigma^3$. We decide to start, as usual, from the simplest case $\hat n // \vec B$. Here, we have
\begin{equation}
    \hat n \cdot \vec B = B \,\,\,,\,\,\, \hat n \cdot \vec \Sigma = \Sigma^3\,.
\end{equation}
Hence, the gyromagnetic related terms \eqref{gyromagnetichamiltonian} give
\begin{equation}
    - \frac{e }{2m_f} (1-\frac{\lambda}{m_f^2}) B \beta \Sigma^3 - \frac{e \lambda}{2 m_f^3} B \beta \Sigma^3 = - \frac{e }{2m_f} B \beta \Sigma^3 \,.
\end{equation}
Defining the Spin operator $ \vec S = 2 \vec \sigma $ and using that for matter particles $\beta \Sigma^3 \to \sigma^3 $, we get
\begin{equation}
    - \frac{e }{2m_f} B \sigma^3 = - (2)  \frac{e }{2m_f} B S^3 = - (2 )  \frac{e }{2m_f}\vec B \cdot \vec S \,,
\end{equation}
from which we can read off the standard LI gyromagnetic factor $\mathsf g =2$, restating the absence of non-trivial VSR effects for the case $\hat n //\vec B$, as in Chapter \ref{ch2}. \\
In the more generic scenario \eqref{genericnorientation}, where instead we take $\hat n = (\sin \theta,0,\cos\theta)$ we have
\begin{equation}
    \hat n \cdot \vec B = B \cos\theta \,\,\,,\,\,\, \hat n \cdot \vec \Sigma = \Sigma^1 \sin\theta + \Sigma^3 \cos\theta\,.
\end{equation}
Following the same steps as before, we end up with
\begin{equation}
    - \frac{e }{2m_f} (1-\frac{\lambda}{m_f^2}) B \beta \Sigma^3 - \frac{e \lambda}{2 m_f^3}  \beta (\Sigma^1 \sin\theta + \Sigma^3 \cos\theta) B \cos\theta \,.
\end{equation} 
Since the leading $ \mathsf g-$factor correction is embodied by the above diagonal term, that turn out to be proportional to $\vec B \cdot \vec S$, in the end we have
\begin{equation}
    - \frac{e }{2m_f} (1-\frac{\lambda}{m_f^2} \sin^2\theta) B \beta \Sigma^3 \,\, \to \,\,-  (2-\frac{2\lambda}{m_f^2} \sin^2\theta)  \frac{e }{2m_f} \vec B \cdot \vec S \,,
\end{equation} 
implying the following VSR correction, in agreement with what found in \eqref{gexp1}
\begin{equation}
    \delta \mathsf g = - \frac{2\lambda}{m^2_f} \sin^2\theta \,.
\end{equation}

\chapter[VSR NEUTRONS IN ACCELERATED FRAMES]{VSR Neutrons in Accelerated Frames} \label{ch3}
In this chapter, we focus on studying the signature of VSR on an Ultracold Neutron (UCN) immersed in a gravitational potential. UCNs are particles with such low energies that their wavelength becomes larger than typical atomic interspacing and can therefore be stored much more easily because they are totally reflected by many materials \cite{ignatovich1986physics}. In recent years, gravitational spectroscopy has received huge attention thanks to its potential to constrain new-physics proposals, ranging from new interactions \cite{sponar2021tests,escobar2022testing} to GR extensions \cite{kostelecky2021searches,ivanov2021quantum,Sung:2023jir,PhysRevD.83.021502} and LV theories \cite{ivanov2019probing}. Moreover, experiments such as qBOUNCE \cite{abele2011qbounce,jenke2009q}, have already managed to effectively measure the gravitational quantum states of the UCN up to a stunning precision of $10^{-6}\,peV$.
\begin{figure}[ht]
	\begin{center}
	\includegraphics[width=0.53\textwidth]{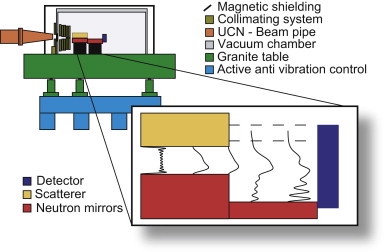}
	\caption[Schematic setup of the qBOUNCE experiment]{Schematic representation of the qBOUNCE experimental setup, which is used for gravitational spectroscopy involving UCNs. \cite{Abele:2012dn}}
	\label{fig:qbouncesetup}
	\end{center}
\end{figure}\\
In these experiments, ultracold neutrons are confined between a bottom mirror and the gravitational potential of the Earth. The standard theoretical framework used to describe them is a basic Schr\"odinger equation with a linear gravitational potential. The spectrum of UCNs is then given by solving the corresponding secular equation.
\begin{equation} \label{secequationneutron}
    \left (-\frac{\hbar^2}{2m} \partial^2_z +m a z \right ) \, \varphi (z) =  E \, \varphi(z) \,,
\end{equation}
with $a$ representing the value of the local acceleration experienced by the neutrons and $z$ being the height coordinate in the laboratory.\\
The presence of the qBOUNCE bottom mirror is simulated by setting the appropriate boundary conditions at the origin $z=0$ of the laboratory coordinate system
\begin{equation} \label{boundaryqbounce}
    \varphi \left (z=0 \right )=0 \,.
\end{equation}
The solution of the above equation \eqref{secequationneutron} is well known and is given by Airy functions 
\begin{equation} \label{varphiairy}
    \varphi_n (z) = C_n \, Ai \left (\frac{z-z_n}{z_0} \right ) ,
\end{equation}
where we defined the quantities \cite{pitschmann2019schr}
\begin{equation} \label{airyparamdef}
    C_n \equiv \frac{z_0^{-\frac12}}{Ai'\left(-\frac{z_n}{z_0}\right )} \, ,\,\,z_0 \equiv \left(\frac{\hbar^2}{2 m^2 a} \right)^{\frac13}\,,\,\, z_n \equiv \frac{E_n}{m a }\,.
\end{equation}
Here, the $Ai'(\zeta)$ represents the derivative of $Ai(\zeta)$ with respect to its argument $\zeta \equiv \frac{z-z_n}{z_0}$.  The $E_n- $values are determined by the quantization condition derived from the boundary condition \eqref{boundaryqbounce} evaluated for the eigenfunctions \eqref{varphiairy}
\begin{equation}
    Ai \left (-\frac{E_n}{m a z_0} \right )=0 \,,
\end{equation}
which allow us to find the leading order energy eigenvalues of the neutron by computing the zeros of the Airy function.
\begin{figure}[h!]
	\begin{center}
	\includegraphics[width=0.4\textwidth]{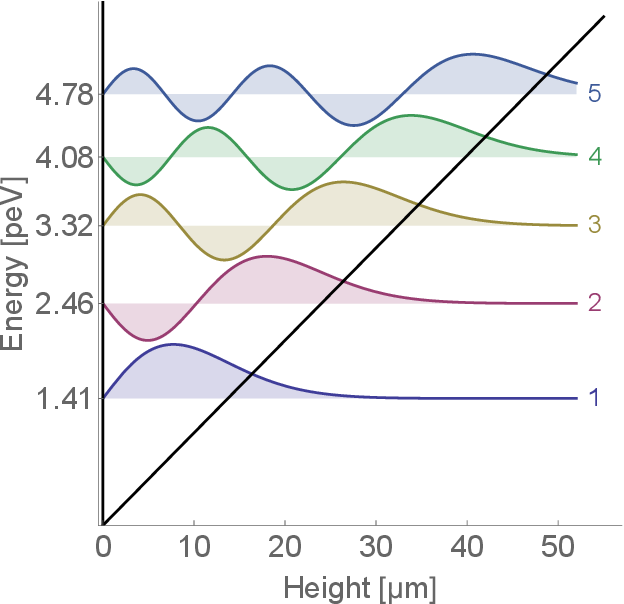}
	\caption[Quantized gravitational energy levels of ultracold neutrons]{First quantum energy levels of an ultracold neutron trapped from a bottom mirror and the Earth's gravitational potential. \cite{sponar2021tests}}
	\label{fig:qbouncelevels}
	\end{center}
\end{figure}\\

\section{Dirac Formulation of the qBOUNCE Scenario}

The above scenario can be easily derived from first principles, if we restrict ourselves to the assumption of a constant and homogeneous acceleration $\vec a$ or equivalently to uniform gravitational fields. This is a good approximation as long as we want to describe small patches of spacetime, that is, small spatial volumes for small time periods. In this context, we can take the spacetime element $ds^2$ and the metric $g_{\mu\nu}$ to be
\begin{equation} \label{rindlerelement}
    ds^2 = \left (1+ \frac{\vec a \cdot \vec x}{c^2} \right )^2 d (ct)^2 - d\vec x ^{\,2} \,\,\to \,\, g_{\mu\nu} = diag \left \{ \left (1+ \frac{\vec a \cdot \vec x}{c^2} \right)^2 ,-1,-1,-1 \right \},
\end{equation}
where we momentarily restored $c-$units and used $x^0 = ct$. Thus, the quantum-relativistic description of the UCN system is given by the Dirac equation in curvilinear coordinates \cite{de1962representations, Pollock:2010zz}, obtained by introducing a set of local observers. That is realized through the tetrad formalism, which allows us to translate from “curved” spacetime indices (here labeled by Greek characters) to “flat” tangent-space indices (labeled by Latin characters). The relevant object for this purpose is the vierbein $e^a_{\;\;\mu}$ and its inverse $e_a^{\;\;\mu}$, encoding the information on the change from the natural coordinate basis $\{ \partial_\mu \}$ to a generic local orthonormal frame $\{\theta_{(a)}(x)\}$
\begin{equation}
    \theta_{(a)}(x) = e_a^{\;\;\mu} (x)\, \partial_\mu  \,.
\end{equation}
The vierbein components satisfy by construction the following relations
\begin{eqnarray}
    && g_{\mu\nu} = e^a_{\;\;\mu} e^b_{\;\;\nu} \eta_{ab} \,,\\
    && e^a_{\;\;\mu} e_b^{\;\;\mu} = \delta^a_b \,.
\end{eqnarray}
Furthermore, from the standard flat gamma matrices $\gamma^a$, we can define their “curved” version $\underline \gamma^\mu $ as
\begin{equation}
    \underline \gamma^\mu \equiv e_{a}^{\;\;\mu} \, \gamma^{ a} \, ,
\end{equation}
which fulfill the consistent spacetime Clifford Algebra 
\begin{equation}
    \{ \underline \gamma^\mu, \underline \gamma^\nu \}= 2 \, g^{\mu\nu} \, .
\end{equation}
Before we move on, let us set up a few additional conventions used just in this chapter. To avoid confusion with vierbeins and their inverse, which are the only objects that intrinsically mix the two types of indices, we will place the (upper or lower) tangent space index as the first one appearing from left to right.\footnote{At the moment of separating spatial and time indices for tangent space and spacetime indices, ambiguities are avoided by simply taking into account the original nature of the objects under consideration.} Dirac indices will instead be often omitted for the sake of simplicity. Since in the gravitational context non-relativistic expansions are often performed in inverse powers of $c$ \cite{chandrasekhar1965post}, we eventually keep explicit the $c-$units in our equations.\\
After the above considerations, the “curved” Dirac equation can be written as follows
\begin{equation} \label{curvdirac} 
    (i \hbar \, \underline {\gamma}^\mu \mathcal D_\mu - m c) \psi =0 \,,
\end{equation}
with $\mathcal D_\mu \psi = \partial_\mu \psi + \Gamma_\mu \psi $ representing the spinor covariant derivative \cite{weinberg1972gravitation} constructed from the Spinor Connection $\Gamma_\mu$. The latter is expressed through the inverse vierbein $e_{a}^{\;\;\mu}$ as 
\begin{equation}\label{Gammamudef}
    \Gamma_\mu \equiv \frac14 \sigma^{ a  b} \, \omega_{\mu\,  ab } = \frac14 \sigma^{ a  b} \, g_{\nu\rho} \, e_{ a}^{\;\;\nu} (\partial_\mu e_{ b}^{\;\;\rho} +\{^{\;\, \rho}_{ \mu\, \alpha}\} \, e_{ b}^{\;\;\alpha}) \,,
\end{equation}
where $\omega_{\mu\,  ab }$ is the Spin connection and $\{^{\;\, \rho}_{ \mu\, \alpha}\}$ are the Christoffel symbols defined in GR
\begin{equation}
    \{^{\;\, \rho}_{ \mu \,\, \alpha}\} = \frac{1}{2} g^{\rho \beta} ( \partial_\mu g_{\alpha \beta} +\partial_\alpha g_{\mu \beta} -\partial_\beta g_{\mu \alpha})\,.
\end{equation}
We highlight that the curved gamma matrices, as well as the vierbein, are constructed to be automatically covariantly constant \cite{Pollock:2010zz}
\begin{eqnarray}
    \mathcal D_\mu \underline \gamma^\nu &=&  \partial_\mu \underline \gamma^\nu + \{^{\;\, \nu}_{ \mu \,\, \rho}\} \underline \gamma^\rho - [ \Gamma_\mu, \underline \gamma^\nu ] = 0 \,, \\
    \mathcal D_\mu e^a_{\;\nu} &=& \partial_\mu e^a_{\;\nu} - \{^{\;\, \rho}_{ \mu \,\, \nu}\} \, e^a_{\;\rho} + {{\omega_{\mu }}^a}_b  e^b_{\;\nu} =0 \,.
\end{eqnarray}
The form of the spacetime element \eqref{rindlerelement} can be deduced from several parallel but different lines of reasoning:
\begin{itemize}
    \item It can be obtained by charting flat space with Kottler–Møller or Rindler coordinates \cite{Rindler1969-RINER,moller1943homogeneous}, which are the natural set of coordinates used by (non-inertial) observers undergoing uniformly accelerated motion.
    \item Upon neglecting $|\vec x |^2$ contributions, it can be regarded as the spacetime element measured by a static observer on Earth's surface when described in its proper reference frame through Fermi-Walker coordinates. \cite{maluf2008construction,klein2008general,Misner:1973prb}
    \item Considering that $(1+ \frac{\vec a \cdot \vec x}{c^2})^2 \sim 1+ \frac{2 \,\vec a \cdot \vec x}{c^2} $, it can be derived from the solution of Einstein field equations in the weak field and low-velocity limits \cite{Ferrari:2020nzo}, truncating the expansion of the Newtonian $\frac{1}{r}-$potential around some point $r=R << |\vec x|$ at first order in the displacement $|\vec x|$.
\end{itemize}
Each of the above perspectives can be seen to reflect, in a first approximation, the laboratory point of view in the context of UCN experiments. The vierbein $e^a_{\;\;\mu}$ which naturally describes an observer at rest with respect to the accelerated coordinate chart \eqref{rindlerelement} is
\begin{eqnarray} \label{rindlertetrads}
   && e^a_{\;\;\mu} = diag\{V,1,1,1\} \,\, \text{with }\,\, V \equiv 1+\frac{\vec a \cdot \vec x }{c^2} \,, \nonumber\\
   && e_a^{\;\;\mu} = diag\{\frac{1}{V},1,1,1\} \, ,
\end{eqnarray}
leading to the following curved gamma matrices
\begin{eqnarray}
    \underline \gamma^0 &=& e^{\;\;0}_a \gamma^a = e^{\;\;0}_0 \gamma^0 = \frac{1}{V} \gamma^0 \,, \\
    \underline \gamma^i &=& e^{\;\;i}_a \gamma^a = e^{\;\;i}_i \gamma^i = \gamma^i\,.
\end{eqnarray}
We include in Appendix \ref{app:rindlerquantities} all the other relevant quantities for our calculations, such as the components of the spinor connection $\Gamma_\mu$ and of the Christoffel symbols $\{^{\;\, \rho}_{ \mu \,\, \alpha}\}$. We should also stress that, since \eqref{rindlerelement} describes an alternative charting for flat spacetime, the associated Riemann tensor is still zero
\begin{equation}
    {{R_{\mu\nu}}^\alpha }_\beta = \partial_\mu \{^{\;\, \alpha}_{ \mu \,\, \beta}\} - \partial_\nu \{^{\;\, \alpha }_{ \mu \,\, \beta}\}+\{^{\;\, \alpha }_{ \mu \,\, \rho}\}\{^{\;\, \rho }_{ \nu \,\, \beta}\} - \{^{\;\, \alpha }_{ \nu \,\, \rho}\}\{^{\;\, \rho }_{ \mu \,\, \beta}\} = 0 \,,
    \label{eq:riemann_tensor_christoffel}
\end{equation}
meaning we should not expect nor look for tidal effects in this scheme. However, in our case, they would not be relevant anyway.

\subsection{Leading-Order LI Hamiltonian}

After multiplying Eq.~\eqref{curvdirac} by $(g^{00})^{-1} \underline \gamma^0$, we can cleverly manipulate it to obtain an equation governing the time-evolution of the spinor $\psi$ \cite{PhysRevD.109.064085}
\begin{equation}
      i \hbar \frac{\partial \psi}{\partial t} = \mathcal{H}^g_D \psi \, ,
\end{equation}
where $\mathcal{H}^g_D$ represents the Dirac Hamiltonian in curvilinear coordinates
\begin{equation} \label{diracH}
    \mathcal{H}^g_D = m c^2 (g^{00})^{-1} \underline \gamma^0  -i \hbar c \, \Gamma_0 - i \hbar c\,  (g^{00})^{-1} \underline \gamma^0 \underline \gamma^i {\mathcal D}_i  \,.
\end{equation}
Specializing to our case \eqref{rindlerelement}, and taking into account the expressions in Appendix \ref{app:rindlerquantities}, we end up with the formula
\begin{eqnarray} \label{diracH2}
      \mathcal{H}^g_D = m c^2 V \gamma^0  -i \hbar c\, \gamma^0\gamma^i \left ( \, V \partial_i+ \frac12 \partial_i V \right )  . 
\end{eqnarray}
Usually, at this point, we should worry about the Hermiticity condition of the Hamiltonian, which can be modified due to the presence of the determinant of the spatial metric \cite{parker1980one,arminjon2006post,huang2009hermiticity}. Nevertheless, here the spatial components of the metric are not curved, and therefore we have no such problems.\\
To tackle the non-relativistic limit, we have to define the scheme of our perturbative expansion. The straight way to go is by performing a formal $\nicefrac{1}{c} \,-$expansion of geometrical objects around inertial flat-spacetime quantities \cite{ulbricht2019gravitational}. In particular, we have
\begin{eqnarray} \label{pNexp1}
    v & \equiv& V-1 \sim {\mathcal{O}}(c^{-2}) \, ,
\end{eqnarray}
leading to the following re-writing of the Hamiltonian up to the leading order $c^0$
\begin{eqnarray} \label{diracH4}
      \mathcal{H}^g_D = m c^2 \gamma^0 + m c^2 v \gamma^0   -i \hbar c\, \gamma^0\gamma^i \partial_i  \, . 
\end{eqnarray}
This approach is similar to the one used for Post-Newtonian (PN) calculations \cite{chandrasekhar1965post,nelson1990post}, which have demonstrated to be very powerful when dealing with gravitational systems. By virtue of this analogy, we borrow some of the PN-vocabulary to classify corrections in a convenient way: in particular, terms of order $\nicefrac{1}{c^N}$ will correspond to $\frac{N}{2}$PN-corrections in our scheme. In this sense, the Hamiltonian \eqref{diracH4} must be interpreted as a 0PN expression.
\begin{table}[ht]
\centering
\begin{tabular}[t]{ccc}
\toprule
$\;\;${$\nicefrac{1}{c}-$order}$\;\;$ & $\;\;$PN-equivalent$\;\;$ \\ 
    \midrule
    {$1$} & 0PN \\
    $c^{-1}$ & 0.5PN \\
    $c^{-2}$ & 1PN \\
    \bottomrule
\end{tabular} 
\caption[Post-Newtonian perturbative scheme]{Correspondence table between our formal expansion and the post-Newtonian one (also look at Fig.1 in \cite{bern2019black}).}
\label{tab:PNcorresp}
\end{table}\\
Thus, using the FW machinery already developed in the previous chapter it is easy to see that, up to the relevant order of approximation, we get 
\begin{equation} \label{HgFWlicase}
    \mathcal{H}^g_{FW} = \gamma^0 \left ( m c^2 + m \vec a \cdot \vec x - \frac{\vec \partial^{\,2}}{2m} \right ) ,
\end{equation}
which, regarding the dynamics along the acceleration direction, is analogous to what we saw already in the Schr\"odinger equation \eqref{secequationneutron} with the Newtonian potential.

\section{VSR Framework for Non-Inertial Observers}

After all these preliminaries, we are ready to also take VSR into account. The standard generalization of the EOM \eqref{diracvsreom} is obtained, as usual, by replacing partial derivatives with covariant ones
\begin{equation} \label{vsreomgrav}
    (i \slashed {\mathcal D} -m +i \lambda \frac{\slashed n}{n \cdot {\mathcal D}}) \psi = 0 \,,
\end{equation}
where we are hiding $c-$units again for simplicity. In principle, the VSR term may suffer from ordering problems, because the non-local operator could also be applied to $\slashed n$. However, for our spacetime \eqref{rindlerelement}, we are free to assume $n^\mu$ to be covariantly constant\footnote{This is actually the natural extension of the constancy condition $\partial_\mu n^\nu=0$ that we is used for inertial frames described by non-curvilinear coordinates.}
\begin{equation} \label{covdevnzero}
    {\mathcal D}_\mu n^\nu = 0\,.
\end{equation}
That is because, in a flat space, we can always solve the set of differential equations defined by \eqref{covdevnzero}. In fact, since the Riemann tensor is already zero, in this geometry we do not run into no-go constraints that would forbid the existence of such an $n^\mu$ \cite{PhysRevD.91.065034,Kostelecky:2021tdf}. Hence, there is no ambiguity in the definition of \eqref{vsreomgrav}. Nevertheless, this also means that $n^\mu$ is now generally position-dependent, while still being lightlike. Therefore, to preserve its constancy condition \eqref{covdevnzero}, we lose, at this stage, the ability to rescale $n^\mu$ as in \eqref{rescalednintro}.\\
Due to the presence of time non-localities, we cannot proceed to directly find the Hamiltonian as in the LI case, but we have to follow a similar path to that of Chapter \ref{ch1}. The first step is to multiply \eqref{vsreomgrav} by $-i n\cdot {\mathcal D}$ from the left 
\begin{equation} \label{vsreomgrav2}
    ( n\cdot {\mathcal D} \slashed {\mathcal D} +i m n\cdot {\mathcal D} + \lambda \slashed n) \psi = 0 \,.
\end{equation}
We should now observe that the commutator of two covariant derivatives acting on $\psi$ is automatically zero in flat space. In fact, we have that in curvilinear coordinates \cite{Pollock:2010zz,Shapiro:2016pfm}
\begin{equation}
    [{\mathcal D}_\mu, {\mathcal D}_\nu]\psi = -\frac{1}{4} R_{\mu\nu\rho\sigma} \underline\gamma^\rho \underline \gamma^\sigma \psi \,\,\to\,\, 0\,,
\end{equation}
because the Riemann tensor vanish identically in our case. Then, we can expand the contractions in \eqref{vsreomgrav2} to obtain
\begin{eqnarray} \label{calculationnDDgrav}
    n\cdot {\mathcal D} \slashed {\mathcal D} &=& n^0 \underline \gamma^0 {\mathcal D}_0 ^2 + (\underline \gamma^0 n^ i +n^0\underline \gamma^ i) {\mathcal D}_ i {\mathcal D}_0 + \underline \gamma^ j n^i {\mathcal D}_ i {\mathcal D}_ j \nonumber \\
    &=& \frac{1 }{V} n^0 \gamma^0 {\mathcal D}_0 ^2 + ( \frac{1}{V} \gamma^0 n^ i +n^0  \gamma^ i) {\mathcal D}_ i {\mathcal D}_0 +  \gamma^ j n^i {\mathcal D}_ i {\mathcal D}_ j \,, \nonumber \\
    i m \, n\cdot {\mathcal D} &=& i m\, n^0 {\mathcal D}_0 + i m \, n^ i {\mathcal D}_i\,.
\end{eqnarray}
Replacing these expressions in \eqref{vsreomgrav2} and multiplying it from the left by $1/(m n^0)$, we get
\begin{eqnarray} \label{almosttheregrav1}
    \left(1 - \frac{i}{m}  (\gamma^0 \frac{ n^i}{V n^0 } + \gamma^i ) {\mathcal D}_i \right) i {\mathcal D}_0 \psi  &=& - \left( \frac{1}{m} \frac{ \gamma^0}{V} {\mathcal D}_0^2 + i  \frac{n^i}{n^0} {\mathcal D}_i \right.  \\
    &&  \left.  \;\;\;\;\;\;\;\; + \frac{1}{m} \frac{n^i}{n^0} \gamma^j {\mathcal D}_i {\mathcal D}_j  + \frac{\lambda}{m} \frac{\slashed n}{n^0} \right) \psi \,. \nonumber
\end{eqnarray}
Observing that
\begin{equation}
    {\mathcal D}_i {\mathcal D}_0 \psi = \partial_i {\mathcal D}_0 \psi + \Gamma_i {\mathcal D}_0 \psi - \{^{\;\, \mu }_{ i \,\,\, 0} \} {\mathcal D}_\mu \psi = \partial_i {\mathcal D}_0 \psi - \frac{1}{V} \partial_i V {\mathcal D}_0 \psi \,,
\end{equation}
the left-hand side of \eqref{almosttheregrav1} becomes
\begin{equation}
   \Delta^{-1}_{{\mathcal D}} \cdot i {\mathcal D}_0 \psi \equiv \left(1 - \frac{i}{m}  (\gamma^0 \frac{ n^i}{V n^0 } + \gamma^i ) \partial_i +  \frac{i}{m}  (\gamma^0 \frac{ n^i}{V n^0 } + \gamma^i ) \frac{1}{V} \partial_i V  \right) i {\mathcal D}_0 \psi \,.
\end{equation}
Inverting the new operator $\Delta^{-1}_{\mathcal D}$ on the right side of \eqref{almosttheregrav1}, we arrive to 
\begin{equation} \label{almosttheregrav2}
    i {\mathcal D}_0 \psi = - \Delta_{\mathcal D} \left( \frac{1}{m} \frac{ \gamma^0}{V} {\mathcal D}_0^2 + i  \frac{n^i}{n^0} {\mathcal D}_i + \frac{1}{m} \frac{n^i}{n^0} \gamma^j {\mathcal D}_i {\mathcal D}_j  + \frac{\lambda}{m} \frac{\slashed n}{n^0} \right) \psi \,.
\end{equation}
The next step is to replace the ${\mathcal D}_0^2$ using the squared EOM. Following the analogous derivations in Chapter \ref{ch1}, it is easy to convince ourselves that, since the Ricci scalar and the commutator of covariant derivatives on $\psi$ vanish, the square of \eqref{vsreomgrav} will be KG-like 
\begin{equation}
    (g^{\mu\nu} {\mathcal D}_\mu {\mathcal D}_\nu + m^2_f)\psi =0 \,,
\end{equation}
implying the relation
\begin{equation}
    {\mathcal D}_0^2 \psi = - (g^{00})^{-1}( {\mathcal D}_i {\mathcal D}^i +m^2_f ) \psi  = -V^2 ( {\mathcal D}_i {\mathcal D}^i +m^2_f ) \psi\,.
\end{equation}
Therefore, the Schr\"odinger-like equation \eqref{almosttheregrav2} becomes 
\begin{equation} \label{almostschemgrav3}
    i {\mathcal D}_0 \psi =  \Delta_{\mathcal D} \left( m V \gamma^ 0 + \frac{V}{m} \gamma^0 {\mathcal D}_i {\mathcal D}^i - i  \frac{n^i}{n^0} {\mathcal D}_i - \frac{1}{m} \frac{n^i}{n^0} \gamma^j {\mathcal D}_i {\mathcal D}_j  +\frac{2\lambda}{m} V \gamma^0 - \frac{\lambda}{m} \frac{\slashed n}{n^0} \right) \psi \,.
\end{equation}
Now, since spatial covariant derivatives of both vectors and spinors do not get here any new contribution with respect to the partial ones
\begin{eqnarray}
    {\mathcal D}_i \psi &=& \partial_i \psi + \Gamma_i \psi = \partial_i \psi\,, \\
    {\mathcal D}_j {\mathcal D}_i \psi &=& \partial_j {\mathcal D}_i \psi + \Gamma_j {\mathcal D}_i\psi -\{^{\;\, \mu }_{ j \,\,\, i} \} {\mathcal D}_\mu \psi =\partial_j \partial_i \psi  \,,\nonumber
\end{eqnarray}
we can re-express the right-hand side of \eqref{almostschemgrav3} with only partial derivatives
\begin{equation} \label{almostschemgrav4}
    i {\mathcal D}_0 \psi =  \Delta_{\mathcal D} \left( m V \gamma^ 0 + \frac{V}{m} \gamma^0\partial_i \partial^i - i  \frac{n^i}{n^0} \partial_i - \frac{1}{m} \frac{n^i}{n^0} \gamma^j \partial_i \partial_j  +\frac{\lambda}{m} V \gamma^0 - \frac{\lambda}{m} \frac{n_i}{n^0} \gamma^i \right) \psi \,.
\end{equation}
At this point, to have a meaningful LI limit, we should have
\begin{equation}
    \Delta_{\mathcal D} \left( m V \gamma^ 0 + \frac{V}{m} \gamma^0 \partial_i \partial^i - i  \frac{n^i}{n^0} \partial_i - \frac{1}{m} \frac{n^i}{n^0} \gamma^j \partial_i \partial_j \right ) = m V \gamma^0 - i V \gamma^0 \gamma^i \partial_i\,,
\end{equation}
such that, when $\lambda\to0$, we get back the LI case \eqref{diracH2}. This can be proven by demonstrating the following statement
\begin{equation}
     m V \gamma^ 0 + \frac{V}{m} \gamma^0 \partial_i \partial^i - i  \frac{n^i}{n^0} \partial_i - \frac{1}{m} \frac{n^i}{n^0} \gamma^j \partial_i \partial_j  = \Delta_{\mathcal D} ^{-1} \left (m V \gamma^0 - i V \gamma^0 \gamma^i \partial_i \right ) ,
\end{equation}
similarly to what is done in the free case \eqref{connectiontodiracfree}. With this consideration in mind, the expression of \eqref{almostschemgrav4} simplifies significantly to
\begin{equation} \label{almostschemgrav5}
    i {\mathcal D}_0 \psi =  \left ( m V \gamma^0 - i \gamma^0 \gamma^i (V \partial_i +\frac12 \partial_i V) +\frac{\lambda}{m} \Delta_{\mathcal D} V ( \gamma^0 - \tilde n_i \gamma^i ) \right ) \psi \,,
\end{equation}
where we defined the spatial vector
\begin{equation}
    \tilde n_i \equiv \frac{n_i}{n^0 V} \,,
\end{equation}
so that
\begin{equation}
    \Delta_{\mathcal D} = \left(1 - \frac{i}{m}  (\gamma^0 \tilde n^i + \gamma^i ) \partial_i +  \frac{i}{m}  (\gamma^0 \tilde n^i + \gamma^i ) \frac{1}{V} \partial_i V  \right) ^{-1} .
\end{equation}

\subsection{Properties of the Vector $\tilde n^i$}

Let us stop for a moment to ponder over the properties of the vector $\tilde n^i$. From the covariant constancy of the spatial components $n^i$ of the VSR four-vector, we find
\begin{equation}
   0 = {\mathcal D}_i n^j = \partial_i n^j + \{^{\;\, j }_{ i \,\,\, \mu} \} n^\mu = \partial_i n^j \,.
\end{equation}
At the same time, the condition \eqref{covdevnzero} for the time component is
\begin{equation}
    0 = {\mathcal D}_i n^0 = \partial_i n^0 + \{^{\;\, 0 }_{ i \,\,\, \mu} \} n^\mu = \partial_i n^0 + \{^{\;\, 0 }_{ i \,\,\, 0} \} n^0 = \partial_i n^0 +\frac{1}{V} \partial_i V n^0 \,.
\end{equation}
Multiplying the last equation by $V$, we get
\begin{equation}
    V \partial_i n^0 +\partial_i V n^0 = \partial_i (V n^0) = 0\,.
\end{equation}
Thus, the combination of $n^\mu-$components in $\tilde n^i$ is such that it does not depend on the spatial coordinates
\begin{equation}
    \partial_ i \tilde n^j = \partial_i (\frac{n^j}{n^0 V}) = 0\,.
\end{equation}
Furthermore, since $n^\mu$ is lightlike
\begin{equation}
    g_{00} (n^0)^2 = V^2 (n^0)^2 = - n_i n^i = n^i n^i = |\vec n |^2 \,,
\end{equation}
the new vector $\tilde n ^i$ is find to actually be a unitary vector
\begin{equation}
    |\vec {\tilde n} |^2 = \frac{|\vec n|^2}{(n^0 V)^2} = 1\,.
\end{equation}
We can therefore identify $\tilde n^i$ as a time-dependent unit vector, which plays a similar role to the one of $n^i$ in inertial frames. To conclude, using the relations
\begin{eqnarray}
    && 0= {\mathcal D}_0 n^i = \partial_t n^i + V\partial_i V n^0\,,\\
    && 0= {\mathcal D}_0 n^0 = \partial_t n^0 + \frac{1}{V} \partial_i V n^i \,, \nonumber
\end{eqnarray}
we highlight the $c-$scaling of its time derivative, observing that
\begin{eqnarray}
    \partial_t \tilde n^i = \frac{1}{V} \partial_t (\frac{n^i}{n^0}) = \frac{1}{V n^0} \partial_t n^i - \frac{n^ i}{V (n^0)^2} \partial_t n^0 = \tilde n^i \tilde n^k \partial_k V -\partial_i V \,\,\propto \,\, \nicefrac{1}{c} \,.
\end{eqnarray}

\subsection{Order of Perturbation} \label{sec:orderofpert}

Before we proceed with the calculations, it is useful to understand what terms we can expect at the end from the perturbative treatment. In fact, while for the LI case, it was possible to easily use $\nicefrac{1}{c}$ as the formal expansion parameter, the presence of VSR corrections, which often go with $c^2$, undermines this approach. Nevertheless, $m$ is still a good expansion parameter and allows us to develop the following argument: In the end, our goal is to obtain the non-relativistic VSR Hamiltonian. If we imagine to divide the Hamiltonian $\mathcal H$ by $mc^2$ and recall that
\begin{equation}
    v \equiv V-1 = \frac{\vec a \cdot \vec x }{c^2} \, \propto \, \nicefrac{1}{c^2} \,,
\end{equation}
symbolically speaking, there are only three dimensionless arrangements of parameters we could expect from the FW procedure
\begin{equation}
    \left \{ \frac{\lambda}{m^2} , \frac{p}{mc} , v \right \} .
\end{equation}
Consequently, the combinations we can construct up to order $\nicefrac{1}{c^2}$ and first order in $\lambda$ are 
\begin{eqnarray}\label{perturbativestructures}
    && \nicefrac{1}{c^2} \,\,\, - \,\,\, \left \{ \frac{\lambda}{m^2} , (\frac{p}{mc})^2, \frac{\lambda}{m^2} (\frac{p}{mc})^2, v, \frac{\lambda}{m^2} v \right \} , 
\end{eqnarray}
where, in the construction of the above terms we used the fact that non-VSR terms (i.e. terms not proportional to $\lambda$) cannot have odd powers of $p$ because there is no way of forming scalars with an odd number of vectors. Furthermore, we note that, since non-trivial VSR effects are absent in the free case, new non-trivial VSR structures (featuring $\tilde n^ i$) should appear with $v$ so that they vanish when taking the zero-acceleration limit.\\
As we can see from \eqref{perturbativestructures}, all $\nicefrac{1}{c^2}-$corrections to $ \mathcal H/mc^2$ are at most of order $\nicefrac{1}{m^4}$. Then, it will be sufficient to perform the FW transformation up to order $\nicefrac{1}{m^3}$, to obtain all the 0PN Hamiltonian contributions. Clearly, during the process, also 0.5PN- and 1PN-corrections can be present, but they are easy to detect and discard. An example is $v^2-$corrections that we can always safely neglect throughout the expansion procedure, as expected from the equivalent LI result.

\subsection{Hamiltonian Expression in Non-Inertial VSR }

At this point, “rationalizing” the denominator of $\Delta_{\mathcal D}$ as in \eqref{freeration}, the last term on the right-hand side of \eqref{almostschemgrav5} becomes
\begin{equation}
     \Delta_{\mathcal D} V( \gamma^0 - \tilde n_i \gamma^i )= \Box_{\mathcal D} \left ( V( \gamma^0 - \tilde n_i \gamma^i ) +\frac{i}{m} V (\tilde n^i  - \gamma^0 \gamma^i - \gamma^0 \gamma^j \tilde n^i \tilde n_j -\gamma^i \gamma^j \tilde n_j) \partial_i \right ) ,
\end{equation}
with the new operator $\Box_{\mathcal D}$ defined to be
\begin{equation}
    \Box^{-1}_{\mathcal D} \equiv 1 +\frac{1}{m^2} (n^i n^j + g^{ij} ) \left (\partial_i\partial_j - \frac{2}{V} \partial_i V \partial_j + \frac{2}{V^2} \partial_i V \partial_j V \right ) .
\end{equation}
Thus, we can identify from \eqref{almostschemgrav5} the non-inertial VSR Hamiltonian as
\begin{eqnarray} \label{Hgvsr1}
    \mathcal{H}^g_{VSR} &=& m V \gamma^0 - i \gamma^0 \gamma^i (V \partial_i +\frac12 \partial_i V) \\
    && +\frac{\lambda}{m} \Box_{\mathcal D} V \left (  \gamma^0 - \tilde n_i \gamma^i  -\frac{i}{m} ( \gamma^0 \gamma^i +\gamma^0 \gamma^j \tilde n^i \tilde n_j +\sigma^{ij} \tilde n_j) \partial_i \right ) . \nonumber
\end{eqnarray}
According to the above considerations, we expand the last result up to order $\nicefrac{1}{m^3}$, while neglecting second order terms in $\lambda^2$ and $O(\nicefrac{1}{c})-$contributions, that would only produce irrelevant corrections, obtaining the following expression
\begin{eqnarray} \label{Hgvsr2}
    \mathcal{H}^g_{VSR} &=& (m+\frac{\lambda}{m}) (1+v)\gamma^0 - i (1+\frac{\lambda}{m^2})\gamma^0 \gamma^i \partial_i  -\frac{\lambda}{m} (1+v) \gamma^i \tilde n_i \\
    &&  - i \frac{\lambda}{m^2} \gamma^0\gamma^i \tilde n^j \tilde n_i \partial_j -i \frac{\lambda}{m^2} \sigma^{ij } \tilde n_j \partial_i - \frac{\lambda}{m^3} (\gamma^0 - \gamma^i \tilde n_i) (\partial_i \partial^i + (\tilde n^i \partial_i )^2 )  \,. \nonumber
\end{eqnarray}

\section{Non-Relativistic Limit for VSR through Foldy-Wouthuysen}

To proceed with the realization of the non-relativistic limit, we first identify the odd and even operators from the Hamiltonian \eqref{Hgvsr2}, which are
\begin{eqnarray} \label{EOvsrcasegrav}
    \mathcal E^g &=& m v \gamma^ 0 + \frac{\lambda}{m} (1+v)\gamma^0 -i \frac{\lambda}{m^2} \sigma^{ij } \tilde n_j \partial_i - \frac{\lambda}{m^3} \gamma^0 (\partial_i \partial^i + (\tilde n^i \partial_i )^2 )  \,, \nonumber \\
    \Theta^g &=& - i (1+\frac{\lambda}{m^2})\gamma^0 \gamma^i \partial_i  -\frac{\lambda}{m} (1+v) \gamma^i \tilde n_i - i \frac{\lambda}{m^2} \gamma^0\gamma^i \tilde n^j \tilde n_i \partial_j \,.
\end{eqnarray}
Once again, we neglected $\nicefrac{1}{m^3}-$terms in $\Theta^g$ as they would not contribute to the end result at our order of approximation. In principle, we should include in the FW Hamiltonian also time derivatives of the transformation $S$, as we have seen in \eqref{HFWcommut}. However, being
\begin{equation}
    \dot S = - \frac{1}{m c^2} \gamma^0 \dot \Theta \sim \frac{\lambda c^2}{m^2 c^2} \dot {\tilde n} \,\,\propto \,\,\nicefrac{1}{c} \,,
\end{equation}
those contribution would already be higher order and are thus safe to ignore. Hence, the relevant formula for the FW expansion is again \eqref{HFWafter}. Calculating each term up to the needed order, we find\footnote{Note that formula \eqref{HFWcommut} is valid even when $S$ is not hermitian (as in VSR), because the FW transformation $U\equiv e^{iS}$ still preserves the original Schrödinger-like dynamics.}
\begin{eqnarray}
    \, [\Theta^g, \mathcal{E}^g] &=& i \frac{2\lambda}{m} \gamma^i \partial_i + 2\lambda v \gamma^0 \gamma^i \tilde n_i -\frac{2\lambda}{m^2} \gamma^0 \gamma^j \tilde n_j \partial_i \partial^i - \frac{2\lambda}{m^2} \gamma^0 \gamma^i \tilde n^j \partial_i \partial_j \nonumber\,,\\ 
    \, (\Theta^g)^2 &=& (1+\frac{2\lambda}{m^2}) \partial_i \partial^i + \frac{2\lambda}{m^2} (\tilde n^i \partial_i) ^2 + i \frac{2\lambda}{m} \gamma^0 \sigma^{ij} \tilde n_j \partial_i \,, \nonumber\\
    \, [\Theta^g, [ \Theta^g, \mathcal{E}^g ]] &=& \frac{4\lambda}{m} \gamma^0 \partial_i \partial^i   \,,\,\, (\Theta^g)^3 = 0 = (\Theta^g)^4 =  \mathcal{E}^g (\Theta^g)^3 \,.
\end{eqnarray}
Putting all together, after a few cancellations, we end up with the following Hamiltonian
\begin{eqnarray}
    \mathcal{H}^g_{FW} = (m+\frac{\lambda}{m})(1+v) \, \gamma^0 + \frac{\gamma^0}{2m} (1-\frac{\lambda}{m^2}) \partial_i \partial^i - i \frac{\lambda}{m^2} \gamma^0 \gamma^i \partial_i + \frac{\lambda}{m} v \gamma^0 \gamma^i \tilde n_i \,.
\end{eqnarray}
Performing a second FW transformation has the only effect of removing the remaining odd terms. Therefore, it is straightforward to write down the final non-relativistic VSR Hamiltonian $H_{VSR }^g $ for UCNs in an accelerated frame
\begin{equation}
    H_{VSR }^g = \gamma^0 \left ( (m+\frac{\lambda}{m})+ (m+\frac{\lambda}{m}) \vec a \cdot \vec x + \frac{1}{2m} (1-\frac{\lambda}{m^2}) \partial_i \partial^i \right ) ,
\end{equation}
which, at this order in $\lambda $, to better expose its analogy with \eqref{HgFWlicase} can be re-formulated as
\begin{equation} \label{finalaccham}
    H_{VSR }^g = \gamma^0 \left ( m_f+ m_f \vec a \cdot \vec x + \frac{1}{2m_f} \partial_i \partial^i \right ) .
\end{equation}
While we probably could have expected the lack of non-trivial VSR terms from dimensional arguments, as mentioned in Section \ref{sec:orderofpert}, this outcome is still very interesting as it reveals at least two fundamental aspects 
\begin{enumerate} 
    \item First, we see that, in the non-relativistic limit, VSR contributions $\propto \lambda$ behave in such a way as to preserve the equivalence between the gravitational” and inertial mass, respectively present in the potential and kinetic term. This feature was not at all evident from the relativistic EOM \eqref{vsreomgrav}.  
    \item Secondly, from the form of \eqref{finalaccham} we deduce that, when describing ultracold neutrons from the perspective of non-inertial frames in a non-relativistic world $c\to\infty$, the VSR preferred direction $n^\mu$ does not play any role.
\end{enumerate}
Furthermore, we did not observe differences between the particle-antiparticle descriptions.

\subsection{Higher Order Corrections}

The results just obtained do not allow us to use the qBOUNCE experiment to directly place constraints on the VSR parameters. That is because, at 0PN-order, $\lambda$ enters the non-relativistic Hamiltonian just to correct the particle mass from its Dirac value $m$ to the effective VSR one $m_f$, and since the neutron mass $m_f$ in this context is just a parameter usually obtained from other measurements, we have no way to disentangle the Dirac and VSR contributions.\\
Nevertheless, when going further into the $\nicefrac{1}{c}-$expansion, the situation is different. 1PN corrections are already extremely tiny in the standard context of the curved Dirac equation \cite{PhysRevD.109.064085}, so we only focus on 0.5PN. In the non-rotating LI case, indeed, there are no 0.5PN contributions, implying that we could safely expect all of them to be VSR related. Realizing a similar dimensional analysis to that of Section \ref{sec:orderofpert} we have the following possible combinations
\begin{eqnarray}\label{perturbativestructures2}
    && \nicefrac{1}{c^3} \,\,\, - \,\,\, \left \{ \frac{\lambda}{m^2} \frac{p}{mc} v , \frac{\lambda}{m^2} (\frac{p}{mc})^3 \right \} \,, 
\end{eqnarray}
where, once again, we excluded non-VSR contributions with an odd number of $p$. Those new structures in \eqref{perturbativestructures2}, if present, would lead to completely novel repercussions originating, for example, from VSR terms such as $\sim \sigma^{ij} \tilde n_i \partial_j v $ that would induce spin-dependent effects. This would readily enable us to use the qBOUNCE observations to constrain VSR and its anisotropic nature from different but complementary viewing angles. For those reasons, at the moment we are extending our analysis to the 0.5PN order, hoping to find some novel implications. \\
Another limitation of our current analysis, which we wish to overcome in the future, is the absence of observer rotation. In fact, when describing laboratories on Earth's surface, rotational effects are certainly worth considering, as they could naturally lead to additional 0.5PN corrections. 


\chapter[LINEARIZED GRAVITY AND GRAVITATIONAL WAVES IN VSR]{Linearized Gravity and Gravitational Waves in VSR} \label{ch4}
Back in 2021, while studying the existing scientific literature on VSR, we noted that no work on spin-2 fields had been realized. At the same time, different attempts to extend VSR to gravity had already been explored \cite{grav1,grav2}. Nevertheless, both classical gravitational waves (GW) and gravitons are theoretically described as spin-2 modes. For this reason, the application of the VSR ideas in the context of linearized gravity seemed a natural extension of the pre-existing research on the topic. We therefore started to work on this problem with the aim of finding and studying a VSR linearized gravity model.\\
As also demonstrated for photons \cite{PhysRevD.100.055029} and non-Abelian gauge theories \cite{Alfaro:2013uva}, one of the universal consequences of VSR extensions is the generation of new masses for the fields involved. Within the gravitational sector, that would imply massive gravitons, a possibility that is generally well restricted by experiments, but that is receiving more and more attention in the last decades due to its potential to solve some of the most puzzling mysteries of our Universe, like its accelerated expansion \cite{hinterbichler2012theoretical}, and cosmological tensions \cite{de2021minimal}. \\
Outside of its recent resurgence, massive gravity has a long story. Back in 1939, Fierz and Pauli presented for the first time a Lagrangian for a massive spin-2 particle \cite{Fierz:1939ix}. Many years later, people realized that the Fierz-Pauli (FP) theory led to a discontinuity with GR in the massless limit, the so-called vDVZ discontinuity \cite{vanDam:1970vg,Zakharov:1970cc}. Furthermore, in addition to the two transverse tensorial modes of GR, the FP gravitons possess extra degrees of freedom (DOF). A direct consequence of that is the manifestation of ghost modes \cite{Boulware:1972yco}, which in turn induce instabilities. Several solutions to these pathologies, such as the Vainshtein mechanism \cite{Vainshtein:1972sx}, have already been studied and implemented in different models \cite{Dvali:2000hr,Dvali:2000rv,deRham:2010kj}. For an exhaustive review on the topic of massive gravity, see \cite{deRham:2014zqa} and the references therein. \\
In general, the pathologies of FP theory seem rooted in LI \cite{Arkani-Hamed:2004gbh,Berezhiani:2009kv, Gabadadze:2004iv}. In fact, it is shown in \cite{Rubakov:2004eb,Dubovsky:2004sg} that, by breaking gauge invariance through a LV mass term, discontinuities and ghost propagation can be effectively avoided. In a similar fashion, breaking Lorentz symmetry to the VSR group allowed us to write a coherent massive linearized gravity model, dubbed “Very Special Linear Gravity” (VSLG), with the additional benefit of preserving gauge invariance under infinitesimal diffeomorphisms. In the following, we will guide you through the construction of VSLG, starting from the very basics and arriving at its application to gravitational waves and binary stars.

\section{Very Special Linear Gravity}

Our first task is to construct a linearized model for gravity in VSR. Linearized refers to two aspects: First, we assume to be in a weak gravity limit where the metric can be expanded around a flat background $g_{\mu\nu}(x)= \eta_{\mu\nu} + h_{\mu\nu}(x)$ with $|h_{\mu\nu}|<<1$. Second, we want the EOM to be linear in the field $h_{\mu\nu}$ or equivalently the Lagrangian to be quadratic
\begin{equation}
    \mathcal{L}_g = \frac12 \, h_{\mu\nu } \, O^{\mu\nu \alpha \beta} \, h_{\alpha \beta } \,.
\end{equation}
The ingredients to build up the operator $O^{\mu\nu \alpha \beta}$ are the Minkowski background $\eta_{\mu\nu}$, partial derivatives $\partial_\mu$ (or equivalently $p_\mu$ in momentum space) and the VSR operator $N^\mu$. Simbolically, we can resume all possible contributions at most quadratic in $\partial_\mu$ and $N^\mu$ as
\begin{equation}
    O =  3 \, \eta \eta + 9 \,  \partial\partial \eta +12 \, \partial N \eta + 12 \, \partial\partial N N \,,
\end{equation}
where the numbers indicate how many different terms we can construct with each subset of objects. At this point, we would have as many as 36 different parameters in our model. However, many of them are related by the index symmetries $\mu \Longleftrightarrow \nu$, $\alpha \Longleftrightarrow \beta$, $\mu \nu \Longleftrightarrow \alpha \beta$ that are inherited from the contraction of $O^{\mu\nu \alpha \beta}$ with the $h-$fields. Additionally, we want our theory, and thus the Lagrangian, to be invariant under linearized diffeomorphisms as in GR. That is realized by requiring invariance under the gauge transformations $\delta h_{\mu\nu} = \partial_\mu \xi_\nu + \partial_\nu \xi_\mu $, that translates into the condition $ O^{\mu\nu\alpha\beta} \partial_\mu = 0 $. Joining all those restrictions, the number of free parameters reduces to two: One can be factored out and taken to be proportional to the Einstein constant $\chi = \frac{c^4}{16 \pi G} \,\to\, O^{\mu\nu \alpha \beta}= \frac{\chi}{2}\mathcal O^{\mu\nu \alpha \beta} $. The other one has the dimension of a squared mass and remarkably plays the role of a graviton mass, as we shall see later. Hence, the complete expression for $\mathcal O_{\mu\nu \alpha \beta}$ is \cite{grav3}
\begin{eqnarray} \label{operatorO}
\mathcal{O}_{\mu \nu \alpha \beta} &=&  \frac{1}{2} \partial_{\mu} \partial_{\alpha} \eta_{\nu \beta} +
\frac{1}{2} \partial_{\mu} \partial_{\beta} \eta_{\nu \alpha}  + \frac{1}{2} \partial_{\nu}
\partial_{\beta} \eta_{\mu \alpha} + \frac{1}{2} \partial_{\nu}
\partial_{\alpha} \eta_{\mu \beta} - \partial_{\mu} \partial_{\nu} \eta_{\alpha
\beta}  -
\partial_{\alpha} \partial_{\beta} \eta_{\mu \nu} \nonumber\\
&& +  ( \eta_{\mu \nu} \eta_{\alpha
\beta} - \frac{1}{2} \eta_{\mu \alpha} \eta_{\nu \beta} - \frac{1}{2} \eta_{\mu \beta} \eta_{\nu \alpha} ) \partial^2 + m^2_g \eta_{\mu
\nu} \eta_{\alpha \beta} - \frac{m^2_g}{2} ( \eta_{\mu \alpha} \eta_{\nu \beta}
- \eta_{\mu \beta} \eta_{\nu \alpha} ) \nonumber\\
&& - m^2_g N_{\mu} N_{\nu}
\partial_{\alpha} \partial_{\beta} + \frac{m^2_g}{2} N_{\mu} N_{\alpha}
\partial_{\nu} \partial_{\beta} + \frac{m^2_g}{2} N_{\mu} N_{\beta}
\partial_{\nu} \partial_{\alpha} + \frac{m^2_g}{2} N_{\nu} N_{\alpha}
\partial_{\mu} \partial_{\beta}  \nonumber\\
&& + \frac{m^2_g}{2} N_{\nu} N_{\beta}
\partial_{\mu} \partial_{\alpha} - m^2_g N_{\alpha} N_{\beta} \partial_{\mu}
\partial_{\nu} + m^2_g \partial^2 N_{\mu} N_{\nu} \eta_{\alpha \beta} -
\frac{m^2_g}{2} \partial^2 N_{\mu} N_{\alpha} \eta_{\nu \beta}  \nonumber\\
&& -
\frac{m^2_g}{2} \partial^2 N_{\mu} N_{\beta} \eta_{\nu \alpha} - \frac{m^2_g}{2} \partial^2 N_{\nu} N_{\beta} \eta_{\mu \alpha} -
\frac{m^2_g}{2} \partial^2 \eta_{\mu \beta} N_{\nu} N_{\alpha} + m^2_g
\partial^2 N_{\alpha} N_{\beta} \eta_{\mu \nu}  \nonumber\\
&& - m^2_g \eta_{\mu \nu}
N_{\alpha} \partial_{\beta} - m^2_g \eta_{\mu \nu} \partial_{\alpha} N_{\beta}
+ \frac{m^2_g}{2} \eta_{\mu \alpha} N_{\nu} \partial_{\beta} + \frac{m^2_g}{2}
\eta_{\mu \alpha} \partial_{\nu} N_{\beta}  \nonumber\\
&& + \frac{m^2_g}{2} \eta_{\mu \beta}
N_{\nu} \partial_{\alpha} + \frac{m^2_g}{2} \eta_{\mu \beta} \partial_{\nu}
N_{\alpha} + \frac{m^2_g}{2} \eta_{\nu \alpha} N_{\mu} \partial_{\beta} +
\frac{m^2_g}{2} \eta_{\nu \alpha} \partial_{\mu} N_{\beta}  \nonumber \\ 
&& + \frac{m^2_g}{2}
\eta_{\nu \beta} N_{\mu} \partial_{\alpha} + \frac{m^2_g}{2} \eta_{\nu \beta}
\partial_{\mu} N_{\alpha}  - m^2_g \eta_{\alpha \beta} N_{\mu} \partial_{\nu} - m^2_g \eta_{\alpha \beta} \partial_{\mu} N_{\nu}  \,.
\end{eqnarray}
The full Lagrangian and respective EOM can now be written as 
\begin{eqnarray} \label{eomfullgrav}
\mathcal L_g = \frac{\chi}{4} \, h_{\mu\nu} \mathcal O ^{\mu\nu\alpha\beta} h_{\alpha \beta}  \,\, \to \,\, \mathcal O_{\mu \nu \alpha\beta } \, h^ {\alpha \beta} =0 \, .
\end{eqnarray}
Sending $m_g \to 0$, the EOM reduces to the standard one for linearized GR
\begin{equation}
   \frac{1}{2} \, ( \, \partial_{\mu} \partial_{\alpha} h^\alpha_{\;\;\nu} + \partial_{\nu}
\partial_{\beta} h_\mu^{\;\;\beta} - \partial_{\mu} \partial_{\nu} h - \eta_{\mu \nu} \partial_{\alpha} \partial_{\beta} h^{\alpha \beta}  + \eta_{\mu\nu }\partial^2 h - \partial^2 h_{\mu\nu})  =0 \,,
\end{equation}
with $ h$ defined as the trace of the metric perturbation $ h \equiv 
 h_\mu^{\;\mu} = h_{\mu\nu } \eta ^{\mu\nu}$.

\subsection{Gauge Choices and Equations of Motion}

The EOM \eqref{eomfullgrav} is not easy to handle. Nevertheless, we have gauge invariance on our side. We can therefore exploit the gauge redundancy we imposed on our theory to drastically simplify our EOM. For convenience, here we work in momentum space and define $\tilde N^\mu \equiv \frac{n^\mu}{n\cdot p}$ in addition to the Fourier-transformed $ \tilde h_{\mu\nu} (p)$. Let us start by exploiting the gauge freedom to impose the first four conditions on $\tilde h_{\mu\nu}$, which will correspond to the Lorenz gauge $p^\mu \, \tilde h_{\mu\nu}=0$. Contracting the EOM with $\eta_{\mu\nu} $ and $p_\mu$, while imposing the Lorenz condition, give us respectively two useful equations
\begin{eqnarray}
   && p^2 \tilde  h= 0\, , \label{eqh}\\
   && p_\nu \,\tilde h + p^2 \tilde N^\mu \tilde h_{\mu \nu} = 0\, \label{eqph} .
\end{eqnarray}
Note that, because of the gauge invariance of $\mathcal L$, these equations will also be valid for transformed fields as long as we preserve the Lorenz condition. However, this gauge choice does not fix the gauge uniquely. In fact, we could still perform, without any inconvenience, transformations that still satisfy the Lorenz constraint
\begin{equation} \label{addgaugecondh}
    p^\mu \tilde h'_{\mu \nu} = 0 = p^\mu \tilde h_{\mu \nu} + p^2 \xi_\nu + p\cdot \xi \, p_{\nu} = p^2 \xi_\nu + p\cdot \xi \, p_{\nu} \, ,
\end{equation}
by choosing for $\xi_\nu$ the following appropriate form
\begin{equation} \label{eqxi}
    \xi_\nu = ( c_\nu- \frac{1}{2}\frac{p\cdot c}{p^2}p_\nu ) \, \delta(p^2) \, ,
\end{equation}
with $c_\nu$ being an arbitrary four-function. Indeed, replacing the expression \eqref{eqxi} in the above condition \eqref{addgaugecondh}, we see that many terms cancel and we just end up with $p^2 \,c_\nu \,\delta(p^2)$ which is automatically zero due to the presence of the $\delta$. \\
The arbitrariness of $c_\nu$ indicates that we still have the freedom to impose at least four more conditions on $\tilde h_{\mu \nu}$. The first condition to impose is the disappearance of the trace $\tilde h$. The general solution of \eqref{eqh} is $\tilde h = h_0 \,\delta(p^2)$. Thus, we have
\begin{equation}
    \tilde h'=0 = \tilde h + 2 p^\mu \xi_\mu = \delta(p^2) (h_0 + p \cdot c) \, ,
\end{equation}
which always has at least the solution
\begin{equation} \label{eqcp}
    c \cdot p = -h_0 \, .
\end{equation}
Such a choice of $c_\nu$, always allow us to impose the traceless condition, without affecting the Lorenz one. With this new constraint, equation \eqref{eqph} for the primed field $\tilde h'_{\mu\nu}$ becomes
\begin{equation}\label{eqNh}
    p^2 \tilde N^\mu \tilde h'_{\mu \nu} = 0 \,\,\, \to \,\,\, \tilde N^\mu \tilde h'_{\mu \nu} = H'_{ \nu} \,\delta(p^2) \, .
\end{equation}
Finally, we would like to impose other four conditions given by $ \tilde N^\mu \tilde h_{\mu\nu}=0 $ or equivalently $H_{ \nu}= 0$. Nevertheless, having already fixed five constraints, we should carefully check if we have enough gauge freedom left to consistently impose also this new condition. Let us make yet another gauge transformation of the form \eqref{eqxi} such that
\begin{equation}
    \tilde N^\mu \tilde h''_{\mu \nu} = 0  = \delta(p^2) (H'_{\nu} +c_\nu+ \tilde N \cdot c \, p_\nu) \, .
\end{equation}
The above equation has at least the solution
\begin{equation} \label{eqrandom1}
    H'_{\nu} +c_\nu+ \tilde N \cdot c \,p_\nu=0 \,\, \rightarrow \,\, c_\nu=- H'_{\nu} - \tilde  N \cdot c \, p_\nu \, .
\end{equation}
Being $\tilde N^\nu c_\nu = - \frac{1}{2} \tilde N^\nu H'_{ \nu} $, the requirement in \eqref{eqrandom1} can be written as
\begin{equation} \label{eqcfin}
    c_\nu= - H'_{\nu} +\frac{1}{2} \tilde N^\mu H'_{ \mu} \,p_\nu\, .
\end{equation}
Note that, while the Lorenz gauge is preserved automatically by choosing $\xi$ according to \eqref{eqxi}, also the traceless condition is satisfied since, contracting \eqref{eqcfin} with $p^\nu$, we get
\begin{equation}
    c \cdot p =  p^\nu H'_{\nu} +\frac{1}{2} p^2 \tilde N^\mu H'_{ \mu} = 0 \,,
\end{equation}
where $p^\nu H'_{\nu} = 0$ due to the Lorenz condition 
\begin{equation}
     p^\nu \tilde N^\mu \tilde h'_{\mu\nu}  = 0 = p^\nu H'_{ \nu } \delta(p^2) \,\,\to\,\, p^\nu H'_{ \nu } =0 \,,
\end{equation}
and $p^2 \tilde N^\mu \tilde H'_{ \mu} = 0$ by default due to \eqref{eqNh}. That concludes our proof of the compatibility of these gauge conditions, having demonstrated that, starting from a Lorenz gauge, it is always possible to exploit the additional gauge freedom to fix the remaining constraints. The above set of gauge choices, which we named VSR gauge, can be summarized as
\begin{eqnarray} \label{finalhgaugecond}
   && p^\mu \tilde h_{\mu \nu}= 0 \, , \nonumber\\
   && n^\mu \tilde h_{\mu \nu} = 0 \, , \\
   && \tilde h= 0 \, .\nonumber
\end{eqnarray}
With all that said, in the VSR gauge, the momentum-space EOM takes the simple form
\begin{equation} \label{KGeq}
(p^2 -m^2_g) \, \tilde  h_{\mu \nu} = 0 \, ,
\end{equation}
that represents an ordinary LI relativistic dispersion relation, with $m_g$ playing the role of a mass for the graviton, as anticipated before. From \eqref{KGeq} we also deduce that for on-shell gravitons $n \cdot p$ can never be zero, allowing for the simplification realized in the second gauge relation of \eqref{finalhgaugecond}. Note also that the position-space equivalent of \eqref{KGeq} would instead be a standard Klein-Gordon equation for the field $h_{\mu\nu}$.

\subsection{Physical Degrees of Freedom}

We are now ready to write our gauge-fixed field $\tilde h_{\mu\nu}$ in function of the physical DOF of the theory. Despite the presence of mass, we will show that VSLG still features only two physical DOFs owing to the preserved gauge invariance. Just for this subsection, we will not assume $n^0=1$ to make the $n^\mu-$scaling invariance of the expressions explicit. From the first two conditions in \eqref{finalhgaugecond} we observe that
\begin{equation} \label{h0hi}
    \tilde h_{0 \beta} = -\frac{n^i}{n^0} \tilde h_{i\beta } = -\frac{p^i}{p^0} \tilde h_{i \beta} \,\,\rightarrow \,\,\left ( \frac{n^i}{n^0}-\frac{p^i}{p^0} \right ) \tilde h_{i \beta} = 0 \, ,
\end{equation}
meaning $\tilde h_{i\beta}$ will have no projection on the direction $\frac{\vec n}{n^0}-\frac{\vec p}{p^0}$. It is then convenient to choose as a three-dimensional spatial basis the orthogonal set $ \{ \vec u \,,\, \vec M \,,\,  \frac{\vec n}{n^0}-\frac{\vec p}{p^0} \}$, where we have defined the dimensionless vectors
\begin{eqnarray}
   \vec u =  \frac{\vec n}{n^0} \times \frac{\vec p}{p^0}  \; , \;\;\; \vec M =
   \vec u \times \left ( \frac{\vec n}{n^0}-\frac{\vec p}{p^0} \right ) \, .
\end{eqnarray}
Hence, we can write the components of $\tilde h$ as linear combinations of the basis' elements
\begin{eqnarray} \label{exprh1}
   && \tilde h_{i \beta} = A_{\beta} u_i + B_\beta M_i \, , \, \, \tilde  h_{0 \beta} = -\frac{n^i}{n^0} \tilde h_{i \beta} = \frac{n_i}{n_0} \tilde h_{i \beta} \, , \nonumber \\
   && \tilde h_{ij } =A_j u_i + B_j M_i = A_i u_j + B_i M_j = \tilde  h_{ji} \, .
\end{eqnarray}
Since we want $( \frac{ n ^i}{n^0}-\frac{ p^i}{p^0}) \,\tilde h_{i j}=0$, thus $A_i$, $B_i$ in our basis must be linear combinations only of the vectors $u_i$ and $M_i$
\begin{eqnarray*}
    A_i = a u_i + b M_i \; ,\;\;\; B_i = c u_i + d M_i \, ,
\end{eqnarray*}
where from the symmetry condition $\tilde h_{ij} = \tilde h_{ji}$ we get $b=c$. The coefficients $A_0$, $B_0$ are constrained by $\tilde h_{0i} = \tilde h_{i0}$, which directly implies
\begin{eqnarray*}
    A_0 = \frac{\vec M \cdot \vec n}{n_0} b \;,\;\;\; B_0 =\frac{\vec M \cdot \vec n}{n_0} b \, .
\end{eqnarray*}
Imposing the traceless condition in \eqref{finalhgaugecond}, we find an expression relating $a$ and $d$
\begin{equation}
    a = \frac{(\vec M \cdot \vec n)^2 - n_0^2\vec M ^2}{ n_0^2 \, \vec u ^2} d  \,.
\end{equation}
Finally, taking $n^0 =1$ again, we obtain the components of $\tilde h_{\mu \nu}$ in function of the VSR physical DOFs $\{b,d\}$, which are just two in contrast to the five of LI massive gravitons
\begin{eqnarray}
    && \tilde h_{00}=\left (\vec M \cdot \hat n\right )^2 d \;,\;\;\; \tilde h_{0i}= \left (\vec M \cdot \hat n \right ) (b \, u_i + d M_i) \, , \nonumber \\
    && \tilde h_{ij}= \left ( \frac{(\vec M \cdot \hat n)^2 - \vec M ^2}{ \, \vec u ^2} u_i u_j +  M_i M_j \right ) d +  (u_i M_j + u_j  M_i) \, b \, .
\end{eqnarray}

\section{Propagation of Gravitational Waves in VSR}

As a first application of VSLG, we study the modifications produced by VSR to the geodesic deviation equations for a monochromatic GW. For this task, we need the expression of the linearized Riemann tensor $R_{\rho \mu \nu \kappa}$
\begin{equation} \label{riemannlinear}
    R_{\rho \mu \nu \kappa} = \frac{1}{2} ( h_{\rho \nu , \mu \kappa} -h_{\mu \nu , \rho \kappa} -h_{\rho \kappa, \mu \nu} + h_{\mu \kappa , \rho \nu} ) \,,
\end{equation}
which we recall to be gauge-invariant. Here, we are making use of the “comma”-notation to indicate partial derivation with respect to indices on the right side of the comma
\begin{eqnarray}
     h_{\alpha \beta , \mu} \equiv \partial_\mu h_{\alpha\beta}  \,.
\end{eqnarray}

\subsection{Plane Wave Ansatz}

Since we want to describe monochromatic GWs, we will take for our spacetime perturbation field $h_{\mu\nu}$ the following ansatz
\begin{equation}
     h_{\mu \nu}(x) =  \Re (A_{\mu \nu} \,e^{i k^\mu x_\mu})= \Re( A_{\mu \nu}\, e^{i(E t - k z)}) \, ,
\end{equation}
where we fixed the propagation direction along the $z-$axis, such that $k^\mu = (E,0,0,k)$ and $A_{\mu \nu }$ is the polarization tensor satisfying the conditions 
\begin{equation}
    k^\mu A_{\mu \nu}=  n^\mu A_{\mu \nu} = A^\mu _{\;\mu} = 0\,.
\end{equation}
By deriving $h_{\mu \nu}$ with respect to $t$ and $z$, we see
\begin{equation} \label{2h0h3}
\begin{cases}
     \partial_0 h_{\mu \nu} = i E h_{\mu \nu} \\ \partial _3 h_{\mu \nu} = -i k h_{\mu \nu}
\end{cases}
    \rightarrow  \;\; \partial_3 h_{\mu \nu} = - \frac{k}{E} \partial_0 h_{\mu \nu } \, .
\end{equation} 
Furthermore, since here $h_{\mu \nu}$ has no dependence on $x$ and $y$, we have $\partial_1 h_{\mu \nu} = \partial_2 h_{\mu \nu}= 0$. Using the Lorenz condition $k^\mu h_{\mu\nu} = 0 $, we also find 
\begin{equation}
    h_{3\nu} = - \frac{E}{k} h_{0\nu} \,.
\end{equation}

\subsection{Geodesic Deviation Equation}

The geodesic deviation equation is an equation that shows how two close geodesics behave one respect to the other due to the tidal effects of gravity. Since its derivation does not take into account the underlying theory of gravity but just geometric arguments \cite{Ferrari:2020nzo, weinberg1972gravitation}, we assume that it would still be valid in our case. Its expression reads
\begin{equation} \label{geoeq}
    \partial_0^2 \delta \xi^\mu = R^\mu_{\;00\gamma } \delta \xi^\gamma = \eta^{\mu \delta} R_{\delta 00\gamma } \delta \xi^\gamma = \eta^{\mu \mu} R_{\mu 00\gamma } \delta \xi^\gamma \, .
\end{equation}
The case $\mu = 0$ is trivial, since from \eqref{riemannlinear} we see that $R_{000\gamma} = 0$ always vanish exactly in the linearized limit. Then, we would have $\partial_0 ^2 \xi^0 = 0$, which, combined with the initial conditions $\delta \xi^0 (t=0) = \partial_0 \delta \xi^0 (t=0) = 0$, implies no temporal displacement $\delta \xi^0 = 0$. Instead, when treating the spatial components of the equation 
\begin{equation} \label{geoeqi}
     \partial_0^2 \delta \xi^i = \eta^{ii} R_{i 00 j } \delta \xi^j = 
     - R_{i 00 j } \delta \xi^j = \frac{1}{2} (h_{00,ij } +h_{ij,00} -h_{0i,0j}-h_{0j,0i}) \, \delta \xi^j \,,
\end{equation}
we must separate the directions perpendicular and parallel to the motion of the wave. For the orthogonal directions $i=1,2$, we have
\begin{equation}
    \partial_0^2 \delta \xi^i=  \frac{1}{2} \partial_0^2 h_{i j }\delta \xi^j -\frac{1}{2} \partial_0 \partial_3 h_{0i} \delta \xi^3 \, .
\end{equation}
Using \eqref{h0hi} and \eqref{2h0h3}, we get
\begin{equation*}
\partial_0^2 h_{i 3} - \partial_0 \partial_3 h_{0i} = \partial_0^2 h_{i 3} (1 - \frac{k^2}{E^2}) = \frac{m_g^2}{E^2} \partial_0^2 h_{i3} \, ,
\end{equation*}
implying that
\begin{equation} \label{x1.1}
     \partial_0^2 \delta \xi^i=   \frac{1}{2} \partial_0^2 h_{i 1 }\delta \xi^1 + \frac{1}{2} \partial_0^2 h_{i 2 }\delta \xi^2 + \frac{1}{2} \frac{m_g^2}{E^2} \partial_0^2 h_{i3} \delta \xi^3 \, .
\end{equation}
Since $h_{\mu \nu}$ is small we can solve the differential equation in a perturbatively by defining $\delta \xi^\mu (t)  = \delta \xi_0^\mu + \delta \xi_1 ^\mu (t)$ where $\delta \xi_1^\mu$ is a small perturbation of $\delta \xi_0^\mu$. In this way equation \eqref{x1.1} becomes
\begin{equation*} 
     \partial_0^2 \delta \xi^i_1=  \frac{1}{2} \partial_0^2 h_{i 1 }\delta \xi^1_0 + \frac{1}{2} \partial_0^2 h_{i 2 }\delta \xi^2_0 + \frac{1}{2} \frac{m_g^2}{E^2} \partial_0^2 h_{i3} \delta \xi^3_0 \, ,
\end{equation*}
the solution of which, with initial conditions $\delta \xi^i_1 (t= 0) = \partial_0 \delta \xi^i_1 (t= 0) = 0$, is 
\begin{equation*}
     \delta \xi_1^i=  \frac{1}{2}  h_{i 1 }\delta \xi_0^1 + \frac{1}{2}  h_{i 2 }\delta \xi_0^2 + \frac{1}{2} \frac{m_g^2}{E^2} h_{i3} \delta \xi_0^3 \, . 
\end{equation*}
Hence, the complete displacement along the $i=1,2$ directions is
\begin{equation} \label{eqxii}
    \delta \xi^i= \delta \xi_0^i + \frac{1}{2}  h_{i 1 }\delta \xi_0^1 + \frac{1}{2}  h_{i 2 }\delta \xi_0^2 + \frac{1}{2} \frac{m_g^2}{E^2} h_{i3} \delta \xi_0^3 \, .
\end{equation}
To work out the case $i=3$, we follow the same procedure as $i =1,2$ but starting from the slightly more involved equation 
\begin{equation*}
     \partial_0^2 \delta \xi^3  = \frac{1}{2} \frac{m_g^2}{E^2} \partial_0 ^2 h_{13} \delta \xi^1 +\frac{1}{2} \frac{m_g^2}{E^2} \partial_0 ^2 h_{23} \delta \xi^2  +\frac{1}{2} \frac{m_g^4}{E^4} \partial_0 ^2 h_{33} \delta \xi^3 \, ,
\end{equation*}
from which, with few more calculations than before, we get to the final expression for the complete displacement in the $z$-direction
\begin{equation} \label{eqxi3}
     \delta \xi^3 = \delta \xi^3_0 +\frac{1}{2} \frac{m_g^2}{E^2}  h_{13} \delta \xi^1_0 +\frac{1}{2} \frac{m_g^2}{E^2}  h_{23} \delta \xi^2_0  +\frac{1}{2} \frac{m_g^4}{E^4} h_{33} \delta \xi^3_0 \, .
\end{equation}
The massless limit of these equations coincides, as expected, with the GR case. More details on this can be found in \cite{grav3}.

\subsection{VSR Effects and their Magnitude}

Looking at equations \eqref{eqxii} and \eqref{eqxi3}, we can deduce at least three direct consequences that we should expect from this VSR extension
\begin{enumerate}
    \item Remarkably, the presence of graviton mass induces a motion also on the propagation direction, in contrast to what happens in the case of GR.
    \item Along the transverse directions, we get new contributions proportional to the graviton mass, modifying the standard GR picture.
    \item Hidden in $h_{ij}$ there are anisotropic effects depending on the direction of $\hat n$
\end{enumerate}
While the exact form of those effects depends on the initial condition vector ${\delta \xi_0^\mu}$, their existence still remains general. But how large can these corrections be? As we can see in the above equations, all the cited effects are proportional to the factor $\frac{m_g^2}{E^2}$. Let us give an approximate estimate of this (perturbative) factor. There exist many different experiments from which we can provide upper bounds to the graviton mass \cite{de2017graviton}. For example, from the time lag measured between the gravitational and electromagnetic signal of the first observed merging of two neutron stars (GW170817), we can infer an upper bound of $m_g \sim 10^{-22} \, eV$ \cite{baker2017strong,will2018solar}. From Binary Pulsars \cite{shao2020new} we get an upper bound of $m_g \sim 10^{-28} \, eV$, while from tests in the Solar system we get $m_g \sim 10^{-24} \, eV$ \cite{will2018solar}. However, one should not forget that the majority of the upper bound's estimates found in literature for the graviton mass are model dependent. Anyway, using $10^{-24} \, eV$ as a cautious average upper bound for the graviton mass, in the range of frequencies spanned by the interferometers LIGO and VIRGO, $10 Hz$ to $10 kHz$, the upper bound for our perturbative parameter will be approximately $\frac{m^2_g}{E^2} \sim 10^{-20}$, making VSR effects probably too small to be detected for our current generation of gravitational wave detectors. However, future generations of interferometers, such as LISA \cite{amaro2017laser}, will explore a much lower frequency range $[0.1 mHz, 1 Hz]$. Thus, the parameter's upper bound increases significantly to $\frac{m^2_g}{E^2} \sim 10^{-10}$, which combined with the larger dimensions of future interferometers and the anisotropic nature of VSR could eventually lead to observable effects.

\subsection{Multipolar Nature of VSR Gravitational Radiation} \label{quadrupolarnature}

Due to gauge invariance, we can only couple the $h_{\mu \nu}-$field to a conserved quantity, which, for gravity theories, is embodied by the energy-momentum tensor (EMT) $T_{\mu\nu}$. In the presence of a source, the EOM for $ h_{\mu\nu} $ relates the two quantities as follows
\begin{equation} \label{abc1}
   \mathcal O_{\mu \nu \alpha \beta} \, h^{\alpha \beta} = \frac{1}{\chi} \, T_{\mu \nu} \, ,
\end{equation}
that reduces to the sourced GR linearized equation when sending $m_g\to0$. Therefore, contracting \eqref{abc1} with one derivative, we have
\begin{equation}
   0=\partial^\mu \mathcal O_{\mu \nu \alpha \beta}\, h^{\alpha \beta} = \frac{1}{\chi} \partial^\mu T_{\mu \nu}  \,\,\,\to\,  \,\, \partial^\mu T_{\mu \nu} = 0 \, ,
\end{equation}
which represents a continuity equation leading to the conservation of total energy $ E $ and momentum $ \vec P $ of the source. In GR, we know that energy-momentum conservation implies the absence of monopolar and dipolar gravitational emission: Let us introduce the gravitational monopole $ M \sim \int_S d^3x' \, \rho(\vec x\,') $ and dipole $ \vec D (\vec x)\sim \int _S d^3x ' \rho(\vec x \,') \,\vec x \,' $, with $\rho(\vec x) $ being the energy density of the system $S$. Since $ M $ and the time derivative $ d_t \vec D$ are proportional to conserved quantities, respectively $ E $ and $ \vec P $, the second time derivative of both the gravitational monopole and dipole is zero $ d^2_t M = d^2_t \vec D= 0 $, leading to null monopolar and dipolar radiation. Using the same arguments for VSR we get to the same conclusions, implying the first non-zero multipolar radiation component for gravitational waves is still the quadrupolar one. That will become even more explicit in the next section.

\section{Emitted Radiation from Binary Systems}

\begin{figure}[h!!]
    \includegraphics[width=4cm]{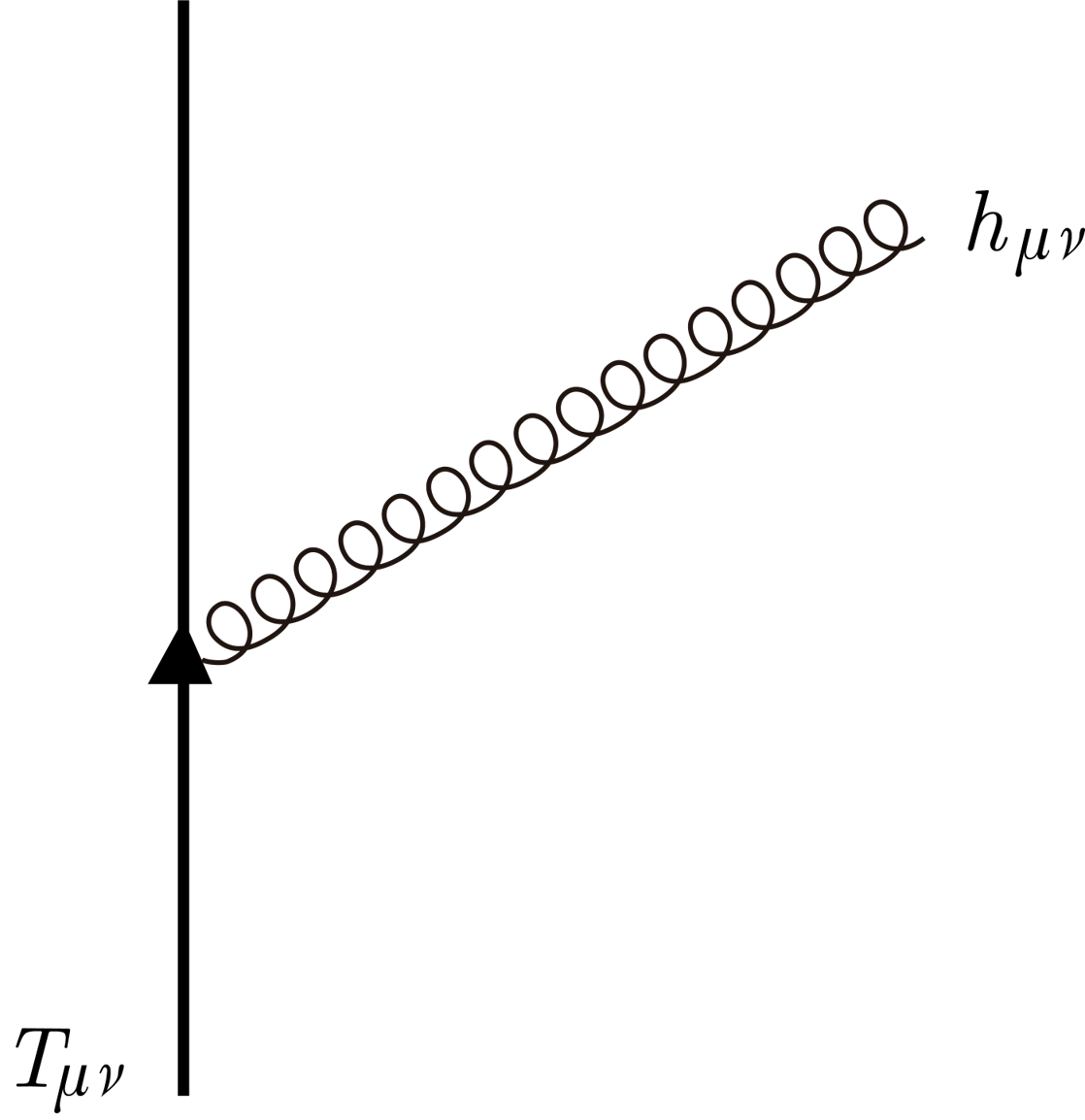}
    \caption[Diagramatic emission of a gravitational wave]{Diagramatic representation of the emission of a gravitational wave, derived from the gauge invariant linear coupling $-\frac{\kappa}{2} T_{\mu \nu}(x) \, h^{\mu \nu} (x)$.}
    \label{fig:diagram-emission}
\end{figure}

We now turn our attention to the calculation of the rate of energy lost by gravitational radiation in binary systems. \cite{Santoni:2023uko} For this purpose, we stick to an EFT approach based on references \cite{goldberger2007effective,goldberger2010gravitational,goldberger2022effective,sturani2021fundamental}. In this context, the spacetime perturbation $h_{\mu\nu}$, which characterizes gravitational waves, is coupled directly to classical sources, represented in the gravitational sector by the energy-momentum tensor $T_{\mu\nu}$. \\
To work in a more canonical framework for QFT, we will redefine the metric perturbation to be dimensionful $h_{\mu \nu } \to \sqrt{\frac{\chi}{2}} h_{\mu\nu }$, so that the new EOM reads
\begin{equation}
    \mathcal O_{\mu \nu \alpha \beta} \, h^{\alpha \beta} =  \frac{1}{\sqrt{ 2 \chi}} \, T_{\mu \nu} = \sqrt{8 \pi G} \, T_{\mu\nu} \,.
\end{equation}
From the above EOM, we can infer the form for our QFT Lagrangian
\begin{equation} \label{lagrdimensionfulh}
    \mathcal L = \frac12 \, h^{\mu\nu} \mathcal O_{\mu \nu \alpha \beta} \, h^{\alpha \beta} + \mathcal L_{int} \,,
\end{equation}
where the interaction is represented by 
\begin{equation}
\label{lagmatt}
\mathcal{L}_{int} = - \frac{\kappa}{2} h_{\mu \nu}T^{\mu \nu} \, , \,\,\, \kappa \equiv \sqrt{ 32 \pi G } \, .
\end{equation}
Within the EFT approach, we shall start from the computation of the Feynman amplitude for a one-graviton emission \cite{Poddar:2021yjd}, represented in Fig.\ref{fig:diagram-emission}. The tree-level amplitude of such a process can be calculated from the interaction term \eqref{lagmatt} and simply reads
\begin{equation}
    {A}_\lambda = - \frac{\kappa}{2} \tilde T_{\mu \nu}(k) \, \epsilon_\lambda^{* \;\mu \nu} (k) \, ,
\end{equation}
with $k^\mu$ being here the momentum of the emitted GW and $\epsilon_\lambda^{\mu\nu}$ its polarization tensor, where $\lambda=1,2$ labels the two different physical polarizations of VSLG.
Starting with the unpolarized squared amplitude $\sum_\lambda |A_\lambda|^2$, we can easily derive the infinitesimal emission probability $d \Sigma$ \cite{goldberger2007effective} as follows
\begin{eqnarray} \label{emrate1}
    d\Sigma = \sum_\lambda |A_\lambda|^2 \frac{d^3 k}{(2 \pi)^3 2 \omega} =\frac{\kappa^2}{ 8 (2 \pi )^3} \sum_\lambda |\tilde T_{\mu \nu}(k) \, \epsilon_\lambda^{* \;\mu \nu} (k)|^2 \,\frac{d^3k}{\omega}\,. 
\end{eqnarray}
Defining the polarization sum $S^{\mu \nu \alpha \beta}$ in the following manner
\begin{eqnarray}
    S^{\mu \nu \alpha \beta} (k) = \sum_\lambda \epsilon_\lambda^{\mu \nu}(k) \epsilon_\lambda^{* \; \alpha \beta} (k) \,,
\end{eqnarray}
we can rewrite Eq.~\eqref{emrate1} in a more compact form
\begin{equation} \label{emrate2}
    d\Sigma = \frac{\kappa^2}{ 8 (2 \pi )^3} \tilde T^{\,*}_{\mu \nu}(k) \tilde T_{\alpha \beta}(k) S^{\mu\nu\alpha\beta}(k) \frac{d^3k}{\omega} \, .
\end{equation}
Being $\omega^2= |\vec k|^2+ m_g^2$ for on-shell physical graviton modes in VSLG, we can replace 
\begin{equation}
    d^3k = |\vec k|^2 d|\vec k| d\Omega_k = \omega^2 \sqrt{1- \frac{m_g^2}{\omega^2} } d\omega d \Omega_k \, ,
\end{equation}
in Eq.~\eqref{emrate2}, obtaining
\begin{equation} 
    d\Sigma = \frac{\kappa^2}{ 8 (2 \pi )^3} \tilde T^{\,*}_{\mu \nu}(k) \tilde T_{\alpha \beta}(k) S^{\mu\nu\alpha\beta}(k) \rho(\omega) \, \omega \, d\omega \, d \Omega_k   \, ,
\end{equation}
where we defined $\rho(\omega) \equiv \sqrt{1- \frac{m_g^2}{\omega^2}}$. To derive the differential rate of the radiated energy, it is now sufficient to multiply the emission probability by the energy $\omega$ of the emitted GW and divide it by the observation time $\mathcal T$. Thus, integrating over all emission angles and energies, we get the total energy loss rate $\dot E \equiv \frac{dE}{dt}$
\begin{eqnarray} \label{dedt1}
    \dot E = \int \frac{\omega}{\mathcal T} \, d\Sigma = \frac{\kappa^2}{ 8 (2 \pi )^3 \mathcal T} \, \int  \rho(\omega) \, \omega^2 \tilde T^{\,*}_{\mu \nu} \tilde  T_{\alpha \beta} S^{\mu\nu\alpha\beta} \, d\omega \, d \Omega_k \,. 
\end{eqnarray}
The explicit expression for $S^{\mu\nu\alpha\beta}$ is included in Appendix \ref{appPolar}. In particular, the relevant formula for the calculations in our case is given in Eq.~\eqref{polsum}. At this point, it is convenient to re-express the $\tilde T^* \tilde T S $ spacetime contraction just as a spatial contraction 
\begin{equation}
    \tilde  T^{\,*}_{\mu\nu} \tilde  T_{\alpha \beta} S^{\mu\nu\alpha\beta} \,\,\to\,\,
     \tilde  T^{\,*\,ij} \, \tilde  T^{\, kl} \Lambda^{ijkl} \, ,
\end{equation}
by exploiting energy-momentum conservation
\begin{equation} \label{timetospatial}
    k_\mu \tilde T^{\mu \nu} = 0 \,\rightarrow \, \left\{\begin{array}{l} 
    \tilde T^{0i} = \frac{k^j}{\omega} \tilde T^{ij} \,,\\
    \tilde T^{00} = \frac{k^i}{\omega} \tilde T^{0i} = \frac{k^i k^j}{\omega^2} \tilde T^{ij}\,.
    \end{array} \right.  
\end{equation}
In this way, Eq.~\eqref{dedt1} transforms to
\begin{equation} \label{dedt2}
    \dot E = \frac{\kappa^2}{ 8 (2 \pi )^3 \mathcal T} \,  \int  \rho(\omega) \, \omega^2d\omega \int \tilde T^{\,*\,ij} \, \tilde T^{\, kl} \Lambda^{ijkl} \,  \, d \Omega_k  \,.
\end{equation}

\subsection{Expression of $\Lambda^{ijkl} (\omega, \hat k ,\hat n)$}

The calculation of $\Lambda^{ijkl} (\omega,\hat k ,\hat n)$ starting from $S^{\mu\nu\alpha\beta}$ is straightforward: as already stated, we must expand the contraction of $\tilde T^* \tilde T S$, separate the time and spatial parts, and finally transform the time indices into spatial ones through \eqref{timetospatial}. Let us give an example of this procedure by realizing it for the term $\sim |\tilde T|^{\,2} = \tilde T^{*} \tilde T \equiv (T^\mu_{\;\mu})^*(T^\nu_{\;\nu})$. We obtain
\begin{eqnarray}
    \tilde T^* \tilde T &=& (\tilde T^{*\, 00} -\tilde T^{*\,ii})(\tilde T^{00}-\tilde T^{jj})=  \tilde T^{*\,00} \tilde T^{00} - \tilde T^{*\,00} \tilde T^{ii} - \tilde T^{00} \tilde T^{*\,ii} + \tilde T^{*\,ii} \tilde T^{jj} \nonumber \\
    &=& (\rho^4 \hat k^i \hat k^j \hat k^k \hat k^l -  \rho^2 \hat k^i \hat k^j \delta^{kl} -\rho^2 \hat k^k \hat k^l \delta^{ij} + \delta^{ij} \delta^{kl}) \tilde T^{*\, i j} \tilde T^{kl} \,,
\end{eqnarray}
where we introduced for convenience the unitary vector $\hat k^i$
\begin{equation}
    \hat k^i \equiv  \frac{k^i}{\omega \rho (\omega)} \,,\,\, \text{ with } \,\,\rho(\omega )\equiv \sqrt{1 -\frac{m^2_g}{\omega^2}} \,,
\end{equation}
so that we can also express the momentum-space VSR operator $\tilde N $ as
\begin{equation}
    \tilde N^\mu = \frac{n^\mu }{n\cdot k} =\frac{n^\mu }{ \omega \, R(\omega,\hat k ,\hat n)} \, ,\,\,\, \text{with } \,\, R(\omega, \hat k,\hat n) \equiv 1- \rho \,\hat n \cdot \hat k  \,.
\end{equation}
Going through all computations, we get the full expression for $\Lambda^{i j k l}(\omega,\hat k ,\hat n)$, which reads
\begin{eqnarray} \label{TTlambda}
    \Lambda^{ijkl} &=& \frac{\rho^4}{2} \hat k^i \hat k^j \hat k^k \hat k^l-  \rho ^2 (\hat k^j \hat k^k \delta^{il} + \hat k^l \hat k^i \delta^{kj}  -  \frac{\hat k^i \hat k^j \delta^{kl} + \hat k^k \hat k^l \delta^{ij}}{2}) +\delta^{ik}\delta^{jl}-\frac12 \delta^{ij}\delta^{kl} + \nonumber\\
    && +\frac{m^2_g}{\omega^2} \frac{1}{R^2}\left [ \rho^4 \hat k^i \hat k^j \hat k^k \hat k^l  - \rho^{3} \hat n^j \hat k^i\hat k^k\hat k^l- \rho^{3} \hat n^l \hat k^k\hat k^i\hat k^j +\frac12 \rho^2 \delta^{ij } \hat k^k \hat k^l +\frac12 \rho^2 \delta^{kl } \hat k^i \hat k^j  \right. \nonumber\\
    && \;\;\;\;\;\;\;\;\;\;\;\;\;\;\;\; +2 \rho^2 \hat n^i \hat n^k \hat k^j\hat k^l - 2 \rho^2 \hat k^i\hat k^k \delta^{jl} - \frac12 \rho^2 \hat n^i \hat n^j \hat k^k \hat k^l - \frac12 \rho^2 \hat n^k \hat n^l \hat k^i \hat k^j \nonumber \\
    &&\;\;\;\;\;\;\;\;\;\;\;\;\;\;\;\; +2 \rho \, \hat n^j \hat k^l \delta^{ik}+2 \rho \, \hat n^l \hat k^j \delta^{ik}  - \rho\, \hat n^i \hat k^j \delta^{ k l}- \rho\, \hat n^k \hat k^l \delta^{ ij}  - 2 \hat n^i \hat n^k \delta^{jl}  \nonumber\\
    && \;\;\;\;\;\;\;\;\;\;\;\;\;\;\;\; \left. + \frac12 \hat n^i \hat n^j \delta^{kl}+ \frac12 \hat n^k \hat n^l \delta^{ij} \right ] \\
    && +\frac{m^4_g}{\omega^4} \frac{1}{R^4}\left [\frac12 \rho^4 \hat k^i \hat k^j \hat k^k \hat k^l - \rho^{3} \hat n^l \hat k^i \hat k^j \hat k^k - \rho^{3} \hat n^j \hat k^k \hat k^l \hat k^i + 2 \rho^2 \hat n^i \hat n^k \hat k^j \hat k^l   \right. \nonumber\\
    &&  \;\;\;\;\;\;\;\;\;\;\;\;\;\;\;\; + \frac12 \rho^2 \hat n^i \hat n^j \hat k^k \hat k^l + \frac12 \rho^2 \hat n^k \hat n^l \hat k^i \hat k^j -  \rho\, \hat n^i \hat n^j \hat n^k \hat k^l -  \rho\, \hat n^k \hat n^l \hat n^i \hat k^j \nonumber\\
    && \;\;\;\;\;\;\;\;\;\;\;\;\;\;\;\; \left. +\frac12 \hat n^i \hat n^j \hat n^k \hat n^l \right] \nonumber \,.
\end{eqnarray}
Note that $\Lambda ^{ijkl}$ is symmetric under the interchange of $i j \Longleftrightarrow k l$. This is because in the contraction $\tilde T^* \tilde T S$ only the symmetrized product $ \frac12(\tilde T^{*\,\mu\nu}\tilde T^{\alpha\beta}+\tilde T^{*\,\alpha\beta}\tilde T^{\mu\nu})$ is relevant due to the symmetries of $S^{\mu\nu\alpha\beta}$. That implies that $\Lambda^{ijkl}$ inherits its same pair-interchange symmetry. Even if not explicit in \eqref{TTlambda}, $\Lambda^{ijkl}$ can also be symmetrized for $i \Longleftrightarrow j$ and $k \Longleftrightarrow l$, since it is contracted with the spatial components of the symmetric EMTs.

\subsection{Energy-Momentum Tensor and Quadrupole Expression}

Due to the linear nature of VSLG and the huge distance from which we observe astrophysical phenomena, the only relevant component that enters in $T^{\mu\nu}$ is the one from the classical source \cite{Poddar:2021yjd}, which in our case describes an inspiraling two-body system. Working in the non-relativistic or low-velocity limit, we can make use of the known EMT formula for a Keplerian binary of total mass $M=m_1+m_2$
\begin{equation} \label{Tdef}
    T^{\mu\nu} (t,\vec x) = \mu \, U^\mu U^\nu \delta^3(\vec x - {\vec r}(t)) \, ,
\end{equation}
where  $U^\mu = (1, \dot r_x , \dot r_y ,0)$ and $ {\vec r}(t)$ are the non-relativistic four-velocity and the trajectory in the $x-y$ orbital plane of the reduced mass $\mu= m_1 m_2 / M $. Defining $b$ and $e$ respectively as the semi-major axis and the eccentricity of the Keplerian orbit, we can parametrize the motion in function of the eccentric anomaly $\phi= \phi(t)$ \cite{celmec} as follows
\begin{eqnarray}
    &&\vec r(t) = b \left ( \cos \phi-e , \sqrt{1-e^2} \sin\phi, 0 \right) \,,\nonumber \\
    && \Omega t = \phi - e \sin\phi \,\,\,\text{with} \,\,\, \Omega \equiv  \frac{2\pi}{P_b}= \sqrt{ \frac{G M}{b^3}}  \,,
\end{eqnarray}
with $\Omega$ representing the fundamental frequency defined by the revolution period $P_b$. At this point let us observe that, since the EMTs in Eq.~\eqref{dedt2} are the momentum space version, we can write them as
\begin{eqnarray}
    \tilde T^{ij} (\omega ,\vec k) = \int dt \int d^3x \, e^{i(\omega t - \vec k \cdot \vec x)} T^{ij}(t,\vec x) = \int d^3 x \,e^{ i \vec k \cdot \vec x} \, \tilde T^{ij}(\omega, \vec x) \,.
\end{eqnarray}
Defining by $d$ the distance between us and the source, we work in the so-called “far zone” or “radiation zone”, which implies the following hierarchy of lengths $ b<< \lambda<< d $, with $\lambda$ being the wavelength of the radiation emitted. Thus, being $\vec k \cdot \vec x \sim  \frac{1}{\lambda} \, b<<1$, we can approximate $e^{i \vec k' \cdot \vec x} \sim 1$ and write
\begin{eqnarray}
   \tilde  T^{ij} (\omega ,\vec k) \simeq \int d^3 x \,\tilde  T^{ij}(\omega, \vec x) \equiv \tilde  T^{ij}(\omega)\,.
\end{eqnarray}
Note that this approximation would also imply an upper limit in the $\omega $-integral. In fact, being $v<<1$ the typical velocity of the system, we have
\begin{equation}
    k << \frac{1}{b} \sim \frac{1}{v}\Omega \,\,\to\,\, \omega << \sqrt{m_g^2 +\frac{1}{v^2}\Omega^2 } \,.
\end{equation} 
Then, considering systems for which $ m_g \lesssim \Omega$, we would have the upper limit
\begin{equation}\label{omegaupper}
    \omega << \frac{1}{v}\Omega \, .
\end{equation}
However, we will not worry about this constraint, because at the end it is already taken care of by the fast convergence of the integrand, as pointed out later in Section \ref{sec:mgbound}.
\newpage
\noindent
Furthermore, let us observe that contracting the conservation equation $\partial_\mu T^{\mu\nu} = 0$ with another $\partial_\nu$, we find $ \partial_i \partial _j T^{ij} (t,\vec x) = \partial_0^2 T^{00} (t,\vec x)$ or equivalently
\begin{equation}
     \partial_i \partial _j \tilde T^{ij} (\omega ,\vec x) = - \omega^2 \, \tilde T^{00} (\omega ,\vec x) \, ,
\end{equation}
that multiplied by $x^k x^l$ and integrated over $d^3x$ becomes, using integration by parts
\begin{eqnarray}\label{Tomega}
   \tilde  T^{kl} (\omega) &=& -\frac{\omega^2}{2} \int dt \, e^{i\omega t} \int d^3 x\, T^{00} (t,\vec x)  x^k x^l \\
    &=& -\frac{\omega^2}{2} \int dt \, e^{i\omega t} Q^{kl}(t) = -\frac{\omega^2}{2} \tilde Q^{kl}(\omega) \,, \nonumber
\end{eqnarray}
from which we see that the first contribution in our expansion comes from the quadrupole moment $Q^{ij}$ of the source $\sim \int d^3x \, T^{00} x^k x^l$, as anticipated in Subsection \ref{quadrupolarnature}. Using the definition \eqref{Tdef} in the expression for $Q^{ij}$, we get
\begin{eqnarray} \label{qtspace}
    Q^{kl} (t) = \mu \int d^3 x \, \delta^3(\vec x -\vec r(t)) x^k x^l = \mu \, r^k (t) \, r^l(t) \,.
\end{eqnarray}
Replacing the expression \eqref{Tomega} for $T^{kl} (\omega)$ back into the energy loss rate and using the definition of $\kappa$, we finally end up with
\begin{equation} \label{dedt3}
    \dot E = \frac{ G}{ 8 \pi ^2 \mathcal T} \, \int d\omega \rho (\omega) \,  \omega^6 \, \tilde Q^{\,*\,ij} \, \tilde Q^{\, kl} \left [\int \, \Lambda^{ijkl} \, d\Omega_k \right ] \,.
\end{equation}

\subsubsection*{System Periodicity and Fourier Analysis}

At this point, we can further simplify our calculations by taking into account the (almost) periodic nature of binaries, which is justified by the smallness of the period decrease rate due to gravitational emission in the initial phase of the binary inspiral (see Table \ref{tab:pulsarsdata}). \\
Because of the periodicity, we can express the quadrupolar moment as a Fourier series
\begin{equation}
    Q^{ij}(t)= \sum_N Q_N^{ij} e^{i \omega_N t} \, , \,\, \omega_N \equiv N\Omega \,.
\end{equation}
Therefore, its Fourier transform will be
\begin{equation}
    \tilde Q^{ij}(\omega) = 2 \pi \sum_N \delta (\omega- \omega_N) \, Q_N^{ij}\, ,
\end{equation}
where we remember the standard definition of the Fourier coefficients
\begin{equation}\label{qfourier}
    Q_N^{ij} = \frac{1}{P_b} \int_0 ^{P_b} dt \, e^{- i \omega_N t} Q^{ij}(t) \, .
\end{equation}
Replacing this expression in the energy loss rate, we can integrate over the frequency $\omega$ to remove the deltas and set $\omega = \omega_N$, since
\begin{equation}
    \int _{m_g}^\infty d\omega \, \delta(\omega-\omega_N)\delta(\omega-\omega_M) \to \delta(\omega_N-\omega_M) = \frac{1}{\Omega} \delta_{NM} \,.
\end{equation}
Using $\delta_{NM}$ to remove one of the sums derived from the two $\tilde Q^{ij}$ and taking the observation time equal to the orbital period $\mathcal T=P_b$, we obtain the average energy loss rate for one complete revolution
\begin{equation} \label{dedt4}
    \dot E = \frac{ G}{ 4 \pi } \sum_{N_{min}} \, \rho (\omega_N) \,  \omega_N^6 \,  Q^{\,*\,ij}_N \, Q^{\, kl}_N \left [\int \, \Lambda^{ijkl} \, d\Omega_k \right ]_{\omega=\omega_N} ,
\end{equation}
with $N_{min} = \lceil \frac{m_g}{\Omega} \rceil$. Here, the brackets $\lceil \, \rceil$ represent the ceiling function. This limitation in the sum's range is due to the fact that, because of the deltas in the $\omega $-integral, the relevant values of $N$ are those for which $\omega_N$ lays in the interior of the integration interval, meaning that they satisfy $\omega_N > m_g$.\\
After all these simplifications, the angular integral $\Pi^{ijkl} (\omega_N,\hat n)$ in \eqref{dedt4} just involves the $ \Lambda^{ijkl}-$structure, now being the single remaining $\hat k-$dependent object 
\begin{equation} \label{angintegralPI}
    \Pi^{ijkl} (\omega_N,\hat n) \equiv \int d\Omega_k \, \Lambda^{ijkl}(\omega_N, \hat k, \hat n) \, .
\end{equation}

\subsubsection*{Non-Zero $Q-$Components}

In this section, we discuss the expressions of the non-zero components of the quadrupolar Fourier coefficients $Q^{ij}_N$ \cite{peters1963gravitational}. Using equations \eqref{qtspace} and \eqref{qfourier}, we have
\begin{equation}
    Q_N^{ij} = \frac{\mu}{P_b} \int_0 ^{P_b} dt \, e^{- i \omega_N t} r^i (t) \, r^j(t) \,.
\end{equation}
Let us observe that since we chose our frame so that the motion is in the $x-y$ plane, the only non-zero components will be $Q_N^{xx}$, $Q_N^{yy}$ and $Q_N^{xy}=Q_N^{yx}$. For example, the $xx-$component is
\begin{eqnarray}
    Q_N^{xx} = \frac{\mu}{P_b} \int_0 ^{P_b} dt \, e^{- i \omega_N t} (r^x (t))^2 = \frac{\mu b^2}{P_b} \int_0 ^{P_b}  dt \, e^{i N \Omega t} (\cos\phi -e)^2 \,.
\end{eqnarray}
The main idea to solve those integrals is to first change the integration variable to $\phi$ and then express everything in terms of Bessel functions $J_N (Ne)$. In the end, all the non-zero components will have the same structure
\begin{equation} \label{QforL}
    Q_N^{ij} =\frac{\mu b^2}{2N} L^{ij}_N \,.
\end{equation}
The explicit expressions of the non-zero $L^{ij}_N$ in functions of the $J_N $ are
\begin{eqnarray} \label{Tcomp}
    && L^{xx}_N= J_{N-2} - 2e J_{N-1} +2e J_{N+1} -J_{N+2}  \,, \\
    && L^{yy}_N= J_{N+2} - 2e J_{N+1}-\frac4N J_N +2e J_{N-1}  -J_{N-2} \, , \nonumber \\
    && L^{xy}_N=-i \sqrt{1-e^2} \left (J_{N-2} - 2 J_{N} +J_{N+2} \right )\nonumber  \,,
\end{eqnarray}
where, we denoted with a prime over $J_N$ its derivative respect to its argument, that here is always the product $N e$. In order to find the expressions in \eqref{Tcomp}, we also made use of the helpful Bessel function identity 
\be 
J_N(t) = \frac{1}{2\pi} \int _0 ^{2\pi} e^{i(N \phi -t \sin\phi )}d\phi \,.
\ee

\subsection{Angular Integral $\int d\Omega_k$}

We are now ready to perform the angular integral $\Pi^{ijkl}$ in \eqref{angintegralPI}. We can break this down as the calculation of the integrals of the following type
\begin{equation} \label{Iijm}
    I^{\{i...j\}} _m \equiv \int d\Omega_k \frac{\hat k^i ... \hat k ^j}{R^m} \,,
\end{equation}
that we encounter when integrating the expression \eqref{TTlambda}. This task can be simplified by exploiting $SIM(2)-$symmetry. Let us give a simple example: we want to calculate the integral $I^{\{i\}}_m$. After the $\hat k-$integration there is only one possible tensorial structure to which the result can be proportional, which is clearly $\hat n^i$. Therefore, we will have
\begin{equation}
        I^{\{i\}}_m = B_m \hat n^i \,.
\end{equation}
Contracting with $\hat n^i$, we get an expression for the coefficient
\begin{equation}
    \hat n^i I^{\{i\}}_m = \int d\Omega_k \frac{\hat n^i \hat k^i }{R^m}= I_m^1 =B_m  \,,
\end{equation}
in which we defined another type of integrals
\begin{equation}\label{Ipm}
    I^p_m= \int d\Omega_k \frac{(\hat n \cdot \hat k)^p}{R^m} = 2 \pi \int_{-1}^1 dx \frac{x^p}{(1 - \rho\, x)^m} \,.
\end{equation}
In the last equation, we used the fact that, since we are integrating over all $\hat k -$directions, the result does not depend on the frame in which we realize the integral. Thus, we chose the frame such that $\hat n \cdot \hat k = \cos \theta = x$. After calculating, with the help of Mathematica, all the needed integrals of the type $I^{\{i...j\}}_n$, we end up with the final explicit formula for the angular integral in Eq.~\eqref{angintegralPI}, which reads
\begin{eqnarray} \label{angInt}
\Pi^{ijkl} &=& \frac{2 \pi}{15} \left(\rho^4+10 \rho^2-15\right) \delta^{ij} \delta^{kl}+\frac{4 \pi}{15} \left(\rho^4-10 \rho^2+15\right) \delta^{ik} \delta^{jl} - 2 \pi (1-\rho^2) \times \nonumber\\
     &&\times \left [{ (  \rho^2+ \frac12-\frac{3 \rho^2+1}{2\rho} \tanh ^{-1}\rho )} \delta^{ij}\delta^{kl} \right.  + ( 2 \rho^2-7+\frac{7-3 \rho^2}{\rho} \tanh^{-1}\rho ) \delta^{ik}\delta^{jl} \nonumber\\
    && \;\;\;\;\; - \frac{1 }{6} \left( 16 \rho^2-3+\frac{9-21 \rho^2}{\rho} \tanh ^{-1}\rho \right) (\hat n^i \hat n^j \delta^{kl} + \hat n^k \hat n^l \delta^{ij}) \nonumber \\
    && \;\;\;\;\; -\frac{2}{3} \left( 16 \rho^2-39 +\frac{33-21 \rho^2}{\rho} \tanh ^{-1}\rho \right) \hat n^i \hat n^k \delta^{jl} \nonumber\\
    && \;\;\;\;\; + \left. \frac{ 1}{6}\left( 80 \rho^2-117+\frac{111-99 \rho^2}{\rho} \tanh ^{-1}\rho \right)  \hat n^i \hat n^j \hat n^k \hat n^l \right ] \,.
\end{eqnarray}
One can easily check that the above expression has the correct limit of linearized GR when sending $\rho \to 1$ or equivalently $m_g\to 0$
\begin{equation}
    \lim_{m_g\to0} \Pi^{ijkl} = -\frac{8 \pi}{15} \delta^{ij} \delta^{kl}+\frac{8 \pi}{5}  \delta^{ik} \delta^{jl}\, .
\end{equation}

\subsection{Radiated Power and Period Decrease Rate}

At this point, the formula for the gravitational radiated power can be written as
\begin{eqnarray} \label{dedt5}
    \dot E_{VSR} &=& \frac{G^4 \mu^2 M^3 }{16 \pi b^5} \sum_{N_{min}}\rho(\omega_N) \, N^4 \, L^{*\,ij} L^{ kl} \Pi^{ijkl}|_{\omega = \omega_N} \nonumber \\
    &=& \frac{32 G^4 m_1^2 m_2^2 M }{5 b^5} \sum_{N_{min}} f(N ,e, \delta , \hat n) \,, 
\end{eqnarray}
where we defined $\delta \equiv m_g /\Omega$ and the function
\begin{equation}\label{fnn0e}
    f(N, e,\delta ,\hat n) \equiv \frac{5 \, \rho }{512 \pi }N^4 \, \, L^{*\,ij} L^{ kl} \Pi^{ijkl} \, .
\end{equation}
From this formula, we can easily derive the orbital period derivative $\dot P$, since we know that in a Keplerian system we must have
\begin{equation}
    \dot P= - 6 \pi \frac{ b^{\frac52 }G^{-\frac32} }{ m_1 m_2 \sqrt{m_1+m_2} } \, \dot E \, .
\end{equation}
After a little manipulation, the rate of period decrease in VSR can be rewritten in an experimentally convenient form
\begin{equation} \label{dpdt}
    \dot P_{VSR} = -\frac{192 \pi \, T_{\odot}^{\frac53}}{5}  \frac{\tilde m_1 \tilde m_2}{\tilde M ^{\frac13}} \left (\frac{P_b}{2\pi} \right )^{-\frac53} \sum_{N_{min}} f(N, e, \delta, \hat n) \,,
\end{equation}
with $T_\odot \equiv G M_\odot / c^3 = 4.925490947 \, \mu s $ and the ‘‘tilde"-masses defined as $\tilde m \equiv m / M_\odot$ , while $M_\odot$ stands for the solar mass \cite{weisberg2016relativistic}. It is easy to see that, when taking the massless limit, Eq.~\eqref{dpdt} reduces to the usual GR formula \cite{peters1963gravitational}
\begin{equation} \label{periodGR}
    \dot P_{GR} = -\frac{192 \pi \, T_{\odot}^{\frac53}}{5}  \frac{\tilde m_1 \tilde m_2}{\tilde M ^{\frac13}} \left (\frac{P_b}{2\pi} \right )^{-\frac53} \frac{1+\frac{73}{24}e^2+\frac{37}{96}e^4}{(1-e^2)^{\frac72}} \,.
\end{equation}
In fact, in this limit the contraction $LL\Pi$ takes exactly the standard GR form, implying 
\begin{equation}
    \lim _{\delta\to 0} f(N,e,\delta,\hat n) = g(N,e) \, ,
\end{equation}
where the expression for $g(N,e)$ is given in Eq.~\eqref{gNe}. Hence, given that the following relation holds \cite{peters1963gravitational} 
\begin{equation}
    \sum_{N=1}^{+\infty} g(N,e) = \frac{1+\frac{73}{24}e^2+\frac{37}{96}e^4}{(1-e^2)^{\frac72}}\,,
\end{equation}
we end up with the correct GR limit for $m_g\to0$.

\subsubsection*{Case $\hat n // \hat z$} \label{nppz}

Let us start by analytically analyzing the simplest scenario, that is, when $\hat n // \hat z$. In this case, all the terms in the $LL \Pi-$contraction proportional to $\hat n$ cancel, since the non-zero $L^{ij}-$components only involve $x$ and $y$. As a result, we are only left with
\begin{eqnarray}
     \frac{L^{*\,ij}_N L^{ kl}_N \, \Pi^{ijkl}}{4 \pi}  &=&  \left (\frac{8}{15} \rho^4 + \frac{1}{12} \rho^2 -\frac{3}{4} + \, \frac{1+2\rho^2-3\rho^4}{4\rho} \tanh^{-1}\rho \right ) L^{*\,ii}_N L^{jj}_N   \\ 
    && + \left (\frac{16}{15} \rho^4 -\frac{31}{6}\rho^2 +\frac{9}{2} \right . \left. - \, \frac{7-10\rho^2+3\rho^4}{2\rho} \tanh^{-1}\rho \right ) L^{*\,ij}_N L^{ij}_N , \nonumber
\end{eqnarray}
while for the contracted $LL-$combinations appearing above we have
\begin{eqnarray}
    L^{*\,ii}_N L^{jj}_N &=& \frac{16}{N^2} J^2_N \, ,\\
    L^{*\,ij}_N L^{ij}_N &=& 2 \left [ (1-e^2)(J_{N-2} -2 J_N +J_{N+2} \,)^2  + \frac{4}{N^2} J^2_N\right. \nonumber \\
    &&  \;\;\;\;\, \left. +(J_{N-2} -2 e J_{N-1} + \frac2N J_N +2 e J_{N+1}-J_{N+2})^2 \, \right ] .\nonumber 
\end{eqnarray}
The argument of all the above Bessel functions is still $N e$. Note that since this contribution to $LL\Pi$ is independent of $\hat n$, it is also present in the more general case where $\hat n$ is oriented differently. The full expression for $f_{//} \equiv f\,|_{\, \hat n// \hat z}$ is given in Eq.~\eqref{f//fperp}. In Appendix \ref{behaviorfpar}, we include a brief description of its properties. 

\subsubsection*{Generic $\hat n$ Case}

Due to the angular dependence, it is much more difficult, for the generic case, to obtain a compact formula. Despite that, we have at least a factorization of the VSR effects generated by the orientation of $\hat n$
\begin{equation}
    f(N,\delta,e, \hat n) = f_{//}(N,\delta,e) + f_{\perp}(N,\delta,e ,\hat n) \,,
\end{equation}
where $f_{//}$ is the contribution previously found and $f_{\perp}$ is the new part. The $\hat n-$dependence can be parametrized in function of two polar angles $\{\theta,\phi\}$  by choosing
\begin{equation}
    \hat n = (\sin\theta \cos\phi, \sin\theta \sin\phi, \cos\theta)\,,
\end{equation}
with $\theta \in (0,\pi) \;,\; \phi \in (0,2\pi)$. Naturally, the case $\hat n // \hat z$ is recovered by taking $\theta = 0$. We include the expression for $f_{\perp}$ in Appendix \ref{genericfcalc} together with a few more comments on its behavior. 

\begin{figure}[h!!]
 \includegraphics[width=14cm]{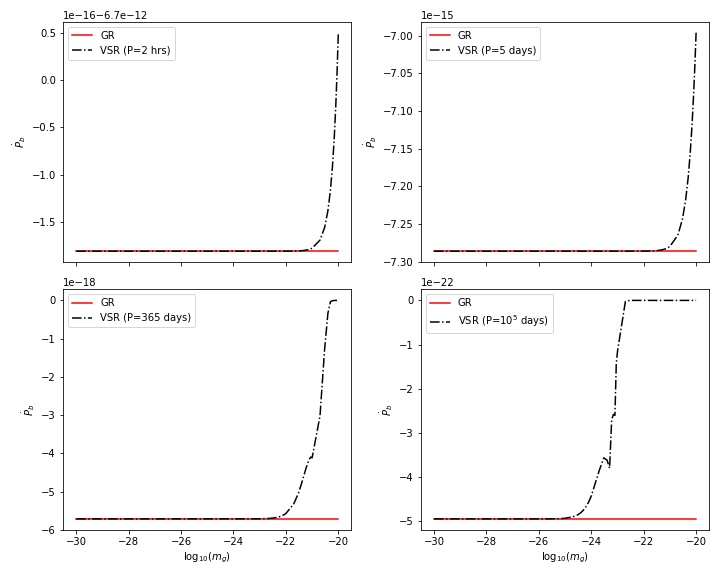}
    \caption[Period decrease rate in GR \textit{vs} VSR for $\hat n // \hat z$]{Period decrease rate in GR and VSR (for $\hat n // \hat z$) in function of $m_g$ and for different values of the orbital period $P_b$. The eccentricity here is set to $e=0.3$. \cite{Santoni:2023uko}}
    \label{fig:P//vsMg}
\end{figure}

\subsection{Graviton Mass Bounds from Data}\label{sec:mgbound}

\begin{figure}[h!!]
    \includegraphics[width=10cm]{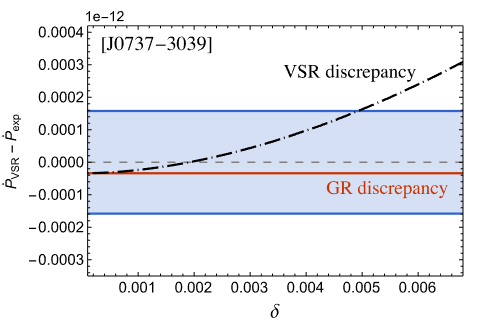}
    \caption[GR and VSR discrepancy in the period time derivative \textit{vs} Experimental range]{Discrepancy in the period decrease rate predicted by GR and VSR calculations with respect to the experimental value measured for the Double Pulsar. The light blue band here represents the experimentally allowed discrepancy region at $95\%$ confidence level. \cite{Santoni:2023uko}}
    \label{fig:massbounds}
\end{figure}

In this section, we want to use the previous findings to experimentally bound the graviton mass parameter in VSR. For this purpose, we will use data from two of the most well-studied binary pulsars: the Hulse-Taylor binary PSR B1913+16, the first binary pulsar ever discovered \cite{hulse1975discovery}, and the Double Pulsar PSR J0737-3039. We include information on these systems in Table \ref{tab:pulsarsdata}.
\begin{table}[ht]
\centering
\begin{tabular}[t]{lcc}
\toprule
{Pulsar} & {B1913+16} \cite{weisberg2016relativistic} & {J0737-3039} \cite{kramer2021strong} \\ \midrule
    {$m_1 (m_\odot)$} & 1.438(1) & 1.33819(1)  \\
    {$m_2 (m_\odot)$}  & 1.390(1)  & 1.24887(1)  \\
    {$P_b (d)$} & 0.322997448918(3) & 0.102251559297(1) \\
    {$e$} & 0.6171340(4)  & 0.08777702(6) \\ \midrule
    {$\dot P_{exp}  (s s^{-1}) \times 10^{12}$}  & {$-2.398$}   & {$-1.247782$}  \\
    {$\sigma_{\dot P}  (s s^{-1}) \times 10^{12}$}  & 0.004  & 0.000079  \\
\bottomrule
\end{tabular} 
\caption[Orbital parameters for binary pulsars B1913+16 and J0737-3039]{Descriptive data for the two binary pulsars selected for this work. The parenthesized numbers represent the $1\sigma-$uncertainty in the last digit quoted. The sources of the data for each pulsar are cited alongside their names.}
\label{tab:pulsarsdata}
\end{table}\\
The fundamental frequency $\Omega$ of both binaries when translated in terms of energies is about $ \sim 10^{-19} eV$. We will then assume to be in the small$-\delta$ regime. This assumption is reasonable since the kinematical bound $m_g \lesssim 10^{-22} eV$ \cite{baker2017strong} obtained from the combined observation of the events GW170817 and GRB170817A should also hold in VSR: In fact, the on-shell graviton's dispersion relation in gauge-fixed VSLG \cite{grav3} takes the usual SR structure independently of the $\hat n- $direction. \\
For the binaries considered, we saw that, when $\delta$ is small, the period decrease rate is always greater (in absolute value) for the $\hat n // \hat z-$scenario. Consequently, we restrict our analysis to that orientation, since it seems to imply for us the greatest VSR discrepancy possible and therefore leads to the strongest bound on graviton mass obtainable taking VSR for granted in this particular context.
\newpage
\noindent
The approach used for the estimation of the $m_g $ upper limit is simple: for small $\delta$, the  $\hat n // \hat z- $contribution increases in absolute value with $\delta$, as also shown in Fig.\ref{fig:F//vsD} in the appendix. Thus, it is sufficient to increase the value of $m_g $ to the point where we saturate the maximal discrepancy allowed by a $95\%$ confidence interval around the experimental rate of period decrease $\dot P_{exp}$ due to GW emission. The value found through this process will represent our constraint. See Fig.\ref{fig:massbounds} for a graphical representation. \\
Let us start from PSR B1913+16. Adhering to the above procedure, we obtain a $\delta $ upper limit of $\sim 0.1102 $, which corresponds to the following bound on the graviton mass
\begin{equation}
    m_g \lesssim 5.2 \times 10^{-20} eV \,.
\end{equation}
For the other studied binary, PSR J0737-3039A, we get
\begin{equation}
    m_g \lesssim 2.3 \times 10^{-21} eV \, ,
\end{equation}
which, while being more stringent than the first one, is still not small enough to be an improvement respect to the kinematical bound obtained from the detection of GW170817.
 \\
However, there are many ways in which these constraints could be improved in the future, apart from simply enhancing the precision of the measurements: One possibility would be to look for binaries with larger orbital size and therefore longer orbital periods. In fact, the longer the period, the larger the $\delta-$ratio at fixed $m_g$, leading to greater VSLG effects.
Furthermore, as already mentioned in subsection \ref{nppz}, small eccentricity values could also be beneficial for these tests since the high-intensity modes in $\sum_N$ would be excluded for smaller values of $m_g$, keeping $P_b$ fixed. Another option would be to focus on high eccentricity binaries due to the enhancement coming from $g(N,e)$ \cite{peters1963gravitational}. Hopefully, thanks to those and further ideas, we will be able to test novel VSR effects or, at least, place better constraints on the VSR origin of a graviton mass.

\chapter[CONCLUSIONS]{Conclusions and Outlook} \label{concl}
In this thesis, we have investigated the repercussions of LV theories in various directions, underscoring the importance of phenomenological approaches in bridging the gap between theoretical predictions and experimental observations, while paving the way for future research in the field. \\
From the theoretical side, we have focused on the framework of Very Special Relativity, which was extensively introduced in Chapter \ref{intro}. Thus, concerning the fermionic sector, we reviewed the VSR modification to the Dirac equation, highlighting some of its most important features and presenting, for the first time, a different perspective on it through the use of field redefinitions. Moreover, in Chapter \ref{ch1} we have developed a way to deal with the Hamiltonian approach, which in the VSR scenario is highly non-trivial due to the non-localities. This formalism was later used in Chapter \ref{chnrlimit} to further explore the non-relativistic limit of the modified Dirac formulation and arrive at the equivalent non-relativistic Schr\"odinger picture. \\
Chapter \ref{ch2} was devoted to the analysis of VSR corrections to the Landau levels of a charged particle immersed in an external constant magnetic field \cite{Koch:2022jcd}. This allowed us to establish a connection between VSR and electron $g-2$ experiments, which was then used to place an upper bound on the electronic VSR parameter of $\sim 1 \,eV$. Importantly, we have also shown how the leading-order correction to the $g-$factor derived in this way and from the Hamiltonian approach of Chapter \ref{chnrlimit} are equivalent.\\
Afterwards, in Chapter \ref{ch3}, we turned our attention to the study of the energy spectrum of Ultracold neutrons in the Earth's gravitational field \cite{PhysRevD.109.064085}. Therefore, we spent some time developing an equivalent of the VSR Hamiltonian formulation for linearized Newtonian potentials or, equivalently, for non-inertial frames. Here, the introduction of curvilinear coordinates further complicates the fundamental picture and the conceptual questions related to the preferred spacetime direction $n^\mu$. Performing a post-Newtonian expansion and keeping only non-relativistic $c^0-$contributions, we found no non-trivial corrections to the fermion energy spectrum apart from the mass redefinition. This suggests that the principle of equivalence between inertial and gravitational mass holds true within the framework of VSR as well. \\
Finally, in Chapter \ref{ch4}, we reviewed the construction of a $SIM(2)-$invariant field theory for a spin-2 field in flat space, possibly representing VSR spacetime perturbations and then gravitational waves \cite{grav3}. While checking for its consistency, we surprisingly found that this formulation accommodates a gauge-invariant graviton mass, which is known to be an important result, especially at cosmological scales. The first application we tackled within this new model was the study of the rate of gravitationally radiated energy in Keplerian binary systems \cite{Santoni:2023uko}. This allowed us to obtain a first upper bound for the VSR graviton mass roughly around $\sim 10^{-21}\,eV$, which is comparable to most of the limits derived from quadrupolar radiation of binary pulsars in the massive gravity literature. 

\section*{Outlook}

Two of the main theoretical results of this manuscript, the formulation of consistent VSR Hamiltonians and of Very Special Linear Gravity, are still in a stage of active research, as both can be further studied and applied in different contexts. In fact, apart from pure conceptual interest regarding possible issues like non-Hermiticity \cite{Bender:2007nj}, the Hamiltonian approach can still be exploited in a wide variety of phenomenological situations. One key example is that of UCN, for which we are already working to further expand our analysis to higher orders. Moreover, it would be appealing to extend the calculations to a full non-inertial picture by also including the possibility of frame rotation, which is clearly relevant on Earth's surface.
Another promising possibility are high-precision spectroscopic experiments with hydrogenic atoms, which have already been widely used to test Lorentz and $CPT$ violations \cite{PhysRevD.92.056002}. Furthermore, a complete analysis of the VSR corrections to the $g-2$ factor of the muon in the context of modified BMT equations is still ongoing. \\
On the other side, our VSR model for linearized gravity strongly calls for a full non-linear extension, which would not only allow to soften problems, such as quantum gravity non-renormalizability, but also be of interest in the cosmological setting where it could serve as an alternative explanation for the Universe's accelerated expansion \cite{DeFelice:2013bxa}. From the phenomenological side, the absence of new polarization modes with respect to GR, makes the study of waveform modifications in VSR intriguing. Moreover, the upcoming gravitational interferometers will be sensitive to a broader spectrum of frequencies \cite{amaro2017laser}, providing greater potential to constrain VSR effects.\\
In summary, numerous unsolved enigmas persist in modern physics, still awaiting solutions. In such a scenario, Lorentz-violation might emerge not only from foundational aspects but also as an effective description of some yet undiscovered puzzle piece in our theories of Nature. Therefore, further experimental tests of Lorentz invariance are needed to continue our quest for new physics. VSR, for his part, has a very rich phenomenology while being, in some sense, the most conservative LV framework. Our investigation has demonstrated that VSR can yield measurable effects in various new contexts, uncovering its potential to explain phenomena beyond the established physical frameworks. Hence, we consider VSR to be worthy of additional investigations and experimental searches for its unique signatures.

\newpage
\pdfbookmark{\contentsname}{toc}



\cleardoublepage


\cleardoublepage
\phantomsection \label{references}
\renewcommand{\bibname}{REFERENCES}


\bibliographystyle{utphys}
\bibliography{./biblio/intro,./biblio/chap2,./biblio/chap3,./biblio/chap4,./biblio/chap5,./biblio/Bmassgrav,./biblio/Bbinary,./biblio/Bgfactor,./biblio/BneutronPN}


\appendix 

\newpage
\section[MISCELLANEOUS OF RELEVANT CALCULATIONS]{Miscellaneous of Relevant Calculations} \label{app1}
Here, we include more details on a set of miscellaneous calculations relevant to this work, especially for Chapter \ref{ch1}.

\subsection{Gauge Invariance of the VSR Lagrangian}\label{gaugeinvarianceVSREM}

To prove the gauge invariance of the VSR Lagrangian we just have to test the transformation properties of the new VSR contribution. In particular, we want to demonstrate that, under local gauge transformations $\psi' = U \psi = e^{i\phi(x)} \psi$, we have
\begin{equation}
    \frac{1}{n\cdot D\,' } \, \psi ' = U \frac{1}{n\cdot D } \psi = e^{i\phi(x)} \frac{1}{n\cdot D } \psi \,. 
\end{equation}
Let us start by representing the non-local operator in the integral representation
\begin{equation}\label{a2transf}
   \frac{1}{n\cdot D\,' } \, \psi ' = \int d\alpha \, e^{-\alpha n \cdot D\,'} \psi' = \int d\alpha \, (1 - \alpha n\cdot D\,' +\frac{\alpha^2}{2} (n\cdot D\,')^2 - ... \,)\, \psi'\,.
\end{equation}
Taking $\delta A^\mu = - \frac{1}{e}\partial^\mu \phi(x)$ as in standard LI theory, we already know that
\begin{equation}
    D\,'_\mu \psi' = e^{i\phi(x)} D_\mu \psi \,\, \to \,\, n\cdot D\,' \psi' = e^{i\phi(x)} n \cdot D \psi \,,
\end{equation}
from the above equation, applying another $n\cdot D'$ on the left we get
\begin{eqnarray}
    ( n\cdot D\,')^2 \psi' &=& n \cdot D\,' (e^{i\phi(x)} n \cdot D \psi) = (n\cdot D - i n \cdot \partial \phi )(e^{i\phi(x)} n \cdot D \psi)  \\
    &=& e^{i\phi(x)} ( n\cdot D + i n\cdot \partial \phi -i n\cdot \partial \phi  ) n \cdot D \psi = e^{i\phi(x)}(n \cdot D )^2 \psi\nonumber \,.
\end{eqnarray}
We could keep applying additional $n \cdot D'$ to end up with
\begin{equation}
    ( n\cdot D\,')^m \psi' =  e^{i\phi(x)}(n \cdot D )^m \psi \,.
\end{equation}
Using this relation into \eqref{a2transf}, we obtain
\begin{equation} 
   \frac{1}{n\cdot D\,' } \, \psi ' = \int d\alpha \, e^{i\phi(x)}(1 - \alpha n\cdot D +\frac{\alpha^2}{2} (n\cdot D)^2 - ... \,)\, \psi = e^{i\phi(x)} \frac{1}{n\cdot D} \psi\,,
\end{equation}
as we wanted to demonstrate.

\subsection{Calculation of $[D_\nu , \frac{1}{n\cdot D}] \psi $} \label{DnDcommutator}

In this section, we include a sketch of the calculation for the commutator in \eqref{commutatorDnDv1}
\begin{equation} \label{commutatorDnDv2}
   [D_\nu , \frac{1}{n\cdot D}] \psi = \sum_{n=0} \frac{(-1)^n}{n!} \int d\alpha \,\alpha^n [D_\nu,(n\cdot D )^n] \psi \,.
\end{equation}
The strategy is to calculate the commutators involved for the first $n-$values. To do that, we should take into account the following relation for the commutator of covariant derivatives
\begin{equation}
    [D_\nu,D_\mu] D_{\alpha_1} ... D_{\alpha_m} \psi = i e F_{\nu\mu}  \,D_{\alpha_1} ... D_{\alpha_m} \psi \,,
\end{equation}
and note that, defining $nF_\nu \equiv n^\mu  F_{\nu\mu}$, we have
\begin{equation}
    n \cdot D \, n F_{\nu} = n \cdot \partial \, n F_{\nu} \,\, \to \,\, n \cdot D \, (n F_{\nu} \psi) = (n \cdot \partial \, n F_{\nu}) \psi + n F_\nu \, n \cdot D \psi \, ,
\end{equation}
reflecting the fact that covariant derivatives satisfy the Leibniz rule just like ordinary partial derivatives. Bearing that in mind, we find the commutators for the first $n-$values to be
\begin{eqnarray}
    \, [D_\nu , n \cdot D] \psi &=& i e \, n F_{\nu} \psi \,, \\
    \, [D_\nu , (n \cdot D)^2] \psi &=& n \cdot D [D_\nu , n \cdot D] \psi + [D_\nu , n \cdot D] n \cdot D \psi \nonumber\\
    &=& 2 i e \, n F_{\nu } \, n \cdot D \psi + i e \, ( n \cdot \partial \,n F_{\nu} ) \psi \nonumber \,, \\
    \, [D_\nu , (n \cdot D)^3] \psi &=& i e (n \cdot D)^2 \, n F_{\nu } \psi + i e \, n \cdot D \, n F_{\nu } \, n \cdot D \psi + i e \, n F_{\nu } (n \cdot D )^2 \psi  \nonumber \\
    &=& 3 i e \, n F_{\nu } \, (n \cdot D )^2 \psi + 3 i e \, (n \cdot \partial \, n F_{\nu} ) n\cdot D \psi + i e \, ((n\cdot \partial)^2 \, n F_{\nu} ) \psi  \nonumber \,.
\end{eqnarray}
We can now group the terms according to the order of the derivative applied on $nF_\nu $. For example, taking into account the sign and numerical factors coming from the sum in \eqref{commutatorDnDv2}, the contributions with respectively zero, one and two derivatives on $n F_\nu$ are
\begin{eqnarray} \label{eqseriesnF}
    && i e \, nF _\nu  (-\alpha ) ( 1 - \alpha \, n \cdot D +\frac{\alpha^2}{2} (n\cdot D)^2 -... \, )\psi = i e \, nF _\nu  (-\alpha ) e^{-\alpha n \cdot D } \psi \,,\nonumber \\
    && \frac{i e}{2} (n \cdot \partial \, nF_\nu ) (\alpha^2) (1 - \alpha n\cdot D+...\,  ) \psi = \frac{i e  }{2} (n \cdot \partial \, nF_\nu ) (\alpha^2) e^{-\alpha n \cdot D } \psi \,,\nonumber\\
    && \frac{i e }{6} ( (n \cdot \partial)^2 nF_\nu) (-\alpha^3) (1 - ...\, ) \psi  = \frac{i e }{6} ( (n \cdot \partial)^2 nF_\nu) (-\alpha^3) e^{-\alpha n \cdot D } \psi \,. 
\end{eqnarray}
The $\alpha-$integral of these terms can be realized using the Feynman trick of differentiating under the integral sign. Consider, for example, the calculation for the first line of \eqref{eqseriesnF}
\begin{eqnarray}
    i e \, nF _\nu \int d\alpha  (-\alpha ) e^{-\alpha n \cdot D } \psi = i e \, nF _\nu \frac{d}{dA} \int d\alpha \, e^{-\alpha A \, n \cdot D } \bigg |_{A=1}  \psi  \,.
\end{eqnarray}
Exploiting the inverse of the integral representation \eqref{intrepr}, we get
\begin{equation}
    i e \, nF _\nu \frac{d}{dA} \left (\frac{1}{ A \, n \cdot D } \right ) \bigg |_{A=1}  \psi = -i e \, nF _\nu \, \frac{1}{( A \, n \cdot D )^2}  \bigg |_{A=1}  \psi = - i e \, nF _\nu \, \frac{1}{( n \cdot D )^2}  \psi \, .
\end{equation}
For the second line of \eqref{eqseriesnF}, instead, we obtain
\begin{equation}
    \frac{i e  }{2} (n \cdot \partial \, nF_\nu ) \frac{d^2}{dA^2} \left (\frac{1}{ A \, n \cdot D } \right ) \bigg |_{A=1} \psi =  i e (n \cdot \partial \, nF_\nu )  \frac{1 }{ ( n \cdot D)^3 } \psi \,.
\end{equation}
Gathering all the contributions together in the expression for the commutator \eqref{commutatorDnDv2}, we finally end up with
\begin{eqnarray}
    [D_\nu , \frac{1}{n\cdot D}] \psi = \left ( - i e \, nF _\nu \, \frac{1}{( n \cdot D )^2}  \psi + i e (n \cdot \partial \, nF_\nu )  \frac{1 }{ ( n \cdot D)^3 } - ...\, \right ) \psi \,,
\end{eqnarray}
which can be condensed as in \eqref{commutator1nDwithD}.

\subsection{Non-relativistic Limit of $\frac{1}{n\cdot D}$} \label{app1divnDlimit}

Our aim for this section is to rewrite the non-local $\frac{1}{n\cdot D}-$term as some kind of expansion in the $m-$parameter. Using Eq.~\eqref{almostfinalschem2}, we have
\begin{equation} \label{a7nDpsi}
    n \cdot D \psi = ( D_0 + n^ i D_i ) \psi = ( -i m \gamma^0 + \mathcal K \, ) \psi \,,
\end{equation}
where we collected all terms $O(m^0)$ in the $\mathcal K- $operator. Now, we multiply \eqref{a7nDpsi} from the left by $\frac{1}{m} \frac{1}{n\cdot D}$, obtaining
\begin{equation}
    \frac{1}{m} \psi \simeq \frac{1}{m} \frac{1}{n\cdot D }( -i m\gamma^0 + \mathcal K \, ) \psi = -i \gamma^0 \frac{1}{n\cdot D} \psi + \frac{1}{m} \frac{1}{n\cdot D} \mathcal K \,\psi \,.
\end{equation}
Re-ordering things and multiplying by $i\gamma^0$, we can write
\begin{equation}
    \frac{1}{n\cdot D} \psi = \frac{i}{m} \gamma^0 \psi - \frac{i}{m} \frac{1}{n\cdot D} \mathcal K \,\psi \,.
\end{equation}
At this point, we can use the fact that\footnote{Actually not all terms in $\mathcal K$ depends on $x$ but that does not matter here.}
\begin{equation}
    \frac{1}{n\cdot D}(\mathcal K \psi) = \sum_{n=0}^\infty ((-n \cdot \partial)^n \mathcal K) \frac{1}{(n\cdot D)^{n+1}}\psi \,,
\end{equation}
to rewrite the above equation in a recursive way, implying
\begin{equation}
    \frac{1}{n\cdot D} \psi = \frac{i}{m} \gamma^0 \psi - \frac{i}{m}  \sum_{n=0}^\infty ((-n \cdot \partial)^n \mathcal K) \frac{1}{(n\cdot D)^{n+1}}\psi = \frac{i}{m}\gamma^0 \psi +O(\frac{1}{m^2}) \,.
\end{equation}
Applying another $\frac{1}{n\cdot D}$ on the left, we finally get to the relevant operator we wanted to expand at leading order
\begin{equation}
    \frac{1}{(n\cdot D)^2} \psi = - \frac{1}{m^2} \psi + O(\frac{1}{m^3})\,,
\end{equation}
leading to the approximation in the main text
\begin{equation}
     - i e \frac{\lambda \Box_D}{m}  \gamma^0 \sigma^{\mu\nu} n_\mu n^\rho \sum_{n =0} ((-n\cdot \partial)^n F_{\nu\rho}) \frac{1}{(n\cdot D)^{n+2}} \,\sim \, \frac{i e \lambda}{m^3} \gamma^0 \sigma^{\mu\nu} n_\mu n^\rho F_{\nu\rho} \,.
\end{equation}

\subsection{Transformation Properties of $\chi$} \label{apptransfproperties}

\begin{figure}[h!!]
    \includegraphics[width=8cm]{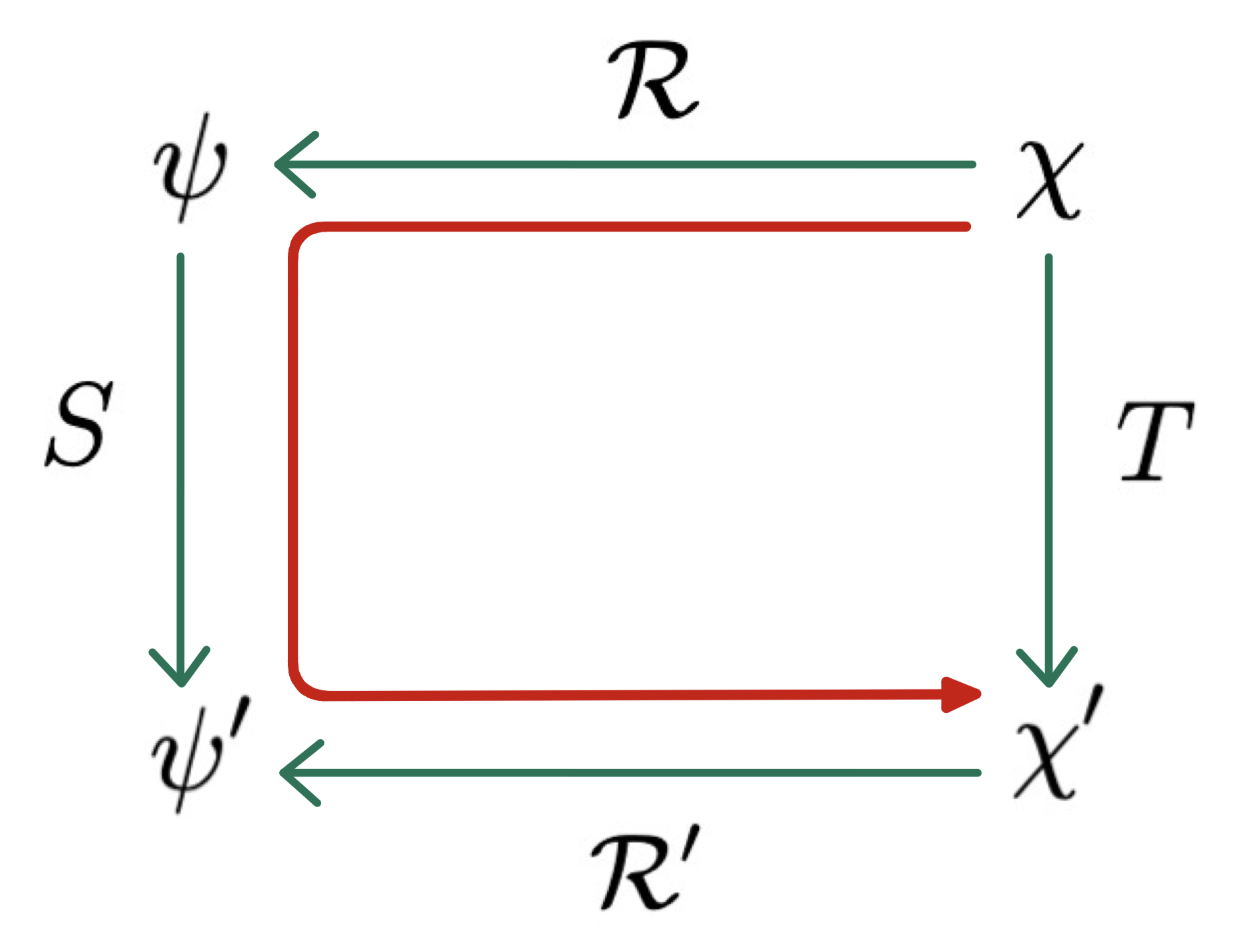}
    \caption[Schematic representations of field redefinitions and transformations]{Schematic representation of the transformations relating the fields $\psi$ and $\chi$ and the Lorentz-transformed $\psi'$ and $\chi'$.}
    \label{fig:chitransformation}
\end{figure}

We want to find the transformation properties of the spinor $\chi$, introduced in Section \ref{sec:fieldredefchi}, under Observer Lorentz transformations. For the VSR Lagrangian to be invariant under OLT we need to assume that $\psi$ transforms as a standard Dirac spinor $\psi' = S(\Lambda) \psi$ with $S$ such that
\begin{equation}
    S(\Lambda) \gamma^\nu S^{-1}(\Lambda) = \gamma^\mu (\Lambda^{-1})^\nu_{\;\mu}\,.
\end{equation}
Thus, if we define as $T(\Lambda)$ the transformation of the new spinor $\chi$ under OLT
\begin{equation}
    \chi' = T(\Lambda) \chi\,,
\end{equation}
we see from Fig.\ref{fig:chitransformation} that we can express $T$ as $T(\Lambda) = (\mathcal R')^{-1} S(\Lambda) \, \mathcal R$, where
\begin{equation}
    \mathcal R' = 1 + i\delta \gamma^\mu N'_\mu = 1+i\delta \gamma^\mu (\Lambda^{-1})^\nu_{\;\mu} N_\nu = 1+i\delta S \slashed N S^{-1} = S( 1+i\delta \slashed N)S^{-1}\,.
\end{equation}
Taking into consideration that $(\mathcal R')^{-1} =S ( 1+i\delta \slashed N)^{-1} S^{-1}$, we have
\begin{equation}
    T(\Lambda) = (\mathcal R')^{-1} S(\Lambda) \, \mathcal R = S ( 1+i\delta \slashed N)^{-1} S^{-1} S \, ( 1+i\delta \slashed N) = S(\Lambda) \,,
\end{equation}
implying that $\chi$ is effectively a vanilla Dirac spinor satisfying both the standard Dirac equation and its transformation properties.

\newpage
\section[USEFUL RELATIONS FOR THE $\mathsf g-$FACTOR CORRECTIONS]{Useful Relations for the $\mathsf g-$Factor Corrections} \label{app2}
In this appendix, we pack some important calculations and useful expressions regarding the computation of the VSR corrections to the $g-$factor developed in Chapter \ref{ch2}.

\subsection{Calculation of the Integrals $I_1( \bar n,k)$}

In the main text, we defined the integrals
\begin{eqnarray}
I_1 (\bar n,k) = \int_{-\infty}^{\infty}d\xi\,e^{-\xi^2/2}H_{\bar n}(\xi)\partial_{\xi}^k\left( e^{-\xi^2/2} H_{\bar n}(\xi)  \right),
\label{eq_In}
\end{eqnarray}
where $H_{\bar n}(z)$ are the Hermite polynomials, defined by
\begin{eqnarray}
H_{\bar n}(z) = (-1)^{\bar n}
e^{z^2}\partial_{z}^{\bar n}\left( e^{-z^2} \right) \,.
\end{eqnarray}
In particular, we remark that $H_{0}(z) = 1$, and hence for the particular case $\bar n=0$, 
Eq.~\eqref{eq_In} reduces to
\begin{eqnarray}
I_1(0,k) = \int_{-\infty}^{\infty}d\xi\,e^{-\xi^2/2}\partial_{\xi}^{k}\left( e^{-\xi^2/2} \right) = 2^{-k/2}(-1)^k
\int_{-\infty}^{\infty}d\xi\,e^{-\xi^2}\,H_{k}(\xi/\sqrt{2}) \,.
\end{eqnarray}
From the standard result (see for instance Gradshteyn-Rhizik \cite{grad}, p. 837)
\begin{eqnarray}
\int_{-\infty}^{\infty}e^{-(x-y)^2}H_{k}(\alpha x)dx  = \sqrt{\pi}(1 - \alpha^2)^{k/2}H_{k}\left(\frac{\alpha y}{\sqrt{1 - \alpha^2}}  \right),\nonumber
\end{eqnarray}
we obtain (setting $\alpha = 1/\sqrt{2}$ and $y=0$)
\begin{eqnarray}
I_1(0,k)= \sqrt{\pi}(-1)^k
2^{-k}H_{k}(0) \,.
\end{eqnarray}
Given that $H_{2n+1}(0)= 0$, while $H_{2n}(0) = (-1)^n \Gamma(2n+1)/\Gamma(n+1)$, we obtain the following 
\begin{eqnarray}
&&I_1(0,2k) = \sqrt{\pi} \,2^{-2k}(-1)^k\frac{\Gamma(2k+1)}{\Gamma(k+1)},\\
&&I_1(0,2k+1) = 0 \,\nonumber .
\end{eqnarray}
Let us now focus on the case $\bar n>0$. For this purpose, we notice that the functions
\begin{eqnarray}
\langle\xi|\varphi_{\bar n}\rangle = \varphi_{\bar n}(\xi) = \frac{1}{\sqrt{2^{\bar n} \bar n! \sqrt{\pi}}}e^{-\xi^2/2}H_{\bar n}(\xi) \,,
\end{eqnarray}
constitute a complete orthonormal set 
\begin{eqnarray}
&& \mathds{1} = \sum_{\bar n=0}^{\infty}|\varphi_{\bar n}\rangle\langle\varphi_{\bar n}| \,, \\
&&\langle\varphi_{\bar n}|\varphi_{\bar n'}\rangle = \int_{-\infty}^{\infty}d\xi \langle\varphi_{\bar n}|\xi\rangle\langle\xi|\varphi_{\bar n'}\rangle = \delta_{\bar n,\bar n'}\nonumber \,.
\end{eqnarray}
With these definitions, along with the momentum operator $ {p}^{\xi}= -i\partial_{\xi}$, we have the rather simple expression
\begin{eqnarray}
I_1(\bar n,k) = 2^{\bar n} \bar n! \sqrt{\pi} (i)^k \langle \varphi_{\bar n}|( {p}^{\xi})^k|\varphi_{\bar n} \rangle \,.
\label{eq_In1}
\end{eqnarray}
Since the momentum (Fourier) eigenbasis is also complete, we have
\begin{eqnarray}
\int_{-\infty}^{\infty}dp|p\rangle\langle p| = \mathds{1} \, ,\,\,\,\langle\xi | p\rangle = \frac{1}{\sqrt{2\pi}}e^{i\xi p}.
\end{eqnarray}
Therefore, we can insert the identity in the momentum basis in Eq.~\eqref{eq_In1} to obtain
\begin{eqnarray}
I_1(\bar n,k) &=& 2^{\bar n} \bar n! \sqrt{\pi} (i)^k \int_{-\infty}^{\infty}dp\langle\varphi_{\bar n}|( {p}^{\xi})^k|p\rangle\langle p|\varphi_{\bar n} \rangle\nonumber\\
&=& 2^{\bar n} \bar n! \sqrt{\pi} (i)^k \int_{-\infty}^{\infty}dp\,p^{k}\left|\langle p|\varphi_{\bar n} \rangle\right|^2 .
\label{eq_In2}
\end{eqnarray}
Finally, notice that the functions $\langle p|\varphi_{\bar n}\rangle$ are related to the functions $\langle\xi|\varphi\rangle$ via Fourier representation, since
\begin{eqnarray}
\langle p|\varphi_{\bar n}\rangle &=& \int_{-\infty}^{\infty}d\xi \langle p|\xi\rangle\langle\xi|\varphi_{\bar n}\rangle = \frac{1}{\sqrt{2\pi}}\int_{-\infty}^{\infty} d\xi \,e^{-i \xi p}\varphi_{\bar n}(\xi) \\
&=& \frac{1}{\sqrt{2^{\bar n} \bar n! \sqrt{\pi}}} \frac{1}{\sqrt{2\pi}}\int_{-\infty}^{\infty}d\xi \,e^{-i \xi p}e^{-\xi^2/2}H_{\bar n} (\xi) = \frac{(-i)^{\bar n}}{\sqrt{2^{\bar n} \bar n! \sqrt{\pi}}}e^{-p^2/2}H_{\bar n} (p) \,,\nonumber
\end{eqnarray}
Substituting this last result into Eq.~\eqref{eq_In2}, we have
\begin{eqnarray}
I_1(\bar n,2k) = (-1)^k \int_{-\infty}^{\infty}dp\, p^{2k} e^{-p^2} \left (H_{\bar n} (p)\right )^2,
\label{eq_In3}
\end{eqnarray}
with $I_1(n,2k+1) = 0$ trivially by parity. Using the Hermite-Fourier series representation for the power $p^{2k}$, we have
\begin{eqnarray}
p^{2k} = \frac{(2k)!}{2^{2k}}\sum_{\ell = 0}^{k}\frac{H_{2\ell}(p)}{(2\ell)!(k-\ell)!} \,, 
\end{eqnarray}
and inserting this expression into Eq.~\eqref{eq_In3} we obtain
\begin{eqnarray}
I_1(\bar n,2k) = (-1)^k\frac{(2k)!}{2^{2k}}
\sum_{\ell=0}^k \frac{1}{(2\ell)!(k-\ell)!} \int_{-\infty}^{\infty}dp\,e^{-p^2} \left(H_{\bar n} (p)\right)^2H_{2\ell}(p) \,. \nonumber
\label{eq_In4}
\end{eqnarray}
We can evaluate this last integral by applying the identity (see  \cite{grad}, p.797)
\begin{eqnarray} \label{rhizikHHH}
\int_{-\infty}^{\infty}dz\,e^{-z^2} H_{k}(z)H_{m}(z)H_{\bar n}(z) = \frac{2^{s}\sqrt{\pi} \, k! \, m! \, \bar n!}{(s-k)!(s-m)!(s-\bar n)!} \,,
\end{eqnarray}
for $s = (\bar n + m + k)/2$. Applying this result to Eq.~\eqref{eq_In4}, we obtain
\begin{eqnarray}
I_1(\bar n,2k) &=& \frac{\sqrt{\pi}(-1)^k (2k)! \, \bar n! \, 2^{\bar n-2k}}{k!}\sum_{\ell=0}^{k}2^{\ell}\left(\begin{array}{c} \bar n \\\ell \end{array}\right)\left(\begin{array}{c}k\\\ell \end{array}\right)\nonumber\\
&=&  \sqrt{\pi} (-1)^k \bar n! \, 2^{\bar n-2k}\frac{\Gamma(2k+1)}{\Gamma(k+1)} F(-k,-\bar n;1;2) \,,
\end{eqnarray}
where $F(a,b;c;z)$ is the Hypergeometric function.

\subsection{Calculation of the Integrals $I_2(\bar n,k)$}

Furthermore, we want to calculate the integrals of the form
\begin{equation}\label{i2def}
    I_2 (\bar n,k) =\int_{-\infty}^{\infty} d\xi \, e^{-\xi^2/2} H_{\bar n}(\xi) \partial^k_\xi \left ( e^{-\xi^2/2} H_{\bar n-1}(\xi) \right ) .
\end{equation}
We can undergo all the previous steps done for the integrals $I_1$ to find
\begin{eqnarray}
    I_2 (\bar n, k) &=& 2^{\bar n-\frac12}  (\bar n-1)!  \sqrt{\pi n} (i)^{k} \langle \varphi_{\bar n}|( {p}^{\xi})^k|\varphi_{\bar n-1} \rangle = \nonumber \\
    &=& i^{k+1} \int_{-\infty}^{\infty} dp \, p^k e^{-p^2} H_{\bar n}(p) H_{\bar n -1}(p)
    \,,
\end{eqnarray}
which, due to the parity of the integrand is easily seen to be 0 for even values of $k$: $I_2 (\bar n,2k) = 0$. Remembering that for odd powers of momentum $p$
\begin{equation}
    p^{2k+1} = \frac{(2k+1)!}{2^{2k+1}} \sum_{\ell = 0}^{k}\frac{H_{2\ell+1}(p)}{(2\ell+1)!(k-\ell)!}, 
\end{equation}
we find
\begin{eqnarray}
    I_2 (\bar n, 2k+1) &=& (-1)^{k+1} \frac{(2k+1)!}{2^{2k+1}} \times \\
     && \;\;\; \times \sum_{\ell=0}^k \frac{1}{(2\ell+1)!(k-\ell)!} \int_{-\infty}^{\infty} dp \, e^{-p^2} H_{2l+1} H_{\bar n} H_{\bar n -1}\,. \nonumber 
\end{eqnarray}
Using again the expression \eqref{rhizikHHH} for the integral in $I_2$, we finally obtain
\begin{eqnarray}\label{i2expr}
    I_2 (\bar n, 2k+1) &=& \sqrt{\pi} (-1)^{k+1}  \frac{ (2k+1)!}{k!} (\bar n -1) ! \, 2^{\bar n -2k-1} \sum_{\ell=0}^k 2^{\ell}\left(\begin{array}{c}  k  \\\ell \end{array}\right)\left(\begin{array}{c}\bar n \\\ell +1\end{array}\right) \nonumber \\
    &=&\sqrt{\pi} (-1)^{k+1}  (\bar n -1) ! \, 2^{\bar n -2k-1} \frac{ \Gamma(2k+2)}{\Gamma(k+1)}  \times \\
    && \;\;\;\;\;\;\;\;\;\;\;\;\;\;\;\;\;\;\;\;\;\;\;\;\;\;\;\;\;\;\times \frac{F(-k-1;-n;1;2)-F(-k;-n;1;2)}{2} \nonumber\, .
\end{eqnarray}
In principle, it may also be useful to calculate another set of integrals, namely
\begin{equation}\label{i3def}
    I_3 (\bar n,k) \equiv \int_{-\infty}^{\infty} d\xi \, e^{-\xi^2/2} H_{\bar n-1}(\xi) \partial^k_\xi \left ( e^{-\xi^2/2} H_{\bar n}(\xi) \right ) .
\end{equation}
However, integrating by parts $k-$times, we trivially have that 
\begin{equation} \label{i3toi2}
    I_3(\bar n ,k )= (-1 )^k I_2(\bar n,k) \,,
\end{equation}
so that, at the end of the day, only the first two integrals are relevant.


\subsection{Calculation of Matrix Elements $V^{\bar n}_{\alpha,\alpha'}$}\label{app3}

In this section, we include some of the important steps for the calculation of the matrix elements shown in \eqref{matrixeldef}. Let us start from 
\begin{eqnarray}
    V^{\bar n}_{-1,-1} &=& \bra{\vec f^{(0)} ,\bar n ,-1}{V}\ket{ \vec f^{(0)} , \bar n ,-1}  .
\end{eqnarray}
Inserting two identities through the completeness relation $\mathds{1}= \int d\xi \ket{\xi}\bra{\xi}$ and remembering the definition \eqref{Pxidef} of the operator $P_\xi$, we get
\begin{eqnarray}\label{a1.1}
V^{\bar n}_{-1,-1}  &= &  \frac{\sin^2\theta}{2^{\bar n}\bar  n! \sqrt{\pi} } \int_{-\infty}^{\infty} d\xi \; e^{-\xi^2/2} H_{\bar n}(\xi) \frac{1}{ P^2_\xi} (e^{-\xi^2/2} H_{\bar n}(\xi))  \\
& =& - \frac{\sin^2\theta}{2^{\bar n} \bar n! \sqrt{\pi} \tilde E^2 } \frac{d}{d A }\int_{-\infty}^{\infty} d\xi \; e^{-\xi^2/2} H_{\bar n}(\xi)  \frac{1}{A+i \eta\sin\theta \partial_\xi} (e^{-\xi^2/2} H_{\bar n}(\xi)) \bigg |_{A=1} \, . \nonumber
\end{eqnarray}
By representing the inverse operator in \eqref{a1.1} with the integral form, we obtain
\begin{eqnarray}\label{a1.2}
V^{\bar n}_{-1,-1}  &=& - \frac{\sin^2\theta}{ 2^{\bar n} \bar n! \sqrt{\pi} \tilde E^2 } \frac{d}{d A}\int_{0}^\infty dt \int_{-\infty}^{\infty} d\xi \; e^{-\xi^2/2} H_{\bar n} \, e^{- A t (1+i \eta\sin\theta \partial_\xi)} (e^{-\xi^2/2} H_n) \bigg |_{A=1}  \nonumber\\
&=& - \frac{ \sin^2\theta}{2^{\bar n} \bar n! \sqrt{\pi} \tilde E^2 } \frac{d}{dA}\int_{0}^\infty e^{-At} dt \int_{-\infty}^{\infty} d\xi \; e^{-\xi^2/2} H_{\bar n}  e^{-i t \eta\sin\theta  \partial_\xi} (e^{-\xi^2/2} H_{\bar n} )\bigg |_{A=1} \,,\nonumber\\
\end{eqnarray}
and expanding the exponential operator
\begin{eqnarray}\label{a1.3}
V^{\bar n}_{-1,-1} &=& - \frac{\sin^2\theta}{2^{\bar n} \bar n! \sqrt{\pi} \tilde E^2 }  \sum_{k=0}^\infty \frac{(i \eta\sin\theta )^k}{k!} \left ( \frac{d}{dA} \int_{0}^\infty e^{-At} (-t)^k dt \right )\bigg |_{A=1} \; I_1(\bar n,k) \nonumber\\
&=&  \frac{ \sin^2\theta}{2^{\bar n} \bar n! \sqrt{\pi} \tilde E^2 }  \sum_{k=0}^\infty (k+1) (-i \eta\sin\theta )^k  I_1(\bar n,k) \, ,
\end{eqnarray}
where we introduced the integral $I_1(\bar n,k)$ defined in \eqref{I1nk}, which vanish for odd $k-$values. Hence, we can rewrite the equation \eqref{a1.3} as
\begin{equation} \label{a1sumk}
    V^{\bar n}_{-1,-1}  = \frac{ \sin^2\theta}{2^{\bar n} \bar n! \sqrt{\pi} \tilde E^2 }  \sum_{k=0}^\infty (2k+1) (-1)^{k}\left(\eta\sin\theta\right)^{2k}  I_1(\bar n,2k) \, .
\end{equation}
Therefore, the final expression for the $V^{\bar n}_{-1,-1}$ matrix element will be
\begin{equation}
    V^{\bar n}_{-1,-1}  = \frac{\sin^2\theta}{ \tilde E^2 }  \sum_{k=0}^\infty \left (\frac{\eta\sin\theta}{2} \right )^{2k} \frac{\Gamma(2k+2)}{\Gamma(k+1)} F(-k,-\bar n;1;2)\, ,
\end{equation}
which for weak magnetic field can be expanded as
\begin{equation}
    V^{\bar n}_{-1,-1}  = \frac{\sin^2\theta}{ \tilde E^2 } \left ( 1 + (2\bar n +1) \, \frac{3}{2} \,\eta^2 \sin^2\theta + O(\eta^4)\right ) .
\end{equation}
The calculation of the other diagonal element $V^{\bar n}_{+1,+1}$ is practically identical to the one above. In fact, we have that
\begin{eqnarray}
V^{\bar n}_{+1,+1}  =  - \frac{\sin^2\theta}{2^{\bar n-1}\bar  (n-1)! \sqrt{\pi} } \int_{-\infty}^{\infty} d\xi \; e^{-\xi^2/2} H_{\bar n-1}(\xi) \frac{1}{ P^2_\xi} (e^{-\xi^2/2} H_{\bar n-1}(\xi)) \,,
\end{eqnarray}
from which we immediately read off the analogy with $V^{\bar n}_{-1,-1}$ and thus the following expression for $V^{\bar n}_{+1,+1}$
\begin{eqnarray}
    V^{\bar n}_{+1,+1} = - V^{\bar n-1}_{-1,-1} = -\frac{\sin^2\theta}{ \tilde E^2 }  \sum_{k=0}^\infty \left (\frac{\eta\sin\theta}{2} \right )^{2k} \frac{\Gamma(2k+2)}{\Gamma(k+1)} F(-k,-\bar n+1;1;2) \,,\;
\end{eqnarray}
or, in the weak field limit, where we have
\begin{equation}
    V^{\bar n}_{+1,+1}  \simeq -\frac{\sin^2\theta}{ \tilde E^2 } \left ( 1 + (2\bar n -1) \, \frac{3}{2} \,\eta^2 \sin^2\theta \right ) .
\end{equation}
We now shift our attention towards the non-diagonal terms. In particular, we start with the calculation of $V^{\bar n}_{-1,+1}$, which reads
\begin{eqnarray}
    V^{\bar n}_{-1,+1}  &= &  \frac{\sin \theta \cos\theta}{2^{\bar n-\frac12} (\bar  n-1)! \sqrt{ \bar n \pi} } \int_{-\infty}^{\infty} d\xi \; e^{-\xi^2/2} H_{\bar n}(\xi) \frac{1}{ P^2_\xi} (e^{-\xi^2/2} H_{\bar n-1}(\xi))  \\
    & =& \frac{-\sin \theta \cos\theta / \tilde E^2}{2^{\bar n-\frac12} (\bar  n-1)! \sqrt{ \bar n \pi} } \frac{d}{d A }\int_{-\infty}^{\infty} d\xi \; e^{-\xi^2/2} H_{\bar n}  \frac{1}{A+i \eta\sin\theta \partial_\xi} (e^{-\xi^2/2} H_{\bar n-1}) \bigg |_{A=1}  \,. \nonumber
\end{eqnarray}
Following the exact same steps as before, we arrive to
\begin{eqnarray}
    V^{\bar n}_{-1,+1} &=& \frac{\sin \theta \cos\theta / \tilde E^2}{2^{\bar n-\frac12} (\bar  n-1)! \sqrt{ \bar n \pi}} \sum _{k=0} (k+1) (-i \eta \sin\theta)^k I_2(\bar n , k) \\
    &=& -i \frac{\sin \theta \cos\theta / \tilde E^2}{2^{\bar n-\frac12} (\bar  n-1)! \sqrt{ \bar n \pi}} \sum _{k=0} (2k+2)  (-1)^k (\eta \sin\theta)^{2k+1} I_2(\bar n , 2k+1)\,,\nonumber
\end{eqnarray}
where here we introduced the integrals $I_2(\bar n ,k)$ defined in \eqref{i2def}, which are non-zero only for odd values of $k$. In the end, using Eq.~\eqref{i2expr}, we have
\begin{eqnarray} \label{vmpexprcompl}
   V^{\bar n}_{-1,+1} &=& i \sqrt{\frac{1}{2 \bar n}} \frac{\sin \theta \cos \theta}{\tilde E^2} \sum_{k=0}^{\infty} \left ( \frac{\eta\sin\theta}{2} \right )^{2k+1}  \frac{\Gamma(2k+3)}{\Gamma(k+1)} \times \nonumber\\
    && \times \left [ F(-k-1;-n;1;2)-F(-k;-n;1;2) \right]  \,. 
\end{eqnarray}
For our purposes, when working in the weak field approximation $\eta\ll1$, it will be sufficient to consider only the first term of the $k-$series, since the next one would already be order $\eta^3$, obtaining
\begin{eqnarray} 
   V^{\bar n}_{-1,+1}  \simeq  i  \eta \, \sqrt{2 \bar n }\,\frac{\sin^2 \theta \cos \theta}{\tilde E^2} \,.
\end{eqnarray}
Finally, emulating this same procedure for the last matrix element $V^{\bar n}_{+1,-1}$, we get
\begin{eqnarray}
    V^{\bar n}_{+1,-1} = \frac{\sin \theta \cos\theta / \tilde E^2}{2^{\bar n-\frac12} (\bar  n-1)! \sqrt{ \bar n \pi}} \sum _{k=0} (k+1) (-i \eta \sin\theta)^k I_3(\bar n , k) \,.
\end{eqnarray}
However, due to the above relation \eqref{i3toi2}, we just end up with the same result as in \eqref{vmpexprcompl} but with an additional minus sign
\begin{eqnarray} \label{vpmexprcompl}
   V^{\bar n}_{+1,-1} &=& - i \sqrt{\frac{1}{2 \bar n}} \frac{\sin \theta \cos \theta}{\tilde E^2} \sum_{k=0}^{\infty} \left ( \frac{\eta\sin\theta}{2} \right )^{2k+1}  \frac{\Gamma(2k+3)}{\Gamma(k+1)} \times \nonumber\\
    && \times \left [ F(-k-1;-n;1;2)-F(-k;-n;1;2) \right]  \\
    &\simeq& - i  \eta \, \sqrt{2 \bar n }\,\frac{\sin^2 \theta \cos \theta}{\tilde E^2} \,.
\end{eqnarray}

\subsection{Borel Regularization}\label{Borel}
In this section, we show in detail the procedure to obtain a Borel regularization for the infinite series Eq.~\eqref{a1n0sumk} defined in the main text
\begin{eqnarray} 
    a^{(1)} _{\bar 0,+1} &=& \frac{ \sin^2\theta}{ \sqrt{\pi} \tilde E^2 }  \sum_{k=0}^\infty (2k+1)(-1)^k \left(\eta\sin\theta \right)^{2k}  I_1(0,2k)\nonumber\\
    &=& \frac{ \sin^2\theta}{ \tilde E^2 } \sum_{k=0}^\infty \left(\frac{\eta\sin\theta}{2}\right)^{2k}\frac{\Gamma(2k + 2)}{\Gamma(k+1)} \, .
\label{eq_a11}
\end{eqnarray}
First, notice that, for $k$ a positive integer, the ratio of $\Gamma -$functions is
\begin{eqnarray}
\frac{\Gamma(2k+2)}{\Gamma(k+1)} = \frac{(2k + 1)!}{k!} \,.
\end{eqnarray}
Let us consider, with a simplified notation $z = \left(\eta\sin\theta/2\right)^2$, the equivalent series
\begin{eqnarray}
A(z) = \sum_{k=0}^{\infty}z^k \frac{(2k + 1)!}{k!}\, ,
\label{Aseries}
\end{eqnarray}
implying Eq.~\eqref{eq_a11} clearly corresponds to
\begin{eqnarray}
a^{(1)} _{\bar 0,+1} = \frac{\sin^2\theta}{\tilde{E}^2}A(\left(\eta\sin\theta/2\right)^2)\,.
\end{eqnarray}
The definition of the Borel transform of this series leads to the expression
\begin{eqnarray}
BA(zt) &=& \sum_{k=0}^{\infty}\frac{(z t)^k}{k!} \frac{(2k + 1)!}{k!} = \frac{1}{\left(1 - 4 z t \right)^{3/2}}\,,
\end{eqnarray}
where the second result directly follows as an identity from the Taylor expansion. The next step is to recover the regularized expression for the series $\tilde{A}(z) \simeq A(z)$ by performing the following integral transform
\begin{eqnarray}
\tilde{A}(z) &=& \int_0^{\infty}e^{-t}
BA(zt) \, dt = \int_0^{\infty}\frac{e^{-t}}{\left(1 - 4 z t \right)^{3/2}} \, dt\,.
\end{eqnarray}
In order to evaluate this integral, we perform the change of variables
$u = t - \frac{1}{4z}$, which implies $-1/(4z) \le u < \infty$, and $du = dt$. Therefore, we have
\begin{eqnarray}
\tilde{A}(z) &=& \frac{e^{\frac{-1}{4z}}}{(-4z)^{3/2}}\int_{-\frac{1}{4z}}^{\infty}u^{-3/2} e^{-u} du = \frac{e^{\frac{-1}{4z}}}{(-4z)^{3/2}}\Gamma\left(-\frac{1}{2},-\frac{1}{4 z} \right),
\end{eqnarray}
where in the final result we applied the definition of the Incomplete Gamma function
\begin{eqnarray}
\Gamma(s,x) = \int_{x}^{\infty}u^{s-1}e^{-u} du \,.
\end{eqnarray}
Note that this regularization method could be applied to all the series treated in Chapter \ref{ch2}, thus ensuring their convergence.

\newpage
\section[GEOMETRIC QUANTITIES FOR ACCELERATED OBSERVERS]{Geometric Quantities for Accelerated Observers}
Here, we include the expressions of several geometric quantities related to the metric in \eqref{rindlerelement}, recalling that we define $V\equiv 1+\frac{\vec a\cdot \vec x }{c^2}$ with $\vec a $ constant and homogeneous. First, we list the non-zero components of the Christoffel symbols
\begin{eqnarray} \label{christsymbols}
    \{^{\;\, 0}_{ 0 \,\, i}\} &=& \frac{1}{V}\partial_i V \,,\\
    \{^{\;\, i}_{ 0 \,\, 0}\} &=& V \partial_i V\,. \nonumber
\end{eqnarray}
As already mentioned in the main text, the components of the Riemann tensor are, instead, all zero. Then we move on to the calculation of the components of the spinor connection \eqref{Gammamudef}, which is straightforward. Starting from the temporal one, we have
\begin{equation}
    \Gamma_0 = \frac{1}{4} \sigma^{ab} g_{\alpha \beta} e_a^{\;\;\alpha} (\partial_t e_{ b}^{\;\;\beta} +\{^{\;\, \beta}_{ 0\, \rho}\} \, e_{ b}^{\;\;\rho}) = \frac{1}{4} \sigma^{ab} g_{\alpha \beta} \{^{\;\, \beta}_{ 0\, \rho}\} e_a^{\;\;\alpha}  \, e_{ b}^{\;\;\rho} \,.
\end{equation}
Using the expressions \eqref{christsymbols}, we obtain
\begin{eqnarray}
    \Gamma_0 &=& \frac{1}{4} \sigma^{0i} g_{00} \{^{\;\, 0}_{ 0\, \,i}\} e_0^{\;\;0}  \, e_{ i}^{\;\;i}+ \frac{1}{4} \sigma^{i0} g_{ii} \{^{\;\, i}_{ 0\,\, 0}\} e_i^{\;\;i}  \, e_{ 0}^{\;\;0} \\
    &=& \frac{1}{4V^2} \sigma^{0i} V^2 \partial_i V - \frac{1}{4V} \sigma^{i0} V \partial_i V \nonumber \\ 
    &=& \frac{1}{2} \sigma ^{0i} \partial_i V \,, \nonumber
\end{eqnarray}
where we also exploited the formulas \eqref{rindlertetrads} for the vierbein and the form of the metric \eqref{rindlerelement}. Finally, using the properties of flat gamma matrices, we end up with
\begin{equation}
    \Gamma_0 = \frac12 \sigma^{0i} \partial_i V = \frac{1}{2} \gamma^0 \gamma^i \partial_i V\,.
\end{equation}
Repeating the same steps for the spatial components of $\Gamma_\mu$ we arrive, instead, at
\begin{equation}
    \Gamma_i = 0\,,
\end{equation}
meaning that only the time component of the spinor covariant derivative $\nabla_0 \psi$ gets a modification from the non-inertial geometry. \label{app:rindlerquantities}

\newpage
\section[IMPORTANT EXPRESSIONS FOR VERY SPECIAL LINEAR GRAVITY]{Important Expressions for Very Special Linear Gravity}
To avoid excessive clutter in the main text, we include in this appendix some of the relevant expressions and calculations presented in Chapter \ref{ch4}.

\subsection{VSLG Propagator} \label{appLagandFeyn}

From the Lagrangian \eqref{lagrdimensionfulh} and an additional gauge-fixing term \cite{Santoni:2023uko}
\begin{equation}
    \mathcal{L}_{GF}=\xi \, \partial_{\mu} \left( h^{\mu \nu} - \frac{1}{2}
\eta^{\mu \nu} h \right) \partial^{\lambda} \left( h_{\lambda \nu} -
\frac{1}{2} \eta_{\lambda \nu} h \right)  ,
\end{equation}
we can derive the momentum-space expression of the graviton propagator, which reads 
\begin{eqnarray}
i \tilde{\mathcal{O}}_{\rho \sigma \alpha \beta}^{- 1} &=& \frac{i P_{\rho \sigma \alpha
\beta}}{k^2 - m^2_g} + \frac{i m^4_g}{2 \, k^4 (k^2 - m^2_g)} \Lambda_{\rho \sigma} \Lambda_{\alpha
\beta} \\
& & - \frac{i \,m^2_g / 2 k^2}{k^2 - m^2_g} \bigg(\Lambda_{\rho \alpha} \eta_{\sigma
\beta} + \Lambda_{\rho \beta} \eta_{\sigma \alpha} + \Lambda_{\sigma \alpha}
\eta_{\rho \beta} + \Lambda_{\sigma \beta} \eta_{\rho \alpha} - \Lambda_{\rho
\sigma} \eta_{\alpha \beta} - \Lambda_{\alpha \beta} \eta_{\rho \sigma} \bigg)\nonumber\\
& & - \frac{i \, m^2_g/ k^2}{k^2 - m^2_g} \bigg[ \frac{(\tilde N_{\rho} k_{\sigma}
+ \tilde N_{\sigma} k_{\rho})(\tilde N_{\beta} k_{\alpha} + \tilde N_{\alpha} k_{\beta})}{2} -
\tilde N_{\rho} \tilde N_{\sigma} k_{\alpha} k_{\beta} - \tilde N_{\alpha} \tilde N_{\beta} k_{\rho}
k_{\sigma} \bigg] , \label{propagator} \nonumber
\end{eqnarray}
where we set $\xi=1$ and defined
\begin{eqnarray}
P_{\rho \sigma \alpha \beta} \equiv \frac{1}{2} (\eta_{\rho \alpha} \eta_{\sigma
\beta} + \eta_{\rho \beta} \eta_{\sigma \alpha} - \eta_{\rho \sigma}
\eta_{\alpha \beta}) \, , \,\,\, \Lambda_{\rho \sigma} \equiv k_{\rho} \tilde N_{\sigma} +
k_{\sigma} \tilde N_{\rho} - \tilde N_{\rho} \tilde N_{\sigma} k^2 \, .\nonumber
\end{eqnarray}
We observe that we recover the conventional propagator in the limit $m_g\to 0$ \cite{Donoghue:2017pgk}.

\subsection{Polarization Sum in VSLG} \label{appPolar}

Here, we include some details on the calculation of the polarization sum $S^{\mu\nu\alpha\beta}$ in VSLG. For that, we exploit its covariance to determine the allowed structures: the available tensorial “building blocks" from which we start are the flat metric $\eta_{\mu\nu}$, the four-momentum $k^\mu$ and the four-vector $\tilde N^\mu $. Simbolically, we obtain
\begin{eqnarray}
    S &=& 3\, \eta\eta  + 6\, \eta \, k k + + 6\, \eta \tilde N \tilde N + 6\, k k \tilde N \tilde N + 12\, \eta \, k \, \tilde N \nonumber \\
    &&+ 4\, k k k \tilde N + 4\, k \tilde N \tilde N \tilde N  + k k k k + \tilde N \tilde N \tilde N \tilde N\, ,
\end{eqnarray}
where the numbers represent the different indices combinations for each possible tensorial structure. At this point, from the gauge conditions already imposed in VSLG, we deduce the following constraints 
\begin{equation}
\begin{cases}
k_\mu S^{\mu\nu\alpha\beta} = 0 \,, \\
n_\mu S^{\mu\nu\alpha\beta} = 0 \,, \\
\eta_{\mu\nu} S^{\mu\nu\alpha\beta} = 0 \, .
\end{cases} 
\end{equation}
Pairing this conditions with the one deriving from normalization $S^{\mu \nu}_{\;\;\;\;\mu \nu}=2$, which ensures the correct limit for $m_g\to 0$, and the index symmetries $\mu \Longleftrightarrow \nu$, $\alpha \Longleftrightarrow \beta$, $\mu \nu \Longleftrightarrow \alpha \beta$, we finally find the expression
\begin{eqnarray} \label{midexprS}
 S^{\mu \nu \alpha \beta} &=& P^{\mu\nu\alpha\beta} +\frac12 \eta^{\alpha\beta} \tilde N^\nu k^\mu -\frac12 \eta^{\nu\beta} \tilde N^\alpha k^\mu -\frac12 \eta^{\nu\alpha} \tilde N^\beta k^\mu + \frac12 \eta^{\alpha\beta} \tilde N^\mu k^\nu -\frac12 \eta^{\mu\beta} \tilde N^\alpha k^\nu \nonumber \\
 &&  -\frac12 \eta^{\mu\alpha} \tilde N^\beta k^\nu -\frac12 \eta^{\nu\beta} \tilde N^\mu k^\alpha  -\frac12 \eta^{\mu\beta} \tilde N^\nu k^\alpha + \frac12 \eta^{\mu\nu} \tilde N^\beta k^\alpha -\frac12 \eta^{\nu \alpha} \tilde N^\mu k^\beta \nonumber \\
 &&  - \frac12 \eta^{\mu\alpha} \tilde N^\nu k^\beta +\frac12 \eta^{\mu\nu} \tilde N^\alpha k^\beta +\tilde N^\mu \tilde N^\nu k^\alpha k^\beta +k^\mu k^\nu \tilde N^\alpha \tilde N^\beta \nonumber +\frac{m^2_g}{2} \eta^{\nu\beta} \tilde N^\mu \tilde N^\alpha  \\
 &&  +\frac{m^2_g}{2} \eta^{\mu\beta} \tilde N^\nu \tilde N^\alpha +\frac{m^2_g}{2} \eta^{\nu \alpha} \tilde N^\mu \tilde N^\beta + \frac{m^2_g}{2} \eta^{\mu\alpha} \tilde N^\nu \tilde N^\beta  - \frac{m^2_g}{2} \eta^{\mu\nu} \tilde N^\alpha \tilde N^\beta\nonumber \\
 &&  -\frac{m^2_g}{2} \eta^{\alpha\beta} \tilde N^\mu \tilde N^\nu -\frac{m_g^2}{2} \tilde N^\mu \tilde N^\nu \tilde N^\alpha k^\beta -\frac{m_g^2}{2} \tilde N^\mu \tilde N^\nu \tilde N^\beta k^\alpha \nonumber \\
 &&   -\frac{m_g^2}{2} \tilde N^\mu \tilde N^\alpha \tilde N^\beta k^\nu -\frac{m_g^2}{2} \tilde N^\nu \tilde N^\alpha \tilde N^\beta k^\mu + \frac{m_g^4}{2} \tilde N^\mu \tilde N^\nu \tilde N^\alpha \tilde N^\beta\,,
\end{eqnarray}
where we also had to consider the on-shell condition $k^2=m^2_g$ for the graviton polarization tensors. This result coincides with what we would have obtained taking $(k^2-m^2_g) \tilde{\mathcal O} _{\rho\sigma \alpha\beta}^{-1}$ on-shell, because the polarization sum is related to the $h-$propagator \cite{Poddar:2021yjd}.
\\ 
Since all indices of $S^{\mu\nu\alpha\beta}$ are contracted with EMTs, the terms proportional to $k^\mu$ are canceled due to energy-momentum conservation $k^\mu \tilde T_{\mu\nu} = 0$. Thus, including only relevant terms, the expression \eqref{midexprS} becomes
\begin{eqnarray}\label{polsum} 
 S^{\mu \nu \alpha \beta} &=& \frac12 (\eta^{\mu \alpha} \eta^{\nu \beta}+ \eta^{\mu \beta} \eta^{\nu \alpha} - \eta^{\mu\nu} \eta^{\alpha\beta}) + \frac{m^2_g}{2} (\eta^{\nu\beta} \tilde N^\mu \tilde N^\alpha + \eta^{\mu\beta} \tilde N^\nu \tilde N^\alpha + \eta^{\nu \alpha} \tilde N^\mu \tilde N^\beta )  \nonumber\\
 &&+ \frac{m^2_g}{2} (\eta^{\mu\alpha} \tilde N^\nu \tilde N^\beta - \eta^{\mu\nu} \tilde N^\alpha \tilde N^\beta - \eta^{\alpha\beta} \tilde N^\mu \tilde N^\nu ) + \frac{m_g^4}{2} \tilde N^\mu \tilde N^\nu \tilde N^\alpha \tilde N^\beta  \,. 
\end{eqnarray}\\

\subsection{Explicit Expressions of $f(N,\delta,e,\hat n)$} \label{genericfcalc}

In this subsection, we collect some details on the final expression assumed by the function $f(N,\delta,e,\hat n)$ defined in \eqref{fnn0e}. We recall that $f$ is generally given by two contributions: $f_{//}$ and $f_{\perp}$
\begin{equation}
    f(N,e,\delta ,\hat n)=f_{//}(N,\delta, e )+ f_\perp(N,\delta, e ,\hat n) \, .
\end{equation}
Being independent of $\hat n$, the $f_{//}-$term is always present. On the other hand, $f_{\perp}$ is relevant only when $\hat n$ is not parallel to the $z-$direction. This fact directly implies that $f_{//}$ should reduce to the GR contribution in the limit of $m_g\to0$, while $f_{\perp}$ would just vanish. Hence, the simplest expression for $f$ is clearly obtained in the case where $\hat n$ is orthogonal to the orbital plane. In the following, we include the explicit expression for both 
\begin{eqnarray} \label{f//fperp}
    \frac{f_{//}(N,e,\delta)}{\sqrt{1-\frac{\delta^2}{N^2}} } &=&  \left [ 1 + \frac{\delta^2}{6 N^2} \left (\frac{91}{2} +\frac{16 \delta^2}{ N^2} -15 \frac{4 + \frac{3\delta^2}{N^2}}{2\sqrt{1-\frac{\delta^2}{N^2}}} \tanh^{-1}\sqrt{1-\frac{\delta^2}{N^2}} \right ) \right ] g_N  \nonumber \\
    && - \frac{ \delta^2 }{48} \left (\frac{25}{6} -\frac{80 \delta^2}{3 N^2} - 5 \frac{ 4 - \frac{15\delta^2}{N^2}}{2\sqrt{1-\frac{\delta^2}{N^2}}} \tanh^{-1}\sqrt{1-\frac{\delta^2}{N^2}}  \right ) J^2_N \,,\\
    \frac{f_{\perp}(N, e, \delta , \hat n)}{\frac{5 N^2 \delta ^2}{384} \sqrt{1-\frac{\delta^2}{N^2}}} &=&   
    - \left (23 +\frac{16 \delta^2}{ N^2} - \frac{12 + \frac{21\delta^2}{N^2}}{\sqrt{1-\frac{\delta^2}{N^2}}} \tanh^{-1}\sqrt{1-\frac{\delta^2}{N^2}} \right ) \hat n^i \hat n^k L^{*\,ij}_N L^{kj}_N \nonumber \\ 
    && 
    +\left (\frac{37}{4} +\frac{20 \delta^2}{ N^2} - \frac{3 + \frac{99\delta^2}{ 4 N^2}}{ \sqrt{1-\frac{\delta^2}{N^2}}} \tanh^{-1}\sqrt{1-\frac{\delta^2}{N^2}} \right )n^i \hat n^j \hat n^k \hat n^l L^{*\,ij}_N L^{kl}_N \nonumber\\
    &&
    -\frac{2  J_N }{N} \left (13 -\frac{16 \delta^2}{ N^2} - \frac{12 - \frac{21\delta^2}{N^2}}{\sqrt{1-\frac{\delta^2}{N^2}}} \tanh^{-1}\sqrt{1-\frac{\delta^2}{N^2}} \right )  (\hat n^i)^2 \, L^{ii}_N  \, , \nonumber 
\end{eqnarray}
with $g_N \equiv g(N,e)$ being the same function defined in \cite{peters1963gravitational} 
\begin{eqnarray} \label{gNe}
    g(N,e) &=& \frac{N^4}{32} \left \{ (J_{N-2} -2 e J_{N-1} + \frac2N J_N +2 e J_{N+1}-J_{N+2})^2 \right. \\
    && \left.\;\;\;\;\;\;\;\;\;+ (1-e^2)(J_{N-2} -2 J_N +J_{N+2} \,)^2 +\frac{4}{3N^2} J^2_N \right \} \,, \nonumber 
\end{eqnarray}
where $N e$ is the argument of all the above $J-$functions, and with
\begin{eqnarray}
    \hat n^i \hat n^j L^{*\,ik}_N L^{jk}_N &=& \sin^2\theta (\cos^2\phi \, (L^{xx}_N)^2 + \sin^2 \phi \,(L^{yy}_N)^2) + \sin^2\theta |L^{xy}_N|^2 \,, \nonumber\\
    n^i \hat n^j \hat n^k \hat n^l L^{*\,ij}_N L^{kl}_N &=& \sin^4\theta (\cos^2\phi \, L^{xx}_N + \sin^2 \phi\, L^{yy}_N)^2 +4 \sin^4 \theta \sin^2\phi \cos^2\phi |L^{xy}_N|^2 \,, \nonumber \\
     (\hat n^i)^2 \, L^{ii}_N  &=& \sin^2\theta (\cos^2\phi \, L^{xx}_N + \sin^2 \phi\, L^{yy}_N) \,.
\end{eqnarray}

\subsubsection{Behavior of $f_{//}(N,e,\delta)$} \label{behaviorfpar}

\begin{figure}[h!!]
\centering
 \includegraphics[width=10.2cm]{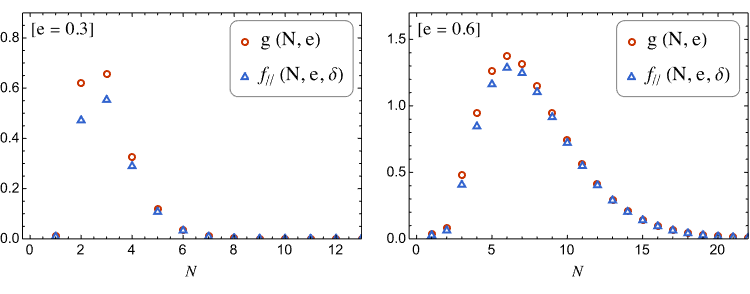}
    \caption[Comparison between $f_{//}(N,e,\delta)$ and $g(N,e)$ in function of $N$]{Comparison between $f_{//}(N,e,\delta)$ and $g(N,e)$ versus $N$, for the values $e=0.3$ and $e=0.6$, taking $\delta=0.5$.}
    \label{fig:f//vsN}
\end{figure}

\begin{figure}[h!!]
 \includegraphics[width=7cm]{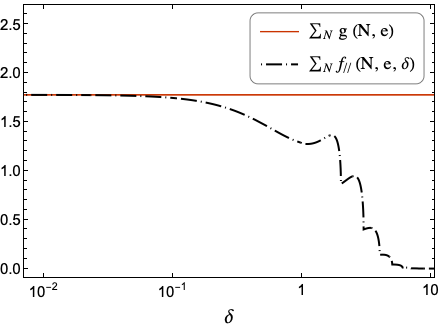}
    \caption[Behavior of $\sum_{N_{min}} f_{//}(N,e,\delta)$ and $\sum_{N=0} g(N,e)$ in function of $\delta$]{Behavior of $\sum_{N_{min}} f_{//}(N,e,\delta)$ and $\sum_{N=0} g(N,e)$ in function of $\delta$, for the value of $e=0.3$. The horizontal axis is represented in logarithmic scale. It is possible to see that already for $\delta \sim 10$ the high-intensity $N$-values have been excluded from the sum, implying a crucial drop in the total emitted radiation.}
    \label{fig:F//vsD}
\end{figure}

When considering only $f_{//}$, the discrepancy obtained with respect to GR is always negative, resulting in a reduction in the final period decrease rate. This effect can also be appreciated in Fig.\ref{fig:f//vsN}, where we compare the behavior of $f_{//}$ and $g$ as functions of $N$, for $\delta=0.5$ and for two different illustrative eccentricity values.
\\
From these graphs, we recognize the fast convergence behavior of $f_{//}$ when increasing $N$. This feature allows us to cut off the $N-$summation (around $ \sim 100$) much before the upper limit given in \eqref{omegaupper}, which for typical binaries' velocities would give $N_{max} \sim 1000$. However, since the $N-$value of the intensity peak of $f_{//}$ increases with $e$, one should be careful about contemplating this approximation with binaries of very high eccentricity. There is also another interesting effect coming from the essence of $N_{min}$ in the sum: if $m_g$ is not exactly zero in Nature, then there should exist binary systems for which $\delta>1 $, implying the modes with $N< N_{min}$ would be excluded from the sum, leading to a drop in the emission intensity, as shown in Fig.\ref{fig:F//vsD}, where to each integer value of $\delta$ corresponds a dip in magnitude. Due to the horizontal shift of the intensity peak of $f_{//}$ when increasing $e$, this effect is more easily noticeable for small eccentricities. This fact may lead to the possibility of placing better constraints on the mass $m_g$ from measurements of the period decrease rate for binaries with longer periods and smaller eccentricities. For example, see Fig.\ref{fig:F//vsD} where, by taking larger values of the period $P_b$ and therefore increasing $N_{min}$, the effect of progressively excluding the first $N-$values from the sum becomes more and more evident.
\\
Note that even if we have not seen any mention about it in literature, this is a phenomenon that should affect not only the VSR realization of linearized massive gravity but other formulations too (like \cite{Poddar:2021yjd}), since it is simply derived from considerations on the $\omega-$integration range.

\subsubsection{Behavior of $f_\perp(N,\delta, e ,\hat n)$}

\begin{figure}[h!!]
 \includegraphics[width=10.5cm]{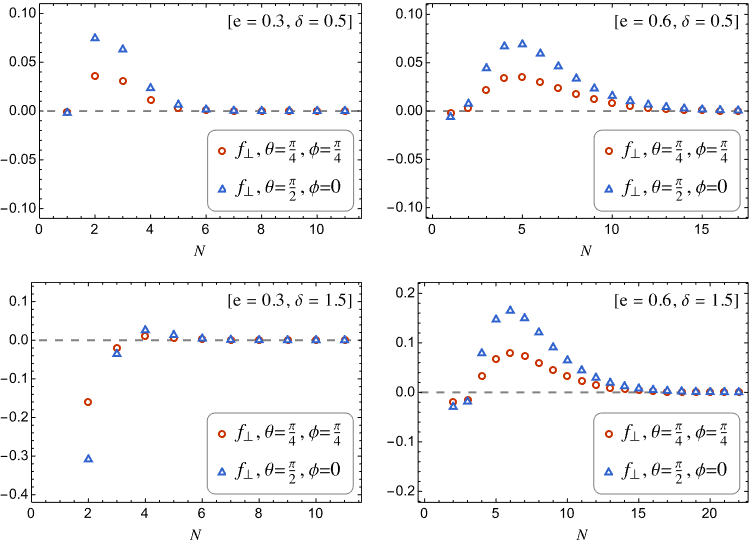}
    \caption[Behavior of $f_{\perp}$ in function of $N$]{Behavior of $f_{\perp}$ in function of $N$, for different values of $\{ e , \delta, \theta , \phi\}$. Here, the qualitative behavior does not change much by varying $\{\theta,\phi\}$, it just scales the magnitude of the VSR effects.}
    \label{fig:fperpvsN}
\end{figure}

\begin{figure}[h!!]
 \includegraphics[width=7.5cm]{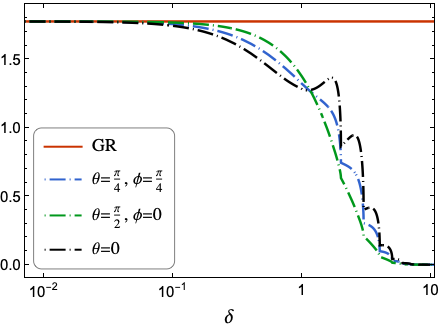}
    \caption[Curves produced by $\sum_{N_{min}} f$ in function of $\delta$]{Comparison of the curves produced by $\sum_{N_{min}} f$ in function of $\delta$, for different values of $\{ \theta , \phi\}$ and with fixed eccentricity $e=0.3$. The horizontal axis is represented in logarithmic scale for convenience.}
    \label{fig:FperpvsD}
\end{figure}

By plotting $f_{\perp}$ versus $N$ for some indicative values of $\{ \delta , e , \theta , \phi\}$, we notice that it is mostly close to zero or positive as long as $\delta$ is small, while for the first few available $N-$values it undergoes a progressive shift to negative values as $\delta$ increases. This feature is shown, for example, in Fig.\ref{fig:fperpvsN}. Thus, in contrast to the VSR contribution $f_{//}-g$, which was always negative, $f_{\perp}$ does not generally have a definite sign.
\\
Another interesting fact that can be graphically appreciated is that, starting from small values of $\delta$, the $f-$contribution in the case $\hat n//\hat z$ is always below its more generic counterpart with $\hat n$ oriented differently. This situation changes when increasing $\delta$, in which case the hierarchy gets turned over, as we can see in Fig.\ref{fig:FperpvsD}.
However, this confirms once again that, for small values of $\delta <1 $, the largest discrepancy with respect to GR is obtained for the case $\hat n // \hat z$.

\newpage

\end{document}